\documentclass{article}

\usepackage{PRIMEarxiv}

\usepackage[utf8]{inputenc} 
\usepackage[T1]{fontenc}    
\usepackage{hyperref}       
\usepackage{url}            
\usepackage{booktabs}       
\usepackage{amssymb}       
\usepackage{amsmath}
\usepackage{nicefrac}       
\usepackage{microtype}      
\usepackage[square, numbers, sort&compress]{natbib}
\usepackage{tabularx}
\usepackage{fancyhdr}       
\usepackage{graphicx}       
\usepackage{longtable}
\usepackage{multirow}
\usepackage{placeins}

\newcolumntype{R}{>{\hsize=\hsize}X}
\newcolumntype{Q}{>{\hsize=.5\hsize}X}
\newcolumntype{Y}{>{\centering\arraybackslash}X}
\title{Modeling multi-agent motion dynamics\\in immersive rooms
}

\author{
  Jerry M. Huang and Stefan T. Radev \\
  Department of Cognitive Science \\
  Rensselaer Polytechnic Institute \\
  Troy, New York, United States\\
  \texttt{\{huangm13, radevs\}@rpi.edu} \\
}

\begin{document}
\maketitle

\begin{abstract}
Immersive rooms are increasingly popular augmented reality systems that support multi-agent interactions within a virtual world. However, despite extensive content creation and technological developments, insights about perceptually-driven social dynamics, such as the complex movement patterns during virtual world navigation, remain largely underexplored. Computational models of motion dynamics can help us understand the underlying mechanism of human interaction in immersive rooms and develop applications that better support spatially distributed interaction. In this work, we propose a new agent-based model of emergent human motion dynamics. The model represents human agents as simple spatial geometries in the room that relocate and reorient themselves based on the salient virtual spatial objects they approach. Agent motion is modeled as an interactive process combining external diffusion-driven influences from the environment with internal self-propelling interactions among agents. Further, we leverage simulation-based inference (SBI) to show that the governing parameters of motion patterns can be estimated from simple observables. Our results indicate that the model successfully captures action-related agent properties but exposes local non-identifiability linked to environmental awareness. We argue that our simulation-based approach paves the way for creating adaptive, responsive immersive rooms—spaces that adjust their interfaces and interactions based on human collective movement patterns and spatial attention.
\end{abstract}


\section{Introduction}
\label{sec:intro}

\begin{figure}[t]
    \centering
    \includegraphics[width=0.99\linewidth]{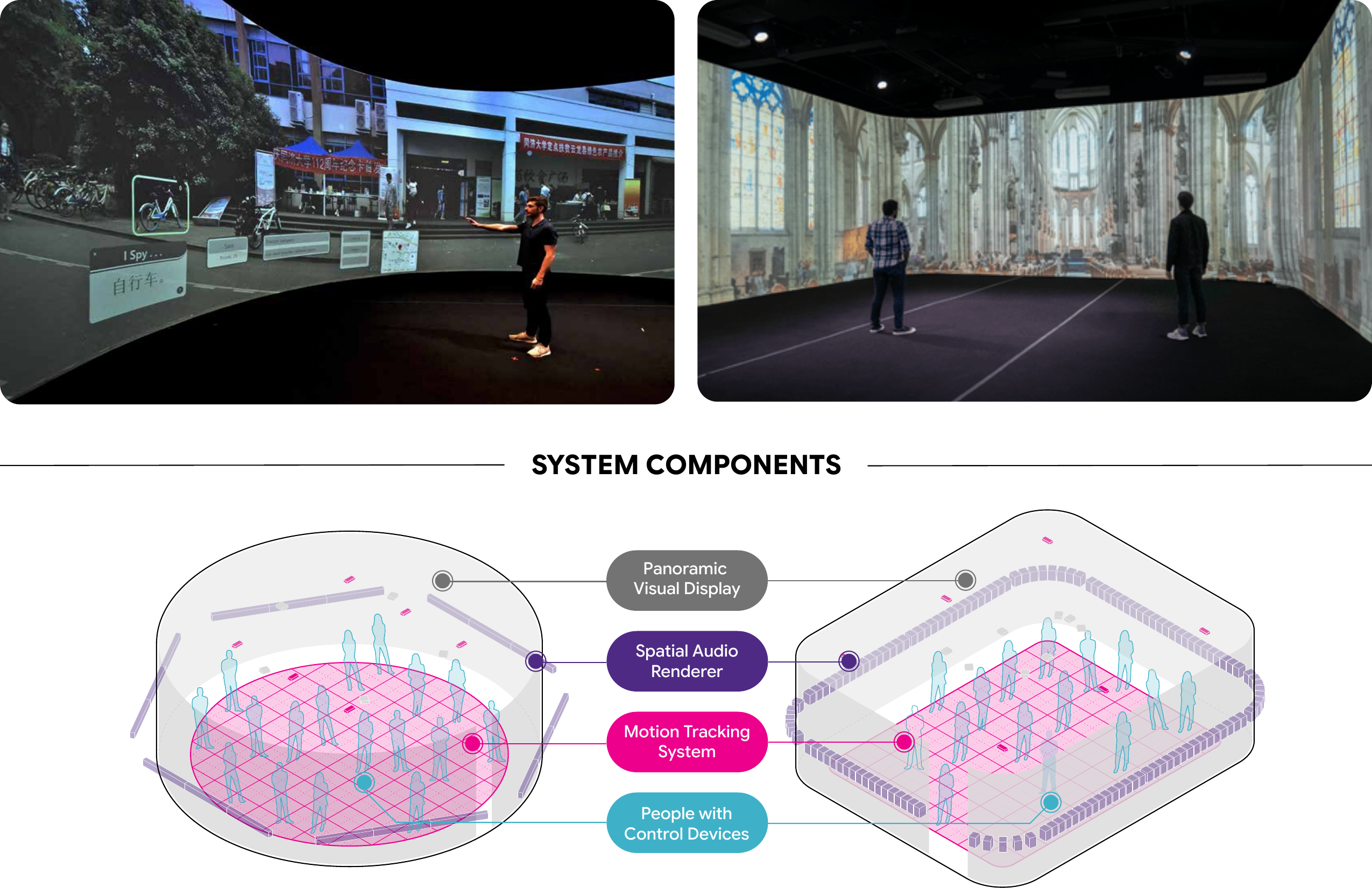}
    \caption{\textit{Two examples of immersive rooms} \cite{huang2023rois}. These immersive rooms share common hardware components, with panoramic visual displays, spatially-situated speaker systems, and motion tracking systems. Both immersive rooms can host multiple agents, but they differ in terms of the spatial configuration of hardware components.}
    \label{fig:rooms}
\end{figure}

\noindent Spatial augmented reality is becoming increasingly commonplace and easier to integrate into the built environment. For example, immersive rooms, a type of virtual reality system evolved from the Cave Automatic Virtual Environment \cite[CAVE;][]{cruzneira1992}, can now support collocated and multi-user virtual world experiences. In some of the latest developments, the system consists of a room-scale panoramic visual display, a multi-channel spatial audio loudspeaker system, an overhead motion tracking system, and controllers that enable navigation and interaction within the virtual world \cite[\autoref{fig:rooms}]{sharma2017interactions}. When multiple people reside in the immersive room, the extended virtual world provides a shared sense of auditory and visual immersion that people can freely experience within the room's enclosure.

At the same time, spatial computing technologies have advanced significantly, expanding the scope and sophistication of interactive environments \cite{kucheramorin2014, chen2015}. 
As a result, immersive rooms now support a wide range of applications, including data visualization \cite{kuhlen2014}, situated learning \cite{chabot2020}, and life-size telepresence \cite{jones2014roomalive, pejsa2016room2room}, demonstrating their potential to reshape how people experience and inhabit built spaces. These capabilities are further extended by modern virtual environment platforms, such as game engines, which traditionally support head-mounted displays (HMDs).~In HMD-based experiences, the virtual environment typically isolates an individual user from the physical world. In contrast, immersive rooms allow multiple users to share the same visual and auditory environment, supporting social interaction without obstructing sensory awareness of their co-present participants.

While spatial data on human movement in immersive rooms can be captured using motion-sensing systems, such data can also be simulated within the virtual environment. Yet, due to the absence of comprehensive modeling frameworks, interpreting these data remains challenging. Most existing applications use motion data to enable interaction but implicitly assume that users will adapt to the technology. Consequently, they seldom leverage these data to infer the behavioral patterns of its users, especially when they are perceptually-driven. Although tracking data can reveal motion dynamics at the individual level \cite{sanz2015}, such insights do not readily generalize to groups, limiting the development of robust human-centered interaction strategies.

Meanwhile, research on complex social dynamics has revealed a lot about how spatially-grounded perception of individual's surroundings shape collective self-organization as physical crowds \cite{warren2024crowds}. Nevertheless, far less is known about how such dynamics unfolds when people engage with a shared built environment, even less so in the shared virtual world. From the perspective of interaction design, behavioral data offers critical information for shaping technologies that can accommodate human action and social engagement. In immersive rooms (or architectural environments more broadly) these data can illuminate the ``activity space'' for human-centered interaction design, especially in multi-user interaction settings \cite{kirsh2019architects, kirsh2025reimagining, huang2020immersion}. Accordingly, a formal modeling approach is necessary to draw connections between raw behavioral data and the higher-level spatial and social features that underpin the dynamics of human interactions. 
\begin{figure}[t]
    \centering
    \includegraphics[width=\linewidth]{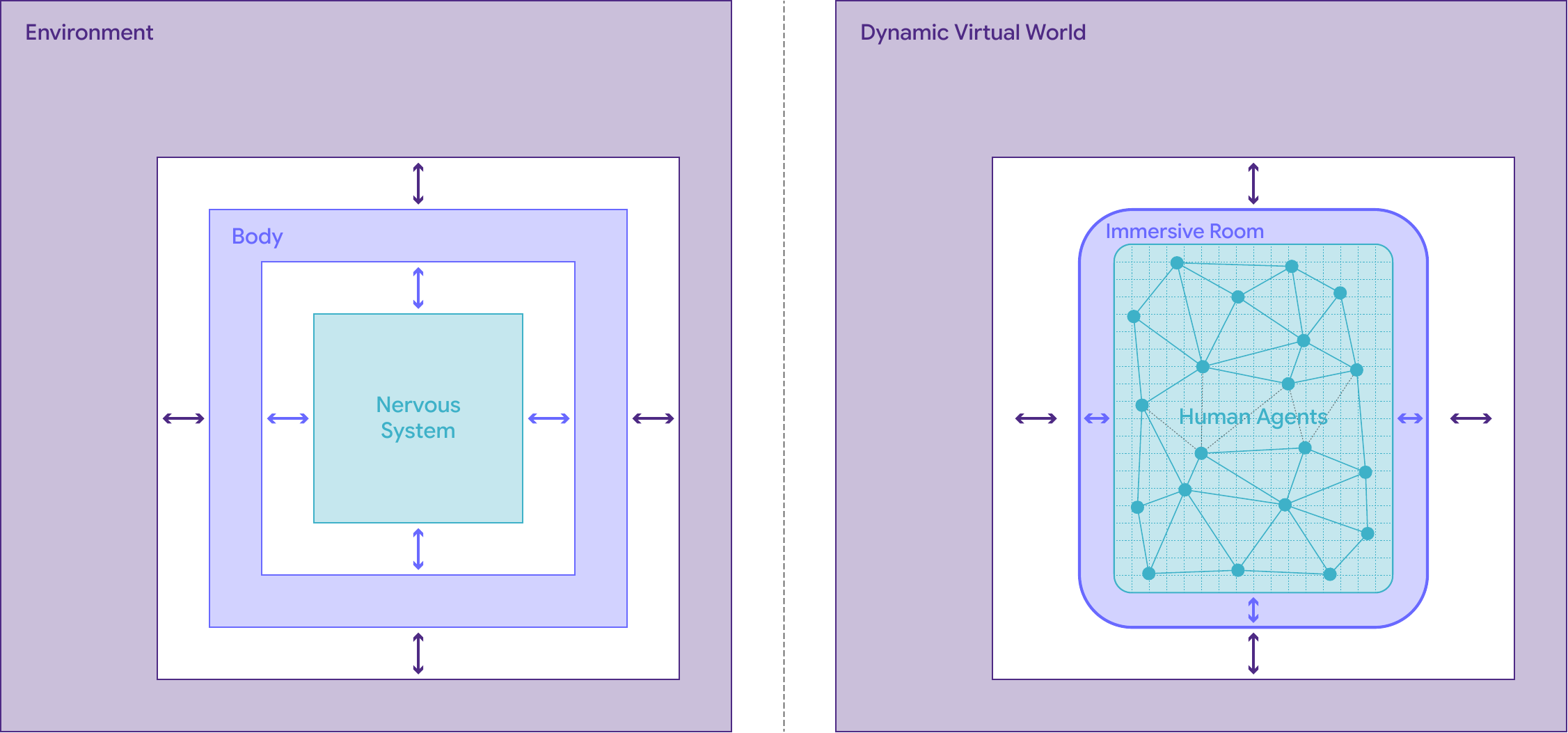}
    \caption{\textit{Dynamical-system analogy of agent-environment interaction in immersive rooms.} \textit{Left}: the original coupled-system perspective reproduced from \cite{beer2000}; \textit{right}: the same perspective applied to immersive rooms, serving as the working hypothesis that interaction between human agents represents the internal state of immersive rooms in dynamic virtual environments.}
    \label{fig:dynamical-analogy}
\end{figure}

In this work, we develop an agent-based model \cite[ABM;][]{bonabeau2002agent, demarchi2014agent, dignum2025simulation} of perceptually-driven motion dynamics in immersive rooms. The rationale behind the model draws on a dynamical system perspective that frames the three-way relationship between the immersive room, its agents, and the extended environment. For a single human agent, the nervous system, the body, and the environment form a coupled dynamical system, from which agent action emerges and stabilizes through its internal states governing perception \cite{beer2000}. By analogy, if we consider the immersive room as a navigational ``body'' in the virtual world, the spatially distributed interaction between human agents within the room and the extended environment may be interpreted as an expression of the room’s internal state (see \autoref{fig:dynamical-analogy}). With this analogy as our working hypothesis, our secondary goal is to identify latent parameters from the time course of agents' spatial movement in the immersive room. These parameters capture how agents process perceptual information about their virtual surroundings and respond both as individuals and as participants in a shared social context.

The present study focuses on systematic model development and estimation within a simulated setting. From a technical standpoint, as the immersive room presents a novel context for computational modeling, simulation serves as a gateway to theory development, model instantiation, and knowledge discovery \cite{lavin2021simulation}. Accordingly, we employ simulation-based inference \cite[SBI;][]{cranmer2020frontier}, and more specifically amortized Bayesian inference \cite[ABI;][]{radev_bayesflow_2023}, to analyze human spatial data in a simulated immersive room. 
We combine an agent-based modeling (ABM) approach with diffusion models \cite[DM;][]{ratcliff2016diffusion} of decision-making to simulate users’ motion patterns in immersive environments, represented as time-series trajectories of virtual-world exploration.
We then conduct an extensive simulation study using ABI to evaluate the identifiability of parameters governing agents’ movement patterns and self-organization, including multiple ways to extend our model for greater simulation realism. Taken together, the novelty of this work is uniquely situated in the integration of ABM in behavioral modeling, the use of ABI for inverse modeling.

\section{Background}
\label{sec:background}

As intrinsically motivated autonomous agents, humans in an immersive room can explore the virtual world independently or by following social cues. For instance, in a virtual museum with digital sculptures and paintings, an agent in the immersive room might approach an exhibit on the basis of their personal interest, or instead follow their peers to view an alternative display. As a result, their locomotion is shaped both via \textit{global} interactions with their surrounding environments and \textit{local} interactions with neighboring individuals. This setting has characteristics of agent-based systems, where global patterns emerge from local interactions. It also links to diffusion-based decision processes that are stimulus-driven and unconstrained by group-level dynamics. Both modeling approaches have well-established foundations in theory and computation. Below, we briefly review the relevant background literature and methods.

\subsection{Agent-Based Models of Collective Motion Dynamics}
\label{ssec:collective-motion}

\noindent
Agent-based models (ABMs) describe emergent properties of self-motivated agents in an environment, capturing the spatiotemporal dynamics of their actions as a social system \cite{bonabeau2002agent}. Agents in the environment follow local computational rules that subject them to the influence of their surrounding agents. This characteristic underscores the essence of collective motion dynamics in natural environments, from crowds \cite{warren2018} to bird flocks \cite{bill2022} to particles under the influence of force fields \cite{vicsek1995, bechinger2016}. 

ABMs have become a popular choice for high-fidelity computer simulations where behaviors of animal groups can be simulated \textit{en masse} and rendered in early computer graphics \cite{reynolds1987}. In addition, ABMs have also been adopted in earlier works to simulate agent behaviors in architectural environments \cite{yanandkalay2006}. Other classical ABMs, such as the social force model \cite{helbing1998social, truong2017toward} and inverse reinforcement learning model \cite{kretzschmar2016socially}, have been used in a navigational context for developing applications from urban design to robotics. In these precedences, the agents are represented as simple geometric abstractions, with their motion rule driven by real-time spatial information of their surrounding neighbors given limited sensing capabilities about the environment. 

In ABMs, an agent's active perception involves sensing the surrounding neighbors as part of the environment. In collective motion, this perceptual coupling allows the agents to 1) align with their neighboring agents' instantaneous headings based on the difference in the available visual information between those who lead and those who follow \cite{dachner2022visual, dachner2014behavioral}, and 2) maintain spatial cohesion by regulating their mutual distances for collision avoidance and movement sustenance \cite{warren2018, wirth2023}. Stable simulation of motion patterns using ABMs require several foundational assumptions. First, the agents are modeled with shared internal properties, often with identical motion speeds and sensing capacities. Second, within the overall agent group, the agents are not self-guided, meaning that motion adjustments depend entirely on their spatial relationship with their immediate neighbors. Finally, the environment in which agents operate is treated as homogeneous, meaning that external influences are uniform across the agents.

For enhanced ecological validity, one or more of the above assumptions can be altered to introduce nuances to simulation. For instance, the agents can be equipped with a structured motion perception of the global environment, from which individual differences emerge from temporal variability in neighbor following and velocity changes \cite{bill2022}. 
Likewise, when agents operate in heterogeneous environments (i.e., an urban neighborhood with many landmarks), we can introduce variability to their local interaction rules, while the environment's global structure influences the agents' perception \cite{toner2018walking}. These cases illustrate the inherently hierarchical structure of ABMs, where agents can share global interaction properties while maintaining individual differences. 

In our model, the virtual world only extends beyond the physical boundaries of immersive room. As a result, the room's interior is free of virtual contents and is treated as a homogeneous spatial environment for agent exploration. Agents interact locally with nearby neighbor, with their physical awareness determined solely by population density and rules of proxemics. Such configuration draws parallel with real-world pedestrians, wherein individuals can notice and respond to different points of interest in their surroundings.
While the room’s physical boundaries ultimately constrains where agents can move, the global navigation trajectory of the immersive room in the virtual world drives their motion, serving as a shared navigational substrate rather than an individualized frame of reference. This content sharing eliminates their tendency of staying at one place in the room, allowing them to interact with different neighbors over time.

\subsection{Diffusion Process as Individual Motion Guidance}
\label{ssec:individual-agent}

As suggested above, agents navigating in the immersive room can respond to salient spatial features in the extended virtual world. Thus, agent movement can be modeled primarily as a bottom-up, stimulus-driven process, where motor behaviors arise from online perception of dynamic auditory and visual cues \cite{warren2006dynamics, dinocera2014}. In this simplified environment, an agent's tendency to approach spatial objects depends solely on spatial relationships such as proximity and perceptual salience. Consequently, multi-agent motion emerges from vectors of self-guidance.

The perceptual motivation for the agents' movement in the immersive room can be derived from many driving factors in the extended virtual world, including motion onset \cite{abrams2003}, cross-modal interaction \cite{prime2010}, and spatial tracking behaviors \cite{carlileleung2016, leung2016head}. The perception of moving spatial objects can influence the direction of people's movement. The influence depends on both the starting position and the direction of movement \cite{royden1996}. Similarly, moving auditory sources produce continuous perceptual streams that can guide head-orientation behaviors \cite{roggerone2019motion}, which in turn influences head movement directions \cite{cohenlhyver2020}. Beyond the mere presence of motion, spatial affordances such as distance \cite{kolarik2016distance}, object salience \cite{lepelley2016}, and temporal ordering \cite{fendrich2001} all shape an agent’s motor response. Together, these factors drive agents to approach environmental objects of interest, a process reflected in their motion trajectories within the room.

In environments with multiple competing spatial stimuli, agents’ independent perceptual decisions can be formalized using spatial variants of the diffusion model \cite[DM;][]{ratcliff2008diffusion, ratcliff2016diffusion}. The DM captures the time course of a decision-making agent’s decisions under uncertainty; here, it can be extended in the context of spatial navigation, where an agent’s initial position corresponds to a starting point, and the spatial objects it approaches serve as multidimensional decision boundaries. This extension naturally generalizes to social settings, where, under specific task constraints in the virtual world (e.g., foraging), the DM can serve as a parallel component of collective decision-making complementing the representation of individuals as reinforcement learning agents \cite{marienhagen2025bridging}. In contrast to reinforcement learning agents, however, our agents do not require explicit goals, and no reward structure is present. Instead, the DM serves to capture individual perceptual tendencies in the environment and introduce perturbation to the emergent collective motion.

\subsection{Simulation-Based Inference}

\begin{figure}
    \centering
    \includegraphics[width=.99\textwidth]{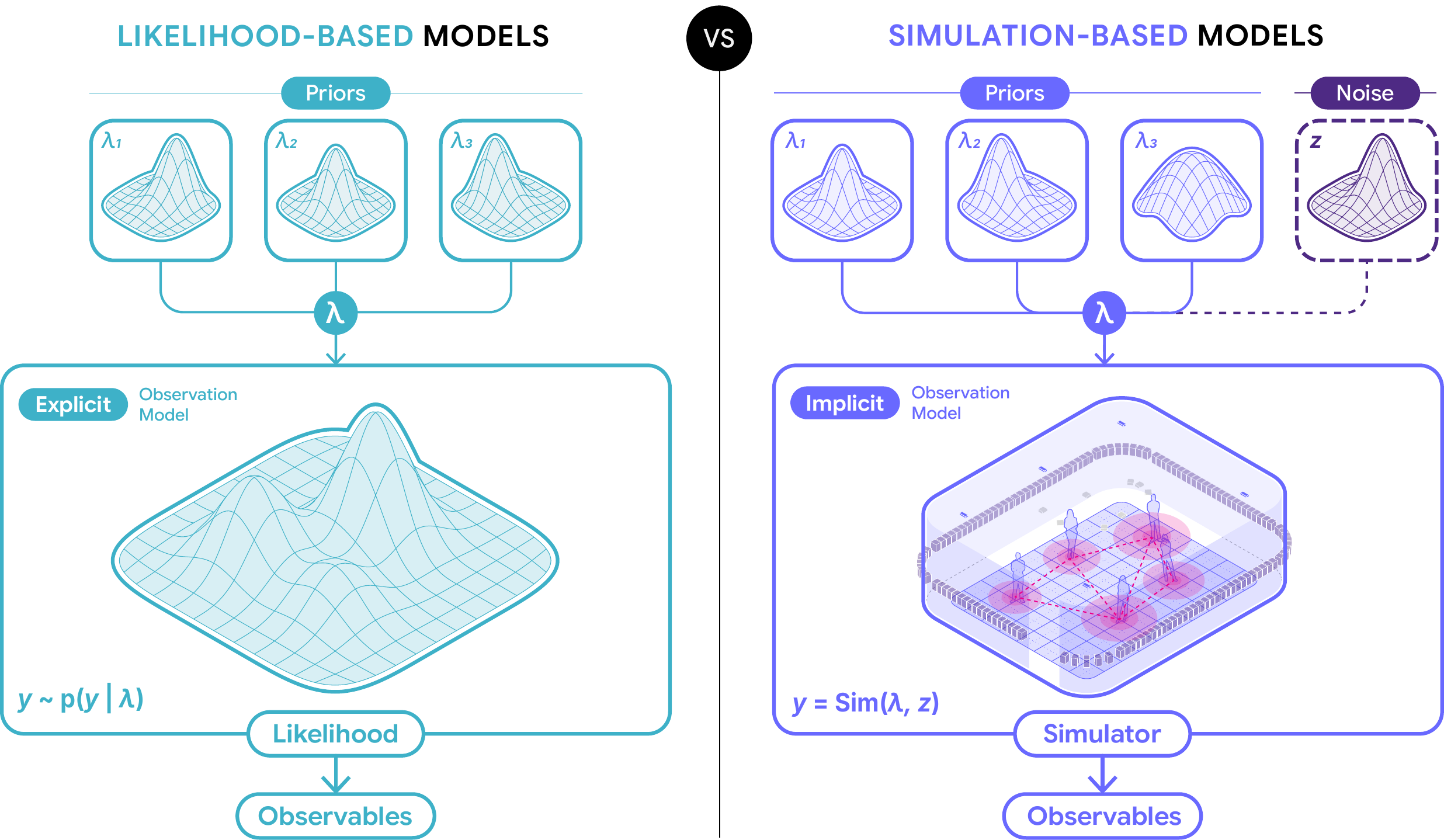}
    \caption{\textit{Flow diagram showing the distinction between likelihood-based and simulation-based generative models.} In likelihood-based inference (\textit{left}), the generative model is specified as a tractable probability distribution (e.g., Gaussian, Poisson, \textit{etc}). In simulation-based inference (\textit{right}, this work), the generative model is specified as a set of computational rules or algorithmic steps which only implicitly define the distribution of observables $y$ given input parameters $\lambda$.}
    \label{fig:sbi}    
\end{figure}

To connect the modeling approaches that bridges individual decision dynamics and collective motion behaviors, we need an inference framework that can recover the latent generative mechanisms from observed motion trajectories. To this end, simulation-based inference (SBI) is a nascent family of statistical method for estimating the parameters of complex stochastic simulators \cite{cranmer2020frontier}. In contrast to traditional statistical inference, where the observation model can be expressed as a known probability distribution (i.e., the likelihood), simulators are defined solely through computational or algorithmic rules (see also \autoref{fig:sbi}). However, the ability to invert simulators (i.e., parameter estimation) can yield important insights into the system's workings, which is why SBI can be an attractive tool for tackling ABMs. 

Applications of SBI to ABMs have grown rapidly, spanning fields such as econometrics \cite{grazzini2017, shiono2021estimation, dyer2024},  competitive sports \cite{youngbloodpassmore2024}, and evolutionary biology \cite{pitocchelli2024}. Methodological advances have kept pace, including the development of surrogate models \cite{angione2022} and novel summary statistics \cite{dyer2021approximate, dignum2025simulation}. The push for data-driven approaches to ABMs within SBI arises from their distinct nature as simulation-first models, specifically, the complex interactions among agents that make parameter estimation difficult. Further complicating matters, simple agent rules can lead to complex emergent behaviors, making ABMs computationally surprising \cite{daly2022}.

Model complexity places additional responsibility on researchers to encode appropriate assumptions about model parameters to ensure simulations are well-behaved. Complex formulations introduce many parameters to estimate, and researchers must carefully select constraints or prior distributions, as maintaining sensible simulation ranges often requires jointly adjusting multiple parameters. This makes results highly sensitive to priors on agent properties. In our case, introducing the spatial DM as a perturbation factor further tightens these conditions. These challenges make parameter estimation with ABMs extremely difficult for traditional methods \cite{said2016applying}. 

A promising solution is amortized Bayesian inference (ABI), which uses synthetic data generated from arbitrary simulators to train a deep learning model for fully Bayesian \cite{radev2020bayesflow} or point estimation \cite{sainsbury2024likelihood}. ABI splits Bayesian estimation into a computationally intensive training phase and a cheap inference phase, allowing fast parameter estimation with real-world data, such as motion-tracking inputs. Moreover, ABI can accommodate the hierarchical structure of ABMs through its via multi-level neural network architectures \cite{elsemuller2024deep}. This makes it an obvious candidate for studying parameter recoverability of agent motion patterns in immersive rooms.

\section{Methodology}
\label{sec:method}

\subsection{Simulation Environments}

\textit{Spatial beacons} are contextually located multimodal stimuli in a navigational environment that anchor the spatial perception of human agents in immersive rooms. For instance, in a digitally reconstructed forest environment, spatial beacons can take the form of a large puffball mushroom on the forest floor or the song of a nearby cardinal. The concept of spatial beacons has applications in developing assistive technologies for real-world human navigation, specifically by incorporating multisensory information with geospatial data for cognitive map formation \cite{clemenson2021}. In the virtual world, we can taxonomize spatial beacons in at least three different categories: \textit{visual beacons} as designated landmark that aims at agents' body movements through viewing experience; \textit{auditory beacons} as near-field virtual sound sources that trigger agents' head movement through listening experience; \textit{bimodal beacons}, which combine visual and auditory information to enhance perceptual salience and navigational guidance. 
\begin{figure}[t]
    \centering
    \includegraphics[width=\linewidth]{figures/nature.pdf}
    \includegraphics[width=\linewidth]{figures/geometric.pdf}
    \includegraphics[width=0.495\linewidth]{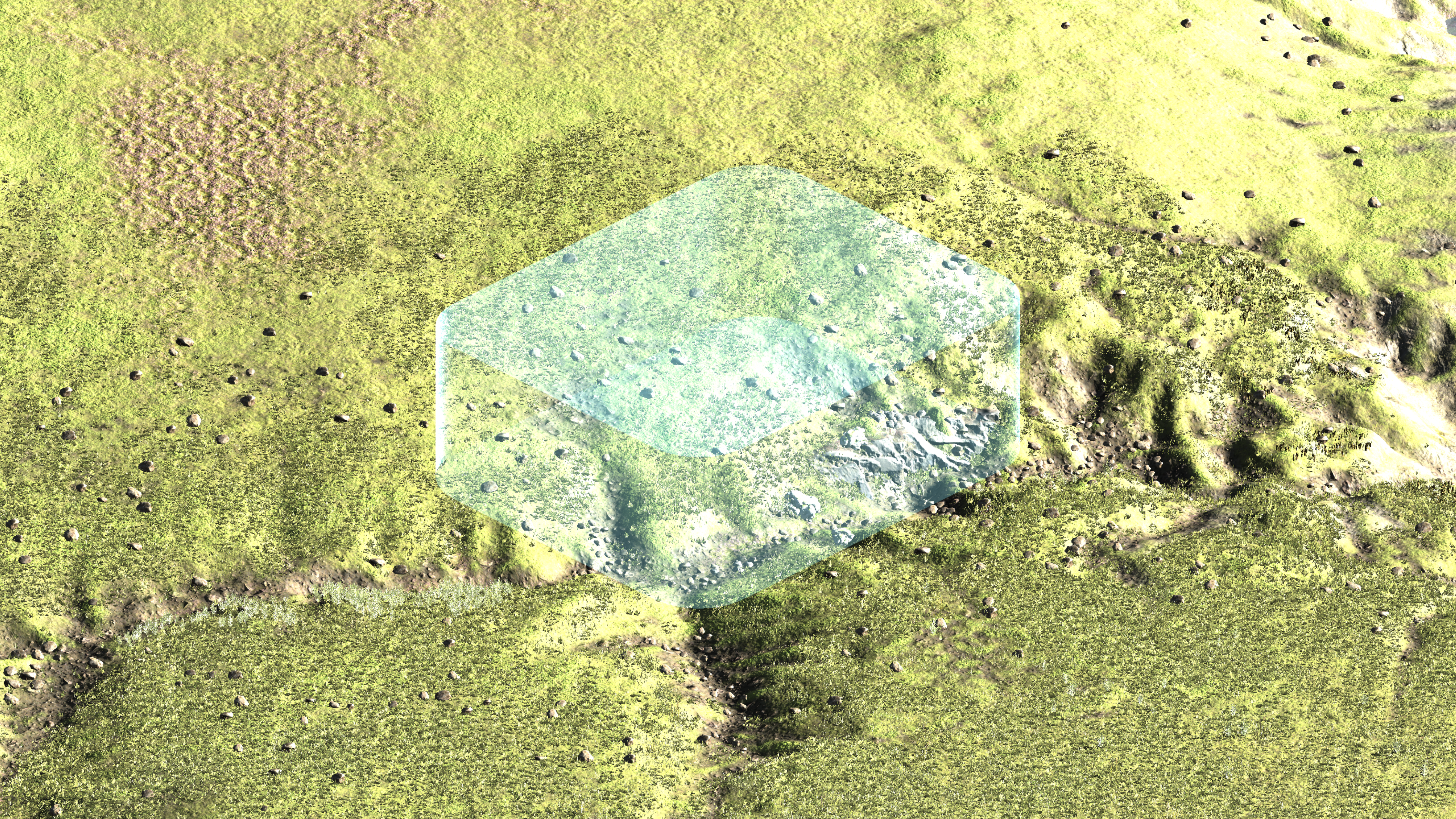}
    \includegraphics[width=0.495\linewidth]{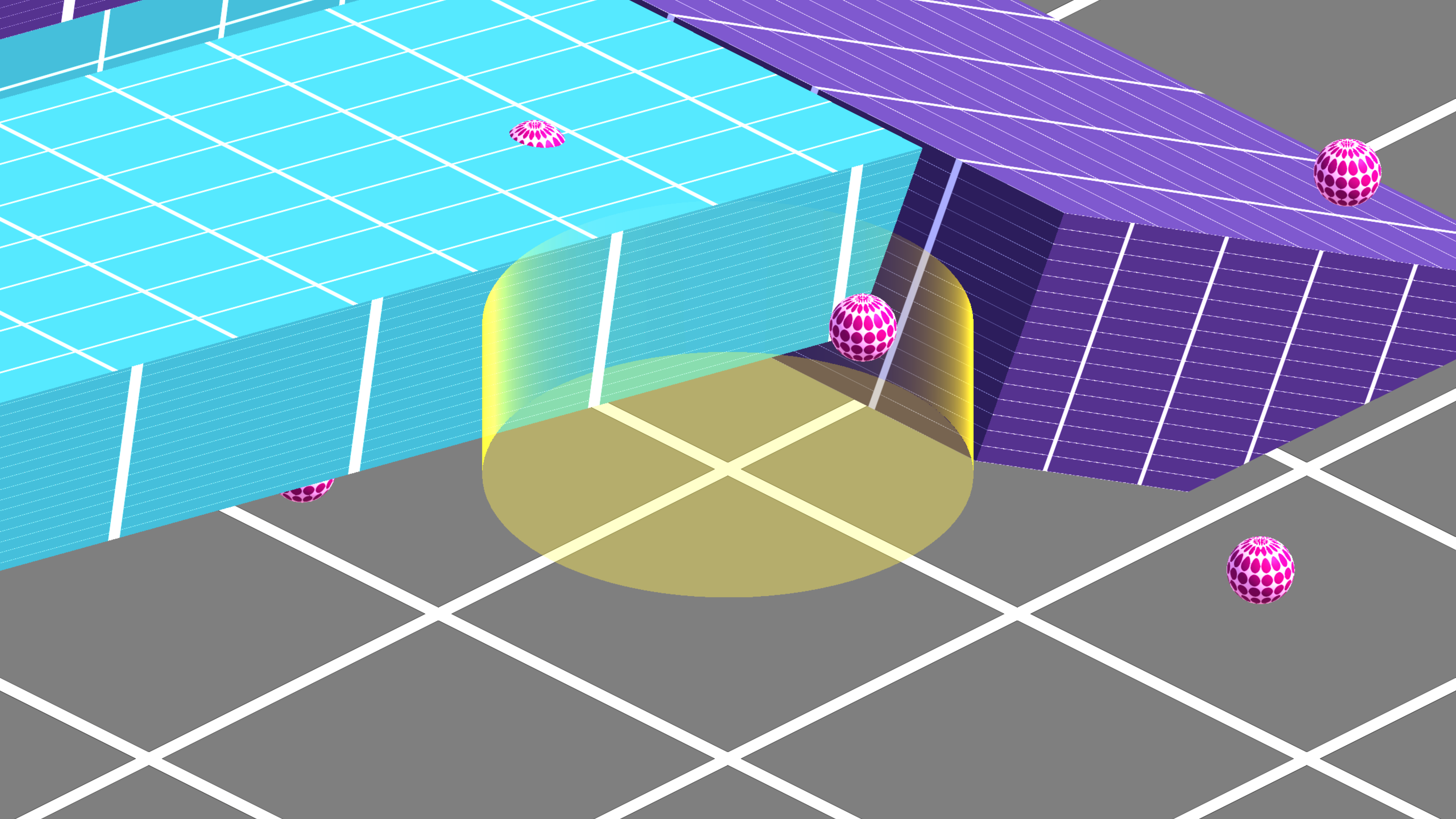}
    \caption{\textit{Two examples of the environments set up for the development of our ABM.} \textit{Top elongated}: a complex natural terrain with landscape features, offering a walkable experience populated with auditory beacons (e.g., birdsongs); \textit{bottom elongated}: a geometric obstacle course for free-form drone navigation populated with visual beacons (e.g., spheres). The panoramic images are rendered using immersive rooms' digital twins (\textit{bottom left}, \textit{bottom right}; see description in \autoref{app:digital-twin}).}
    \label{fig:environment}
\end{figure}

We constructed several test environments for navigable experiences that can be deployed in the immersive room. They include realistic scenarios, such as complex natural terrains with landscape features like mountains and lakes, to synthetic environments, such as geometric obstacle courses for free-form drone navigation (see \autoref{fig:environment}). At an abstract level, these environments share key properties: they provide spaces for free navigation within the room and spatial beacons distributed throughout. The use of game-based platforms in immersive rooms enables these beacons to be rendered in a spatially congruent manner. In general, we emphasize spatial rather than semantic congruence: for a bimodal beacon, we focus less on whether its audio and visual contents share meaning and more on whether they spatially coincide in the same location.

We simulate collective navigation using a digital twin of the immersive room in the virtual world \cite{huang2024digitaltwin}. At the most fundamental level, this digital twin is a modified first-person controller with a panoramic camera attached. In addition, the digital twin represents the spatial footprint of the immersive room in a 1:1 scale mapping to the virtual world, with human agents represented as a simple point with a 2D orientation and a defined sensing range for its neighbors. This set of representations allows the spatial information of the room and the agents to be projected onto the virtual environment. Conversely, as the room moves within the virtual world, positions and orientations of the spatial beacons are tracked by the agents relative to their egocentric reference frame.

As the room moves within the virtual world, the agents start perceiving and tracking the motion of its surrounding beacons \cite{neuhoff2018, hayhoe2005eye}. Agent movements are not planned in advance: they act by moving and reorienting themselves without knowing beforehand which beacon to move and orient themselves to. The agents freely move within the room under this navigation scenario, where they may, for example, spontaneously take ``mental notes'' of the spatial objects that drive their curiosity. Different agents may be inclined to approach different beacons, and some agents may be more likely to follow suit with other agents than engage with the environments by themselves. This process combines individual and collective motion, producing emergent behavior as a collection of room-centered agent motion trajectories.

\subsection{Model Formulation}
\label{sec:model}

\begin{figure}[t]
    \begin{center}
    \includegraphics[width=\textwidth]{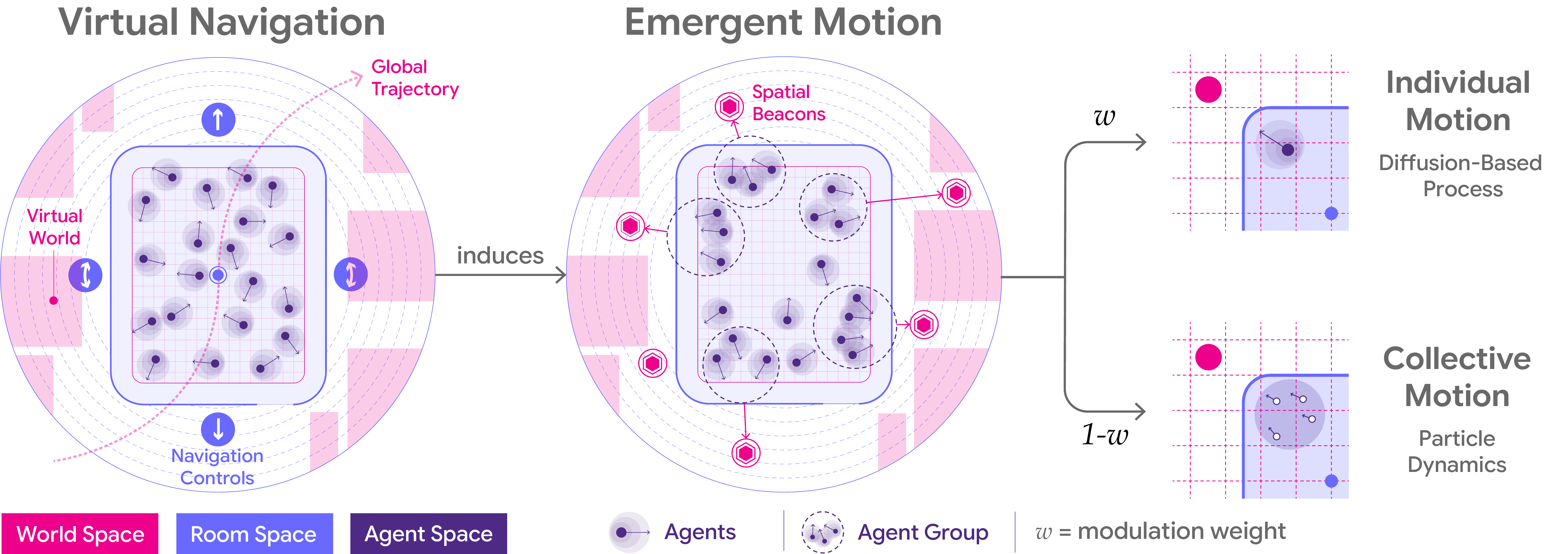}
    \end{center}
    \caption{\textit{Conceptual diagram of our agent-based model (ABM) deployed in the immersive room.} Motion perception during collective virtual navigation in the immersive room induces sensorimotor response to the virtual beacons. For motion patterns, individual (external) influence is modeled as a 2D drift diffusion process, and collective (internal) influence is modeled using particle dynamics. At a given point in time, a stationary weight $w_k$ modulates the switch between internal and external influences. By extension, clustering of agents from the partitioned movement pattern can encode distributed spatial memory among the immersed human agent group.}
    \label{fig:concept}
\end{figure}

The conceptual illustration of our motion dynamics model is shown in \autoref{fig:concept}. As discussed in \autoref{sec:background}, the agents' emergent motion pattern is activated by the individual motion-induced response to the virtual navigation process of the immersive room in the virtual world. Therefore, we formally define \textit{external influence} as agents' tendency to be directed to a spatial beacon, a proxy representation of their active perception; and \textit{internal influence} as the agents' inclination to move along with their surrounding neighbors. In addition, we have implemented several directions to extend the models, including collision avoidance, beacon salience, and time-varying modulation weights for the agents. Their respective formulation and simulation studies outlined in \autoref{sec:model-extensions}.

For each agent $a = 1, \ldots, A$ moving within a room $M$ bounded by $\Omega_M$, we represent the physical aspect of collective motion of all agents by their positions 
$\mathbf{X} = (\mathbf{x}_1, \ldots, \mathbf{x}_A)$ 
and orientations $\boldsymbol{\Theta} = (\theta_1, \ldots, \theta_A)$.
In addition, we also collect the individual agents' number of neighbors $\mathbf{N} = (N_1, ..., N_A)$
as well as the average neighbor distances $\mathbf{D} = (D_1, ..., D_A)$.
Using the immersive room's digital twin, an agent $a$ in the room would collectively navigate through a virtual world that is populated with multiple spatial beacons, $b = 1, ..., B$, along with their neighbors in the same room, $n = 1, ..., N$. We consider the agents' locomotion and reorientation as parallel events with separate action states. In other words, the agents can move and rotate in both concurrent and parallel fashions. It is possible for the agents to approach a beacon without reorientation, or simply reorient themselves without moving. \autoref{tab:notation} summarizes the notation used for formulating the model.

\begin{longtable}{p{0.3\textwidth}p{\dimexpr0.7\textwidth-4\tabcolsep\relax}}
\caption{Symbols and notation for the proposed agent-based motion of emergent human motion in immersive rooms.}\label{tab:notation}\\
\toprule
\textbf{Notation} & \textbf{Meaning}\\
\midrule
\endfirsthead

\toprule
\textbf{Notation} & \textbf{Meaning}\\
\midrule
\endhead

\bottomrule
\endfoot

\bottomrule
\endlastfoot

\multicolumn{2}{l}{\textit{Indices}}\\
$a = 1, \ldots, A$ & Indices of $A$ agents\\
$b = 1, \ldots, B$ & Indices of $B$ beacons\\
$n = 1, \ldots, N$ & Indices of $N$ neighboring agents\\
\midrule

\multicolumn{2}{l}{\textit{Positions}}\\
$\mathbf{x} = (\mathbf{x}_1, \ldots, \mathbf{x}_A)$ & The set of positions for all agents\\
$\mathbf{x}_a = (x_a, y_a)$ & Room-centered position of agent $a$\\
$\mathbf{x}_n = (x_n, y_n)$ & Room-centered position of neighboring agent $n$\\
$\mathbf{x}_b = (x_b, y_b)$ & Room-centered position of spatial beacon $b$\\
$\mathbf{x}_M = (x_M, y_M)$ & World-centered position of the room $M$\\
$\mathbf{x}_{a|b}$ & Position of agent $a$ as influenced by spatial beacon $b$\\
$\mathbf{x}_{a|n}$ & Position of agent $a$ as influenced by neighboring agent $n$\\
\midrule

\multicolumn{2}{l}{\textit{Orientations}}\\
$\theta_{a|b}$ & Orientation of agent $a$ as influenced by spatial beacon $b$\\
$\theta_{a|n}$ & Orientation of agent $a$ as influenced by neighboring agent $n$\\
\midrule

\multicolumn{2}{l}{\textit{Agent properties}}\\
$w_a$ & External influence weight for agent $a$\\
$r_a$ & Neighbor sensing range of agent $a$\\
$v_a$ & Locomotive drift rate of agent $a$\\
$\omega_a$ & Rotational drift rate of agent $a$\\
$\sigma_a$ & Diffusion coefficient for position noise of agent $a$\\
$\eta_a$ & Rotational noise variance of agent $a$\\
\midrule

\multicolumn{2}{l}{\textit{Observables}}\\
$\mathbf{X} = (\mathbf{x}_1, \ldots, \mathbf{x}_A)$ & Positions for all agents\\
$\boldsymbol{\Theta} = (\theta_1, \ldots, \theta_A)$ & Rotations for all agents\\
$\mathbf{N} = (N_1, \ldots, N_A)$ & Number of neighbors for all agents\\
$\mathbf{D} = (D_1, \ldots, D_A)$ & Average distances of neighbors for all agents\\
\midrule

\multicolumn{2}{l}{\textit{Other properties}}\\
$\Omega_M$ & Boundary of the room $M$\\
$R_M$ & Beacon detection range for the room $M$\\
\end{longtable}

We formulate an agent $a$'s motion dynamics as a modulation between external and internal influences, namely,
\begin{align}\label{eq:weight}
    \theta_{a, t} &= w_a \theta_{a|b, t} + (1 - w_a) \theta_{a|n, t} \\
    \mathbf{x}_{a, t} &= w_a \mathbf{x}_{a|b, t} + (1 - w_a) \mathbf{x}_{a|n, t},
\end{align}
\noindent where $a$'s influences at time $t$ are given by their change in orientation $\theta_{a, t}$ as well as their change in position $\mathbf{x}_{a, t}$. The agent's external influence, denoted as $\theta_{a|b, t}$ and $\mathbf{x}_{a|b, t}$\footnote{One should read $a|b$ as ``agent $a$ given beacon $b$'' or ``agent $a$ as influenced by beacon $b$.''}, is given by its approached beacon; by contrast, its internal influence, denoted as $\theta_{a|n, t}$ and $\mathbf{x}_{a|n, t}$, is given by its surrounding neighbors. While $\theta_{a|b, t}$ and $\mathbf{x}_{a|b, t}$ are updated through a spatial diffusion-based model with the agent $a$'s positions of approached beacon as decision boundary, $\theta_{a|n, t}$ and $\mathbf{x}_{a|n, t}$ are updated through an ABM of self-propelling particle dynamics. All influences are specified in 2D, following both the sensing capability of the motion tracking system and the rendering capability of the multi-channel spatial audio speaker system\footnote{The spatial sound sources are typically rendered in the virtual world using wave field synthesis \cite{berkhout1993acoustic}.}.

\begin{figure}[t]
    \begin{center}
    \includegraphics[width=\textwidth]{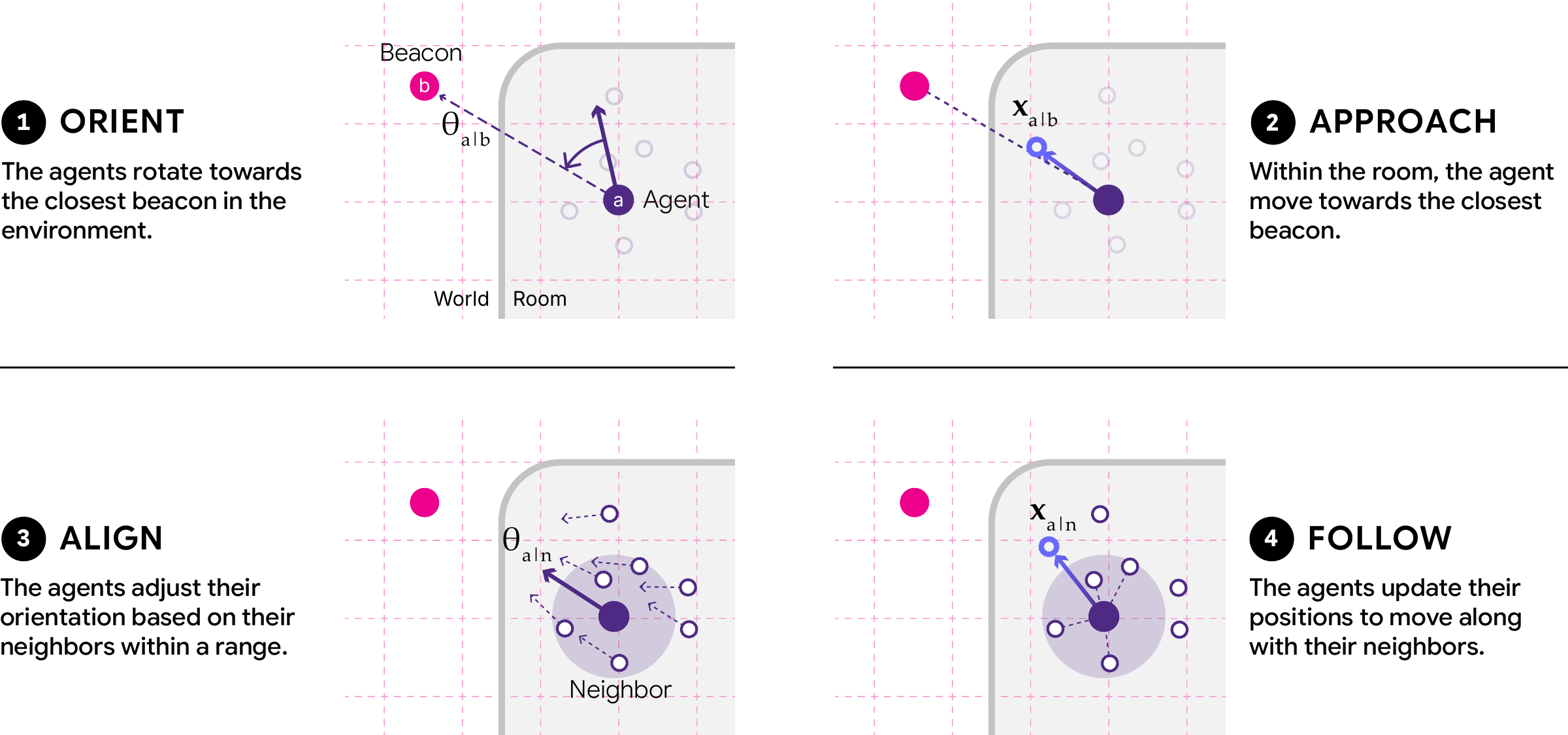}
    \end{center}
    \caption{\textit{Influence vectors driving agent motion in an immersive room.} These influence vectors are separated into external (individual, \textit{top half}) and internal (collective, \textit{bottom half}) components, each exhibiting separate rotational (\textit{left half}) and translational (\textit{right half}) motion. The gray boundary and gray area represents the immersive room's spatial footprint, the solid purple dot represents an agent $a$ of focus, the white circle dots depict neighboring agents (within the shaded purple circle), and the solid magenta dot represents a spatial beacon. Individual orientation and approach to the spatial beacon both follow a diffusion model in 2D (external influence), whereas collective (internal) alignment and following influences conform to a Vicsek model.}
    \label{fig:influences}
\end{figure}

A visual breakdown of the influences can be found in \autoref{fig:influences}. Below, we formulate the components of these influences in greater detail.

\subsection{External Influence: Individual Motion as Spatial Drift Diffusion}
\label{ssec:external-influence}

\noindent We use a spatial diffusion model in 2D space to provide a baseline representation of the agent's external influence. As the room travels in the virtual world, it detects a subset of spatial beacons within its detection range $R_M$, where $||\mathbf{x}_{b} - \mathbf{x}_{M}|| < R_M$. Within this subset of beacons, the agents approach the one that is closest to them, namely:
\begin{equation}
    b_{a,t} = \arg \min_{b \in B} ||\mathbf{x}_{a, t} - \mathbf{x}_{b, t}||, 
\end{equation}
\noindent so that, within the room's boundary $\Omega_M$, agent $a$ approaches $b_a$ based on $b_a$'s relative position in the agent-centered coordinates $\mathbf{x}_{a,b} = \mathbf{x}_{b} - \mathbf{x}_{a}$, as well as the relative orientation, $\theta_{b|a} = \mathbf{x}_{a,b} - \theta_{a|b,t}$. The orientation update for the agent $a$ can be expressed as 
\begin{equation}
    \mathrm{d}\theta_{a|b, t} = (\omega_a + \phi_{a, t})\mathrm{d}t ,
\end{equation}
\noindent where $\omega_a = \tan^{-1}(\mathbf{x}_{a,b})$ is the rotational drift rate (speed) of the agent $a$, and $\phi_{a, t} \sim \mathcal{U}(-\kappa_a, \kappa_a)$ is a non-cumulative rotational noise component that follows a uniform distribution, where $\kappa_a < 0.01$ controls the variability of the agent's reorientation at each time step. In addition, the positional update for the agent $a$ can be expressed as:
\begin{equation}\label{eq:individual}
    \mathrm{d}\mathbf{x}_{a|b, t} = v_{a, t}  
    \begin{bmatrix}
        \cos \theta_{a|b, t} \\
        \sin \theta_{a|b, t} 
    \end{bmatrix}
    \mathrm{d}t
    + \sigma_{a|b} \mathrm{d} \mathbf{W}_t,
\end{equation}
where $\mathbf{W}_t$ is a two-dimensional Wiener process, and 
$\sigma_{a|b} > 0$ controls the diffusion amplitude (i.e., the scale of positional noise). 
The scalar $v_{a,t}$ denotes the instantaneous speed (drift magnitude) of agent~$a$, and the unit vector \smash{$[ \cos \theta_{a|b,t},\, \sin \theta_{a|b,t}]^\top$} specifies its current heading relative to the beacon~$b$.
This means that, at each time step, the agent $a$ moves approximately in the direction of the beacon $b$ with adjustment to noise perturbation. The drift magnitude $v$ may vary between agents and between simulations, and therefore becomes a candidate parameter to capture the influence of the virtual world on the agents.

\subsection{Internal Influence: Collective Motion as Active Particle Dynamics}
\label{ssec:internal-influence}

To express collective motion, we follow the Vicsek model \cite{vicsek1995} as our baseline formulation of internal influence. The Vicsek model is an ABM that considers the alignment dynamics of individual agents and their local interaction with neighborhood agents, where the agent $a$ takes into account the average orientation of neighboring agents $n = 1, ..., N$ under a sensing radius $r_a$, and adjust its orientation $\theta_{a|n, t}$ accordingly. We can formalize the angular update for the internal influence of agent $a$ as
\begin{equation}\label{eq:vicsek}
    \theta_{a|n, t} = 
    \langle \theta_{n, t}\rangle_{||\mathbf{x}_{a,t} - \mathbf{x}_{n, t}|| < r_a, n \in N_t}  + 
    \gamma_{a, t},
\end{equation}
where $\theta_a, \theta_n$ are the orientations of agent $a$ and their neighbor $n$ within a sensing distance $r_a$, $\langle \theta_n \rangle$ denotes the average of all neighbor orientations within the agent's sensing radius, and $\gamma_a \sim \mathcal{N}(0, \eta_a)$ represents rotational noise for agent $a$ under variance $\eta_a$. Both $r_a$ and $\eta_a$ may vary between agents and between simulations, and therefore are candidate parameters that capture the agents' local rules in the immersive room The motion update under internal influence is expressed as:
\begin{equation}\label{eq:collective}
    \mathrm{d} \mathbf{x}_{a|n, t} = 
    v_{a, t}
    \begin{bmatrix}
        \cos \theta_{a|n, t} \\
        \sin \theta_{a|n, t} 
    \end{bmatrix}
    \mathrm{d}t,
\end{equation}
where $\mathbf{x}_a, \mathbf{x}_n$ are the orientations of agent $a$ and their neighbor $n$ within a sensing distance $r_a$, and $\langle \mathbf{x}_i \rangle$ denotes the average of all neighbor positions. The sensing distance $r_a$ for alignment and cohesion dynamics are identical since the number of agents is finite and the interaction boundary is finite and non-periodic.

\section{Implementation}

\subsection{Forward Model and Simulation}
\label{ssec:forward-model}

The forward model detailed in \autoref{sec:method} is implemented in two ways. First, a playable interactive simulator visualizing the real-time motion dynamics is implemented using the Unity game engine and integrated as part of the digital twin \cite{huang2024digitaltwin}. This simulator assumes an infinite time horizon, where a switching regime between beacons is incorporated for the agents to update the beacon to approach. 

\begin{figure}[t]
    \centering
    \includegraphics[width=0.32\linewidth]{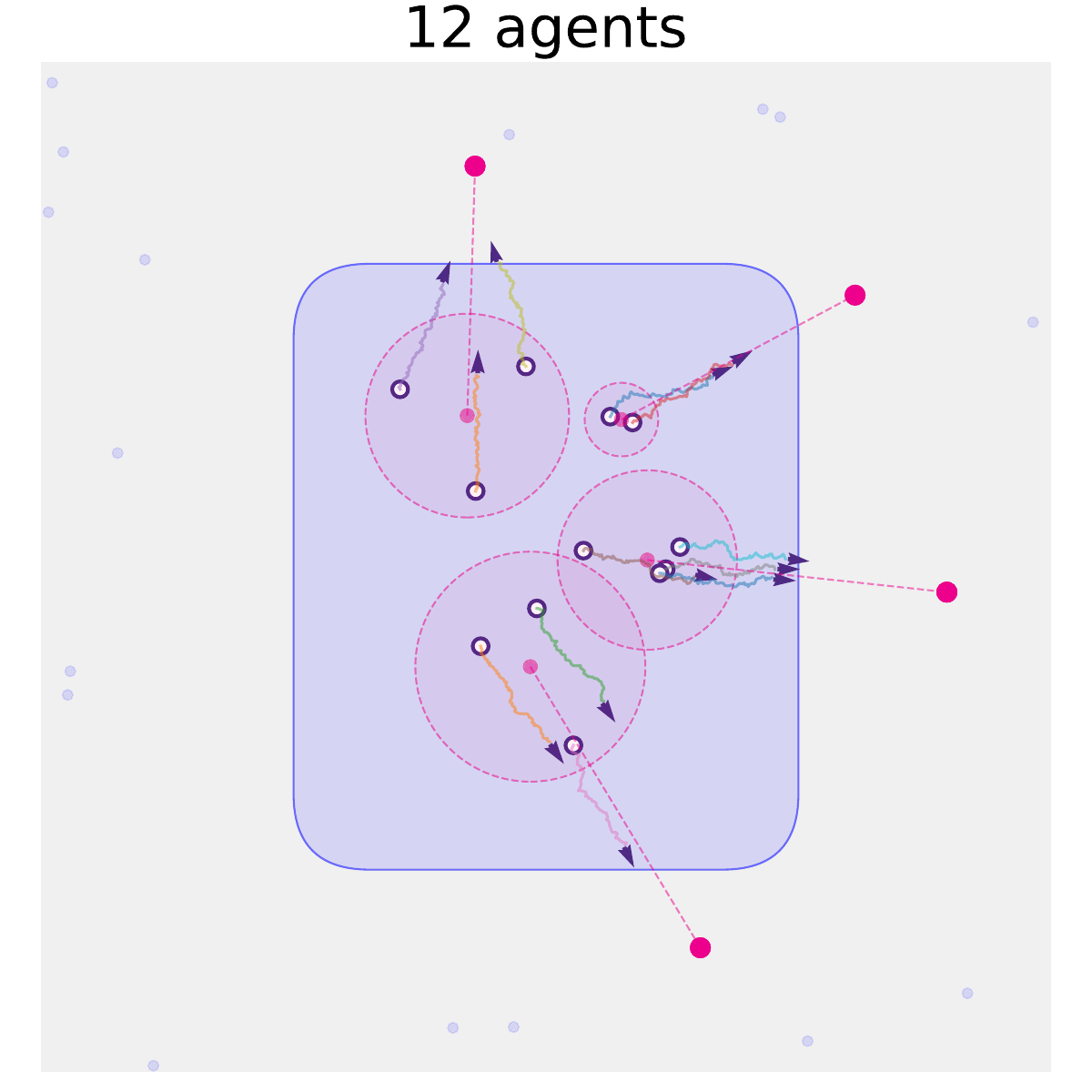}
    \includegraphics[width=0.32\linewidth]{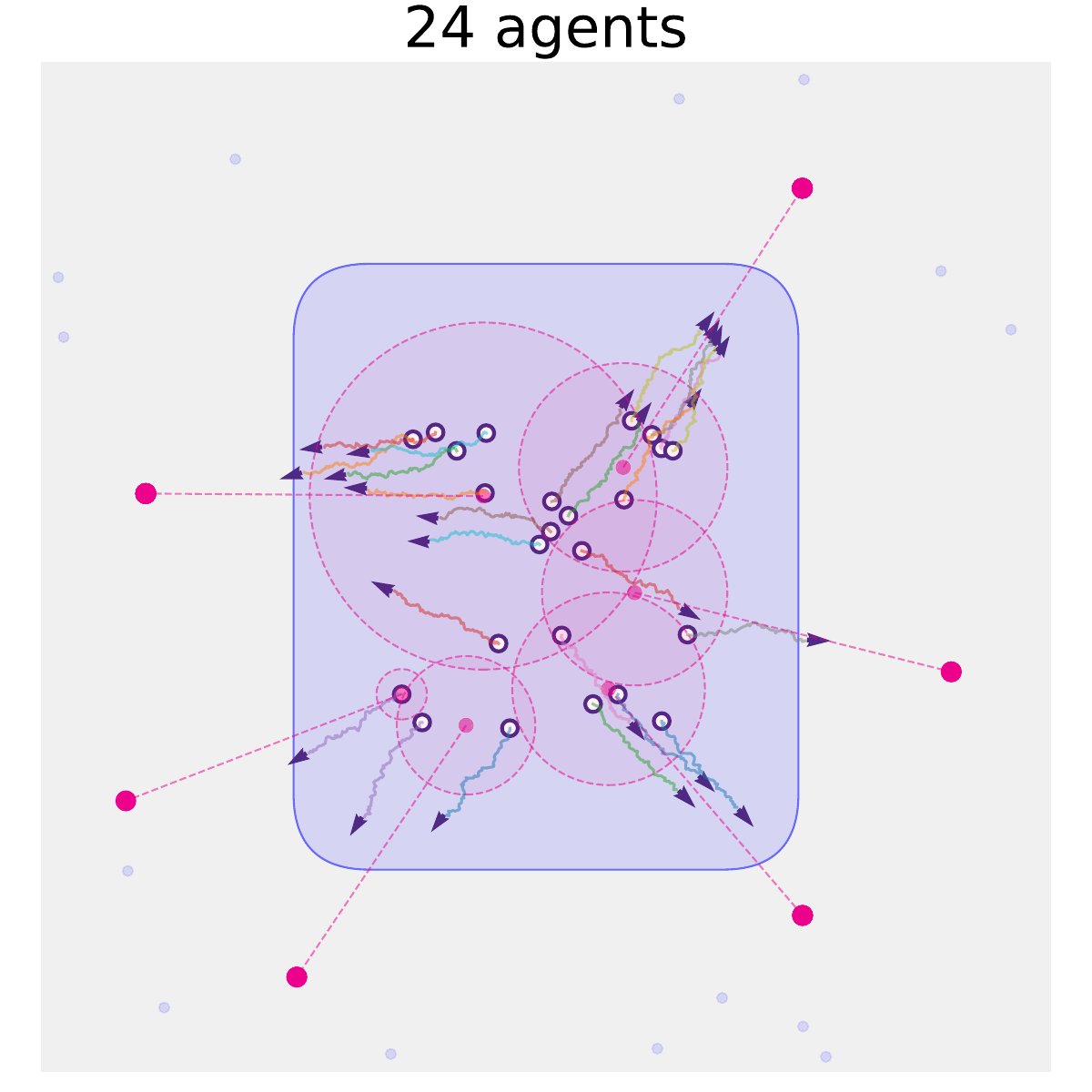}
    \includegraphics[width=0.32\linewidth]{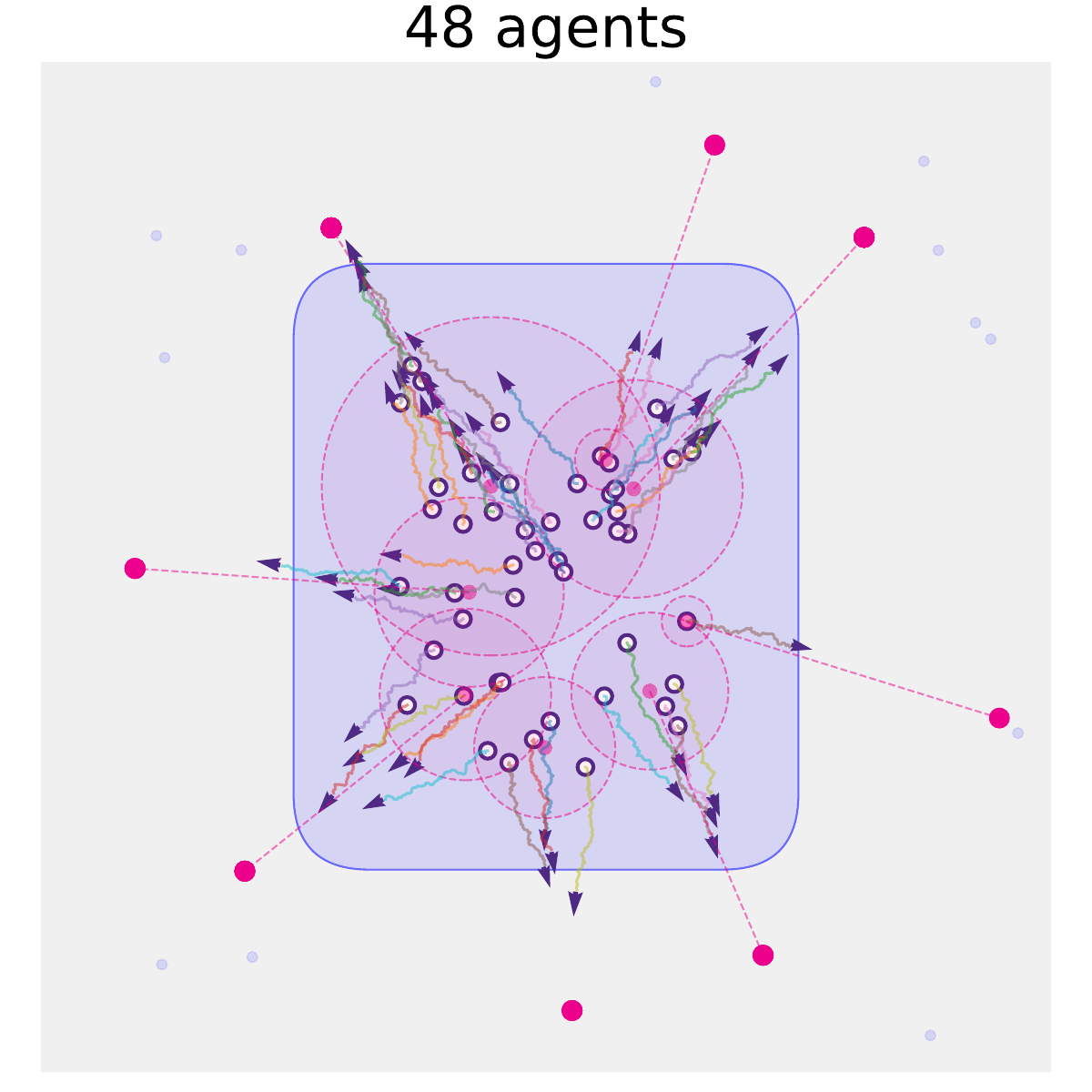}    
    \caption{\textit{Simulated onset trajectories for three different agent and beacon configurations}. The blue-shaded rectangle represents the immersive room's spatial footprint in the virtual world, surrounded by beacons that are either approached (magenta dots) or not approached (blue dots) by different agents in the room (white circle dots with colored paths). Each beacon has a cluster of agents approaching it (magenta dashed circles).}
    \label{fig:sim-data}
\end{figure}

To generate our training data for the neural estimator, the same simulator is also implemented in Python assuming a finite time horizon without a switching regime. We use a baseline configuration of $A = 49$ agents in the same room, initially distributed evenly, and $B = 8$ beacons within its sensing range traveling in $T = 600$ time steps with time resolution $\mathrm{d}t = 0.1$, lasting for $60$ seconds in records. Example visualization for simulation onset can be found in \autoref{fig:sim-data}. This baseline configuration is closely aligned with the real-world infrastructural capability for the same task design. The observables for the simulator is a a collection of agent time series, including their positions $\mathbf{X} = (\mathbf{x}_1, \ldots, \mathbf{x}_A)$, rotations $\boldsymbol{\Theta} = (\theta_1, \ldots, \theta_A)$, the number of neighbors $\mathbf{N} = (N_1, \ldots, N_A)$, and average neighbor distances $\mathbf{D} = (D_1, \ldots, D_A)$. For data augmentation, we also compute the angular velocity from~$\boldsymbol{\Theta}$ and the change in neighbor count from $\mathbf{D}$. In both simulators, we confine the agents to the room's spatial boundary $\Omega_M$, so that they realistically reflect the real-world experience of the immersive rooms.

\subsection{Prior Specification}
\label{ssec:prior}

We are interested in estimating four key parameters: the influence weight~$w$, sensing radius~$r$, movement speed~$v$, 
and internal noise variance~$\eta$. 
These parameters jointly describe the agent's action capability and environmental awareness. 
The model can be implemented with different degrees of parameter sharing: \emph{no pooling} (each agent has distinct parameters), 
\emph{complete pooling} (all agents share the same parameters), 
or \emph{partial pooling} (hierarchical structure capturing individual deviations from global means). 
In this proof-of-concept study, we consider complete pooling and partial pooling. For the former, we assign the parameters with the following priors:
\begin{align}
w &\sim \mathrm{Beta}(2,2), &
r &\sim \mathrm{LogNormal}(0,0.5), \nonumber\\
\eta &\sim \mathrm{Beta}(2,5), &
v &\sim \mathrm{Beta}(2,2).
\end{align}

For partial pooling, we draw individual weights and radii for each agent from shared population-level distributions. Let $z_a=\operatorname{logit}(w_a)$ and $u_a=\operatorname{logit}(r_a)$ denote the individual weights and radii on their logit scales. We further define the population-level mean weights and radii on the logit scale $m_w=\operatorname{logit}(\mu_w)$ and $m_r=\operatorname{logit}(\mu_r)$, where $(\mu_w, \mu_r)$ are the population-level means on the physical scale. Similarly, $(\sigma_w, \sigma_r)$ are the corresponding population-level variance for the weights and radii. We can assign the hyperpriors
\begin{align}
\label{eq:partial-pooling-hyperpriors}
    \mu_w \sim \mathrm{Beta}(2,2), \quad
    \sigma_w &\sim \mathrm{Beta}(1,3),\\
    \mu_r \sim \mathrm{LogNormal}(1,0.1), \quad
    \sigma_r &\sim \mathrm{Beta}(2,2).
\end{align}
Conditional on these population parameters, individual weights and radii are generated independently for $a=1,\ldots,A$:
\label{eq:partial-pooling}
\begin{align}
    z_a
    &\overset{\mathrm{iid}}{\sim}
      \mathcal{N}(m_w,\sigma_w^2),
    & w_a&=\frac{1}{1+\exp(-z_a)},\\
    u_a
    &\overset{\mathrm{iid}}{\sim}
      \mathcal{N}(m_r,\sigma_r^2),
    & r_a&=\frac{1}{1+\exp(-u_a)}.
\end{align}
Here, the logit transformation treatment ensures that the values of $w_a$ and $r_a$ are bounded within the realistic range for our simulation ($0<w_a<1$ and $0<r_a<R$). Thus, we infer $(\mu_w,\sigma_w,\mu_r,\sigma_r,v,\eta)$ and all individual $(w_a,r_a)$ jointly. This gives a partial pooling model given that $v$ and $\eta$ remain global.

\subsection{Neural Estimator for Simulation-Based Inference}
\label{ssec:neural-estimator}

\begin{figure}
    \centering
    \includegraphics[width=\linewidth]{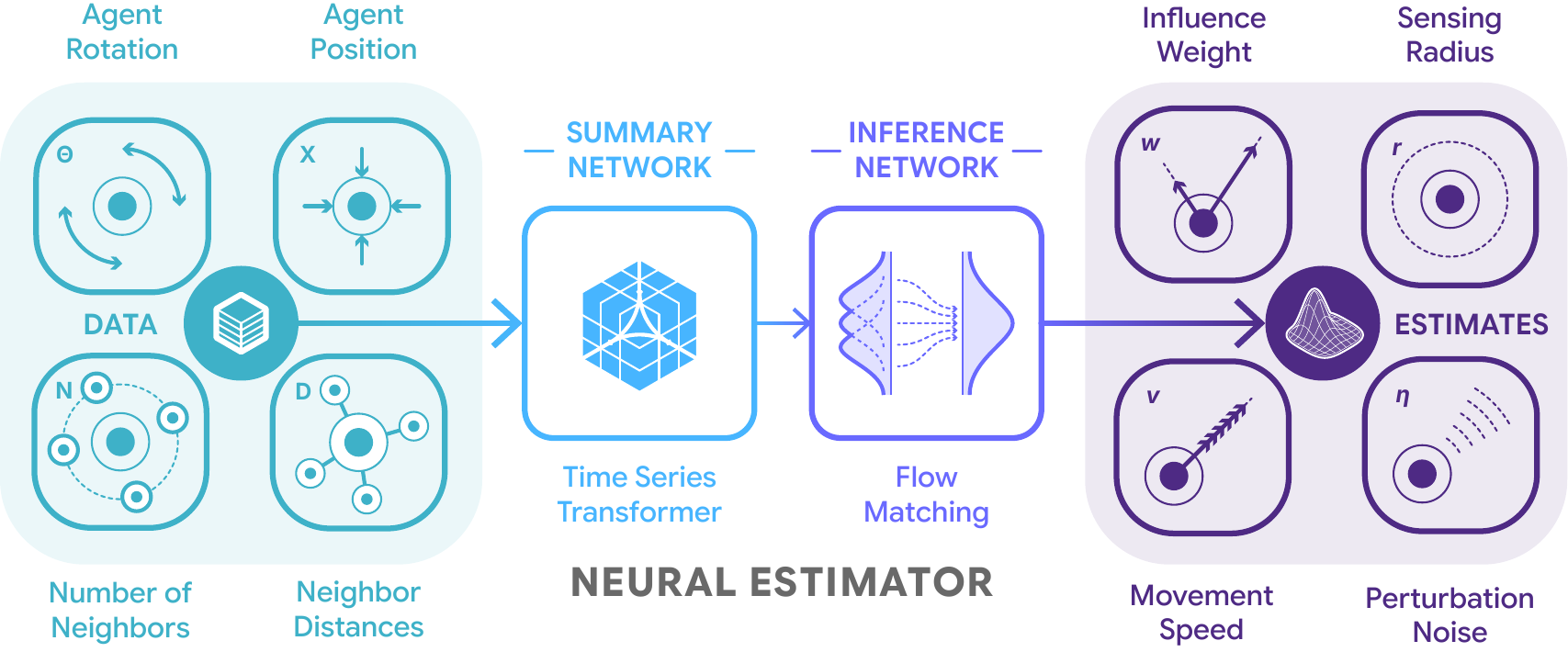}
    \caption{\textit{Schematic diagram of our SBI workflow for complete pooling.} The agent data, which consists of position, rotation, adjacent neighbor counts, and average neighbor distances, is used to train a two-stage neural network architecture with 1) a time-series transformer summary network, and 2) a flow matching inference network \cite{lipman2023flow}. The inference network estimates four target parameters: influence weight $w$, sensing radius $r$, movement speed $v$, and perturbation noise $\eta$. For partial pooling, we use a two-level conditional flow matching networks for population-level and agent-level parameters, respectively.}
    \label{fig:network}
\end{figure}

Our neural estimator, design with the BayesFlow library for amortized SBI \cite{kuhmichel2026bayesflow}, is used to perform posterior estimation of the model parameters specified in \autoref{ssec:prior}. A schematic diagram of the neural estimator and its relationship with the simulated data can be found in \autoref{fig:network}. 

Given the simulated trajectories of agent positions, orientations, adjacent neighbor counts, and average neighbor distances as inputs, the summary network is a time series transformer \citep{wen2022transformers}. We also provide additional results showing the use of a bidirectional LSTM as an alternative summary network (see results in \autoref{app:architecture}).

For parameter estimation, the embedded data needs to be passed through an inference network that can estimate the parameter posterior $p(w, r, \eta, v \mid \mathbf{X}, \mathbf{\Theta}, \mathbf{N}, \mathbf{D})$. We choose flow matching \cite{lipman2023flow} as a flexible and lightweight generative architecture. To evaluate parameter recoverability as a function of the simulation budget, we generate synthetic trajectory data from the implemented simulator in Python training sets with  $B \in \{3{,}000, 30{,}000\}$ simulations based on the configurations specified in \autoref{ssec:forward-model}, where we sample the parameters from their respective prior distributions specified in \autoref{ssec:prior}for each simulation. For each of the simulation budgets, we train a neural estimator for $100$ epochs with a batch size of $32$ and assess parameter recoverability on $S = 300$ unseen simulations as our validation ground truth. 

\section{Simulation Study}
\label{sec:sim-study}

To assess the global performance of our neural estimator as part of the SBI workflow, we validate the model from two different perspectives. First, we use simulation-based calibration \cite[SBC;][]{talts2018validating, sailynoja2022} to inspect the computational faithfulness of the estimator. Then, we evaluate whether agents' latent properties (i.e., parameters) can be recovered under idealized conditions by assessing parameter recovery. We provide both graphical checks and numerical summaries of these results.

\subsection{Computational Faithfulness: Testing Calibration}

\begin{figure}
    \begin{center}
    \includegraphics[width=\textwidth]{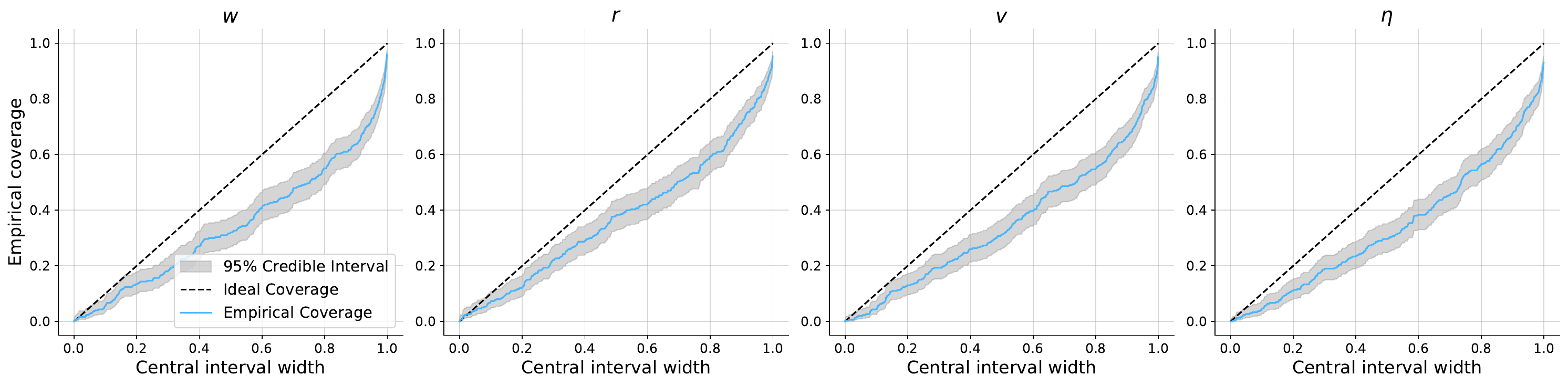}
    \includegraphics[width=\textwidth]{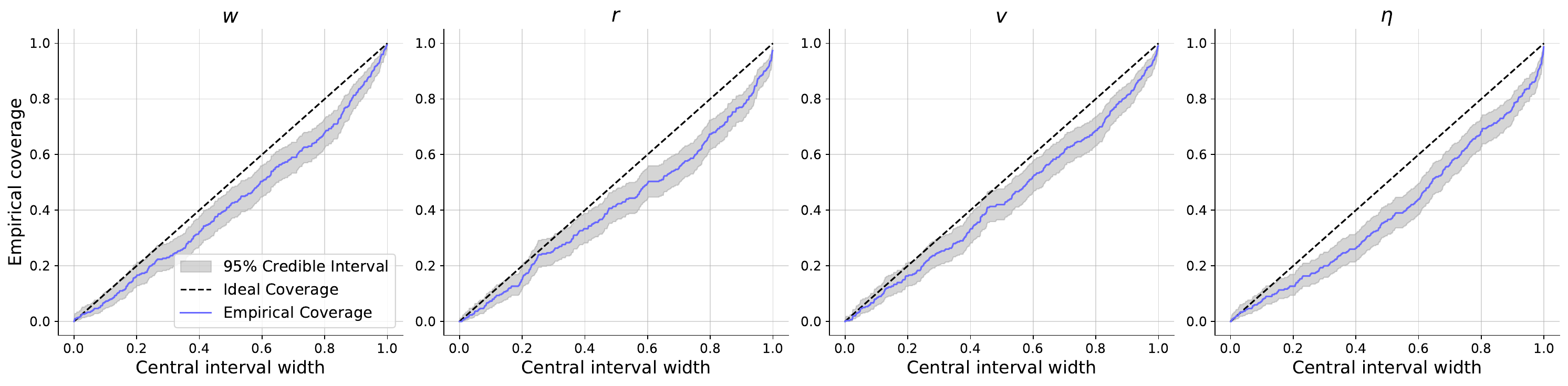}
    \includegraphics[width=\textwidth]{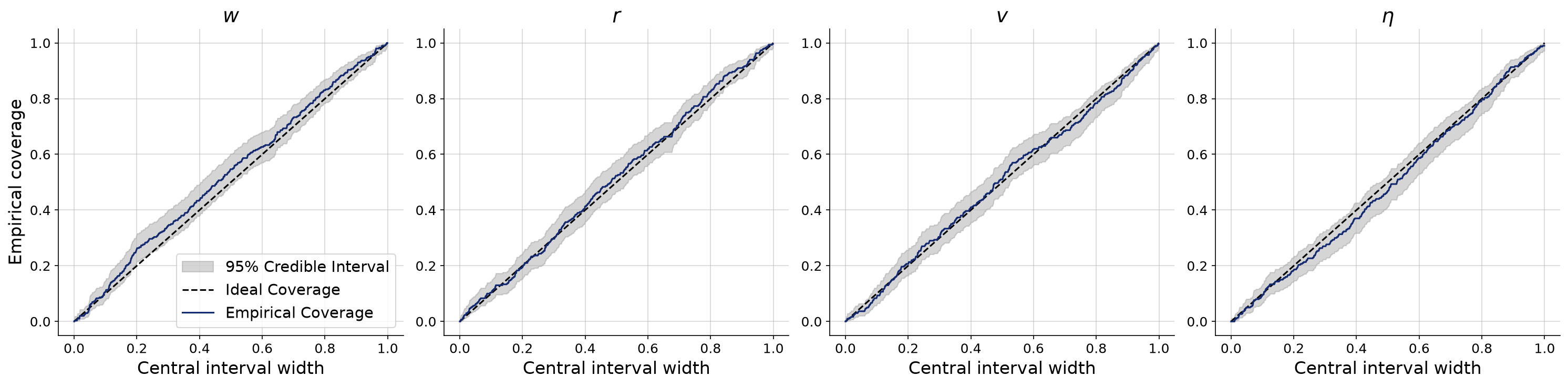}
    \end{center}
    \caption{
    \textit{Empirical coverage across the three training regimes as an indicator of computational faithfulness}. \textit{Top}: coverage of the neural estimator trained on $3{,}000$ model simulations. \textit{Middle}: coverage of the neural estimator trained on $30{,}000$ simulations. \textit{Bottom}: coverage of the neural estimator with online training.}
    \label{fig:sbc}
\end{figure}

First, we check whether the approximate posteriors are properly calibrated, that is, whether they faithfully reflect the epistemic uncertainty of the estimation problem. We leverage SBC, which evaluates properties of an ensemble of posteriors estimated from simulated data relative to the prior.
Here, we evaluate the empirical coverage of credible intervals obtained from posterior samples, rather than relying on the more common inspection of ranks or other sophisticated diagnostics \cite{yao2023discriminative}. Since any proper posterior has correct coverage under the true model, any deviation from the ideal coverage indicates approximation problems. The empirical coverage $c(\alpha)$ of a credible interval of size $\alpha$ (e.g., $\alpha = 0.95$) is computed from $S$ test simulations and corresponding posterior samples
\begin{equation}
c(\alpha) = \frac{1}{S} \sum_{s=1}^S \mathbf{1}[\lambda_s \in \left[q_{s, \beta}, q_{s, 1-\beta}\right]],
\end{equation}
where $\mathbf{1}$ is the indicator function, $\beta = (1-\alpha)/2$ and $1 - \beta$ define the bounds of the credibility interval, and $q_{s, \beta}$ is the $\beta$-quantile derived from the posterior samples $\{ \hat{\lambda}_s \}$ of the $s$-th test simulation.

Coverage can be checked graphically (\autoref{fig:sbc}) or summarized via the Expected Calibration Error (ECE) metric for a range of widths $\alpha_k$:
\begin{equation}
    \mathrm{ECE} = \frac{1}{K} \sum_{i=1}^K |c(\alpha_k) - \alpha_k|.
\end{equation}
The parameter-wise empirical coverage of our neural estimator is shown in \autoref{fig:sbc}, and the parameter-wise ECE is provided in \autoref{tab:metrics}. From the diagnostics, we observe that globally, model calibration systematically improves with increasing number of simulations. The offline estimators trained on 3,000 and 30,000 simulations show undercoverage: empirical coverage generally falls below the ideal coverage (black dashed line). Calibration aligns highly with the ideal for online training (480,000 simulations total) even though it has otherwise identical training protocol with the offline training regimes. 

\begin{figure}
    \centering
    \includegraphics[width=\linewidth]{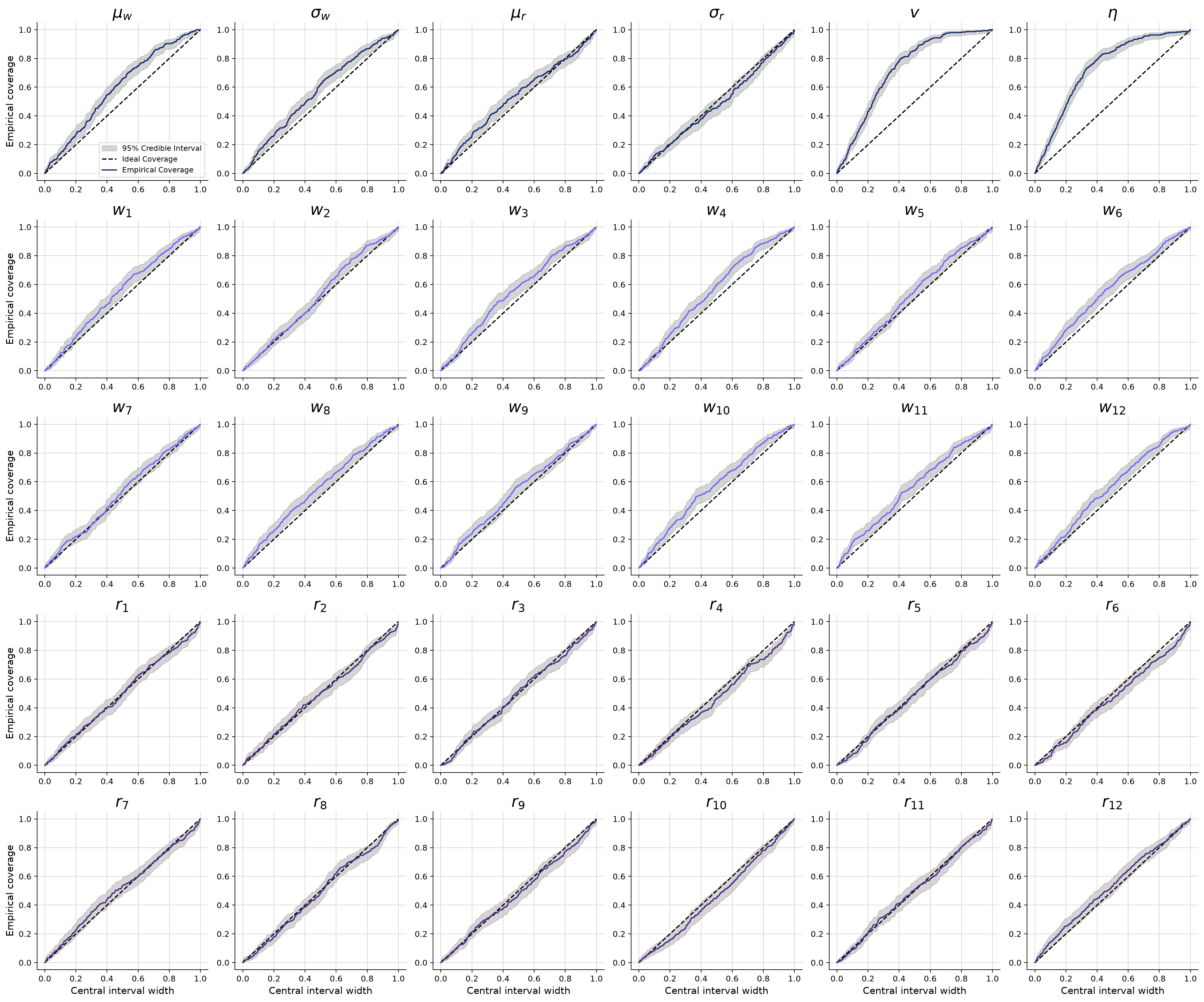}
\caption{\textit{Empirical coverage for the partial pooling model with 12 agents.} \textit{Row 1}: coverage for shared parameters including population-level means and variances for modulation weights ($\mu_w,\,\sigma_w$) and sensing radii ($\mu_r,\,\sigma_r$), as well as drift $v$ and noise amplitude $\eta$; \textit{Row 2-3}: coverage for the modulation weights $w_a$ for individual agents; \textit{Row 4-5}: coverage for the sensing radii $r_a$ for individual agents.}
    \label{fig:coverage-heterogeneity}
\end{figure}

\autoref{fig:coverage-heterogeneity} shows the empirical coverage for the partial pooling model. We can observe that agent-level parameters are generally well-calibrated, with the ideal coverage falling on or within the 95\% credible interval of the majority of the agents' sensing radii $r_a$, while slightly overconfident in their modulation weights $w_a$. At the population-level, mild overconfidence is also observed for the mean and variance of the agents' modulation weight $\mu_w$ and $\sigma_w$, while significant overconfidence is observed in $v$ and $\eta$.

\subsection{Model Sensitivity: Testing Parameter Recovery}

\begin{figure}[t]
    \begin{center}
    \includegraphics[width=\textwidth]{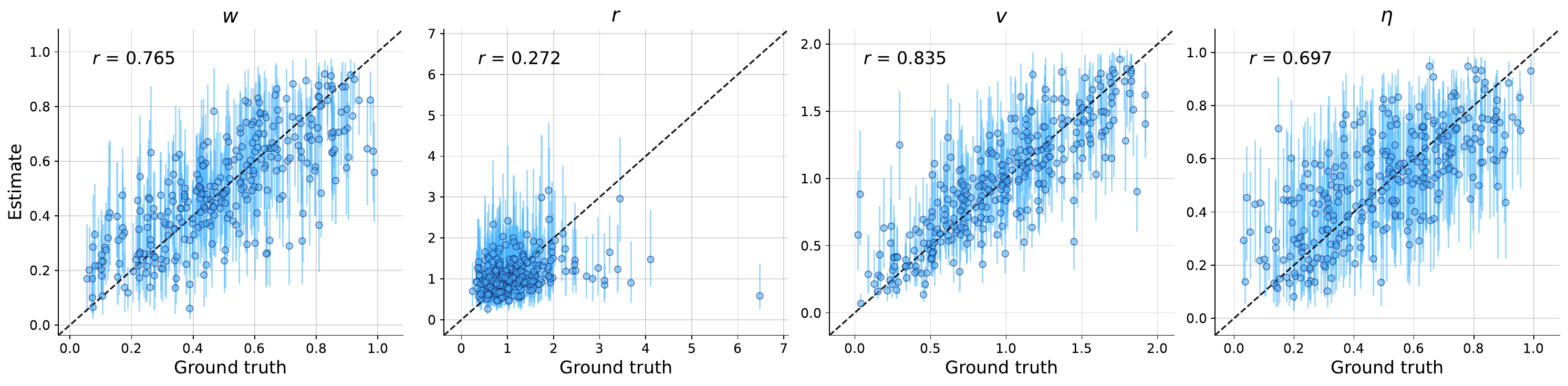}
    \includegraphics[width=\textwidth]{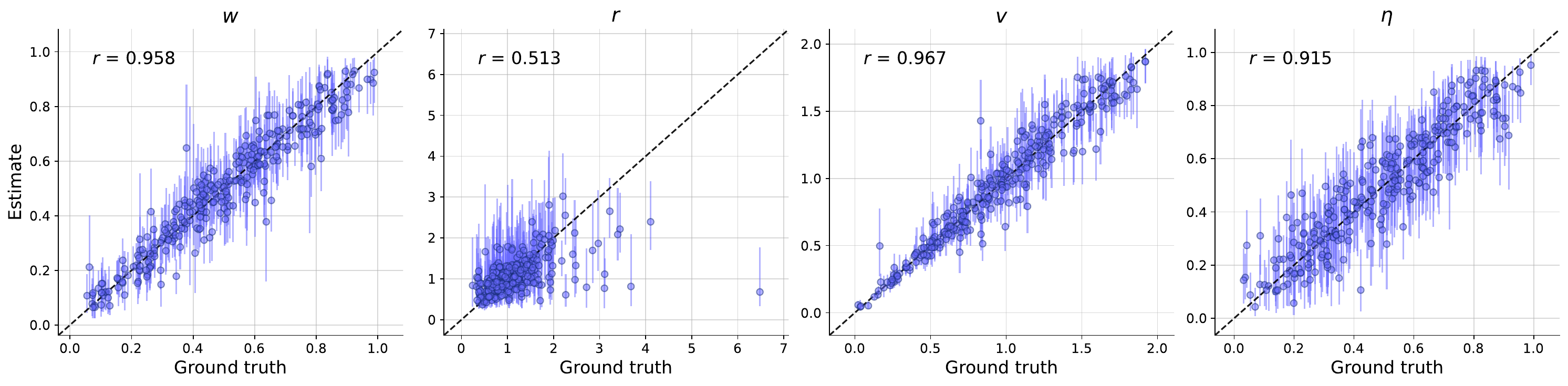}
    \includegraphics[width=\textwidth]{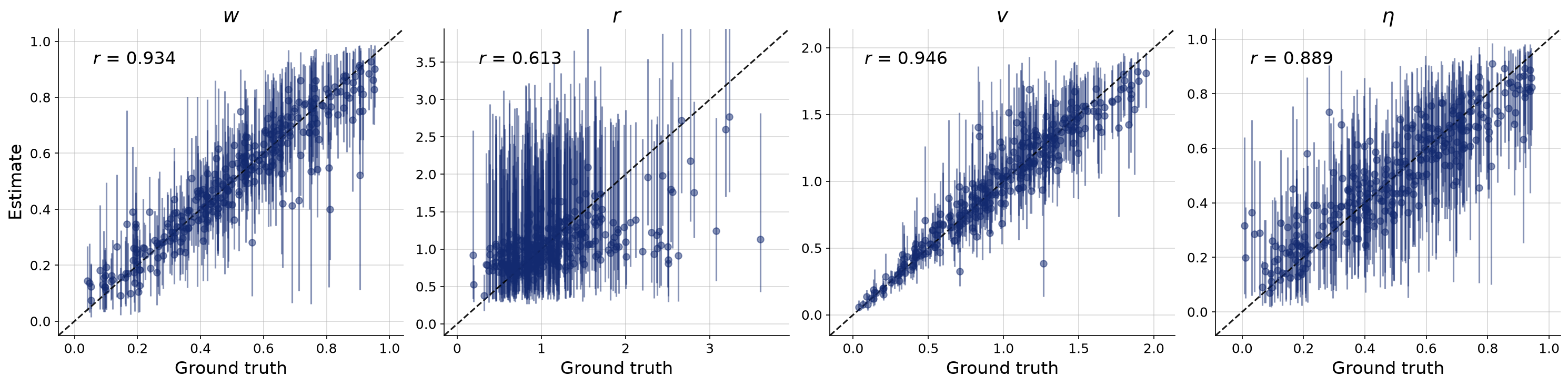}
    \end{center}
    \caption{\textit{Parameter recovery across the three training regimes}. \textit{Top}: recovery from the neural estimator trained on $3{,}000$ simulations. \textit{Middle}: recovery from the neural estimator trained on $30{,}000$ simulations. \textit{Bottom}: recovery from the neural estimator with online training. Dots indicate posterior medians and vertical lines denote 95\% credible intervals.}
    \label{fig:recovery}
\end{figure}

\begin{table}[htbp]
  \centering
  \caption{Parameter recovery metrics for the baseline model under three training regimes. NRMSE is normalized by prior-only prediction error; NRMSE and posterior contraction are medians across test simulations.}
  \label{tab:metrics}

  \begin{tabularx}{\textwidth}{@{} l c *{3}{Y} @{}}
    \toprule
    \multirow{2}{*}{\textbf{Training}} & \multirow{2}{*}{\textbf{Parameter}}
      & \textbf{Calibration} & \multirow{2}{*}{\textbf{NRMSE}} & \textbf{Posterior} \\
    & & \textbf{Error} & & \textbf{Contraction} \\
    \midrule
    \multirow{4}{*}{\begin{tabular}[c]{@{}l@{}}Offline\\ $N = 3{,}000$\end{tabular}}
      & $w$    & 0.149 & 0.453 & 0.869 \\
      & $r$    & 0.120 & 0.660 & 0.702 \\
      & $v$    & 0.170 & 0.362 & 0.901 \\
      & $\eta$ & 0.177 & 0.520 & 0.802 \\
    \midrule
    \multirow{4}{*}{\begin{tabular}[c]{@{}l@{}}Offline\\ $N = 30{,}000$\end{tabular}}
      & $w$    & 0.078 & 0.231 & 0.958 \\
      & $r$    & 0.076 & 0.491 & 0.839 \\
      & $v$    & 0.071 & 0.180 & 0.970 \\
      & $\eta$ & 0.121 & 0.310 & 0.916 \\
    \midrule
    \multirow{4}{*}{Online}
      & $w$    & 0.015 & 0.306 & 0.908 \\
      & $r$    & 0.040 & 0.726 & 0.496 \\
      & $v$    & 0.011 & 0.246 & 0.939 \\
      & $\eta$ & 0.011 & 0.416 & 0.817 \\
    \bottomrule
  \end{tabularx}
\end{table}

Next, we assess the global inferential adequacy of the neural estimator—how well it can recover model parameters from observable quantities. We compare the estimated posteriors to the true simulator parameters to evaluate both accuracy and confidence. Specifically, we quantify parameter recovery error and posterior contraction (uncertainty reduction relative to the prior) as summary measures of inferential performance.

Similar to empirical coverage, parameter recovery can be inspected graphically (\autoref{fig:recovery}). Effective parameter recovery is characterized by posterior distributions centered around the true parameters with reduced variance relative to prior distributions. As a summary metric, we also compute the Pearson correlation coefficients $r$ between the ground-truth simulator parameter and the median posterior estimates produced by the neural estimator, with better performance indexed by higher correlations.

The uncertainty of the neural estimator can be quantified through posterior contraction, which measures how much the posterior distribution narrows around the true parameter as more data become available. Greater contraction indicates improved estimator performance and a more information gained. Formally, posterior contraction (PC) is defined as
\begin{equation}\label{eq:pc}
    \text{PC} = 1 - \frac{\operatorname{Var}(\lambda_i^{\mathrm{post}})}{\operatorname{Var}(\lambda_i^{\text{prior}})},
\end{equation}
\noindent where $\mathrm{Var}(\lambda_i^{\mathrm{post}})$ is the posterior variance and $\operatorname{Var}(\lambda_i^{\text{prior}})$ is the prior variance \cite{schad2021toward}.
The total PC is the average of \eqref{eq:pc} across $S$ test simulations.
For completeness, we also compute the Bayesian normalized root-mean-square error (NRMSE), given by
\begin{equation}
    \mathrm{NRMSE} = \frac{1}{\lambda_{\max} - \lambda_{\min}}    
    \sqrt{
      \frac{1}{S}
      \sum_{s=1}^S \frac{1}{J}\sum_{j=1}^J
      (\hat{\lambda}_{j,s} - \lambda_s)^2
    },
\end{equation}
\noindent where $\lambda_s$ is the ground-truth parameter, $\lambda_{\max}$ and $\lambda_{\min}$ are the maximum and minimum parameter values, and $\hat{\lambda}_{j,s}$ is the $j$-th posterior sample obtained for the $s$-th simulation.
Together, these measures indicate whether something about the model parameters can be learned under idealized conditions (i.e., \textit{in silico}).

The parameter-wise results for recovery are shown in \autoref{fig:recovery}, and the parameter-wise PC and NRMSE are summarized in \autoref{tab:metrics}. We observe that the model can recover $w, v$, and $\eta$, but is limited in its recovery of $r$. This is apparent in graphical inspection of recovery, where uncertainty most notably reduced for $v$, followed by $w$ and $\eta$. Recovery of these parameters improves when more simulations are available, a typical behavior of SBI methods \cite{lueckmann2021benchmarking}.
Single-valued metrics also confirm this observation, where we observe higher PC and lower NRMSE for $w$, $v$, and $\eta$ than for $r$. These results suggest that the neural estimator can detect local non-identifiability of $r$, but is otherwise capable of identifying most model parameters based on the simulated data.

\begin{figure}[!htbp]
    \centering
    \includegraphics[width=\textwidth,height=0.62\textheight,keepaspectratio]{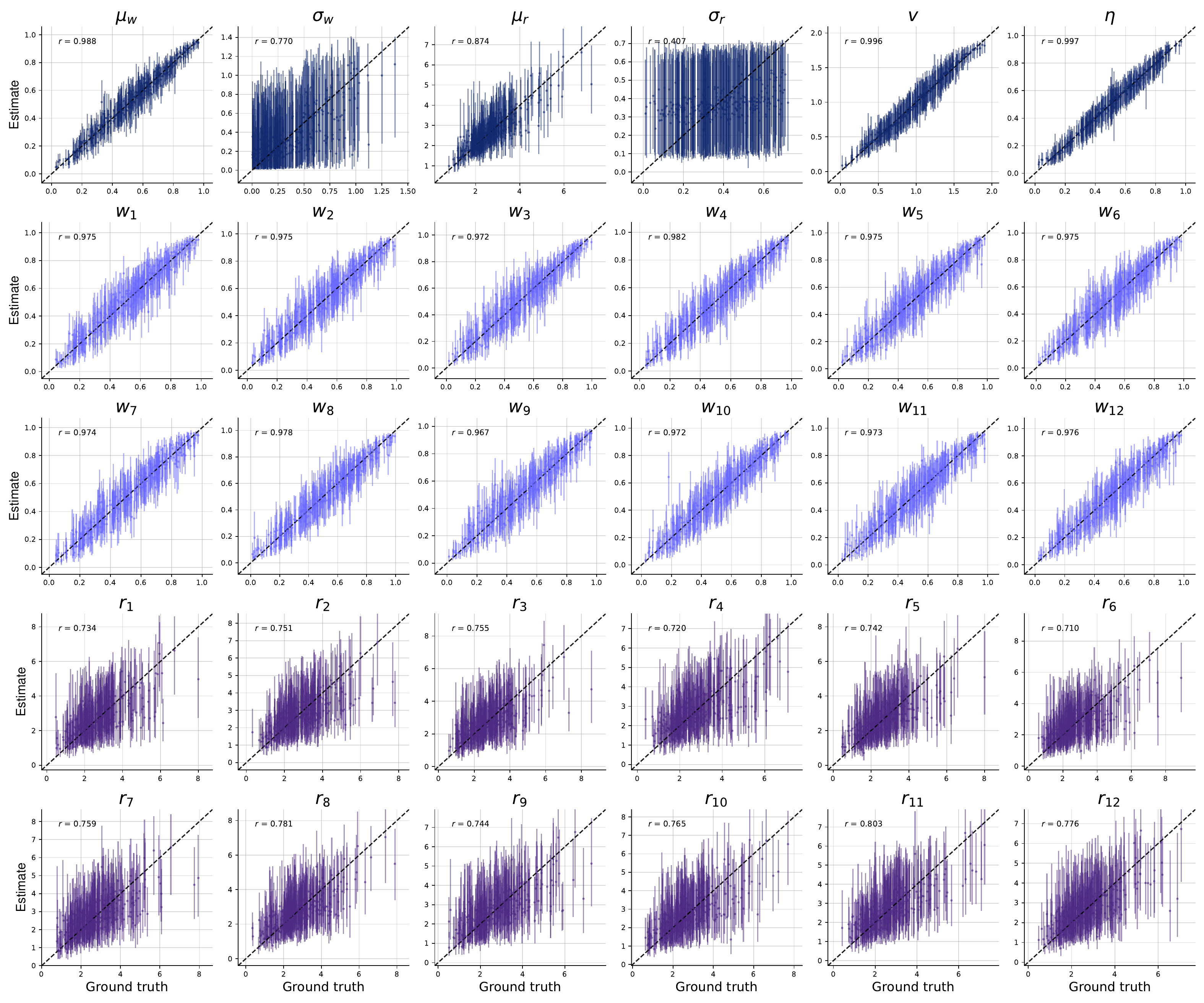}
\caption{\textit{Parameter recovery for the partial pooling model with 12 agents.} \textit{Row 1}: recovery for shared parameters including population-level means and variances for modulation weights ($\mu_w,\,\sigma_w$) and sensing radii ($\mu_r,\,\sigma_r$), as well as drift $v$ and noise amplitude $\eta$; \textit{Row 2-3}: recovery for the modulation weights $w_a$ for individual agents; \textit{Row 4-5}: recovery for the sensing radii $r_a$ for individual agents.}
    \label{fig:recovery-heterogeneity}
\end{figure}

\begin{figure}[!htbp]
    \centering
    \includegraphics[width=\textwidth]{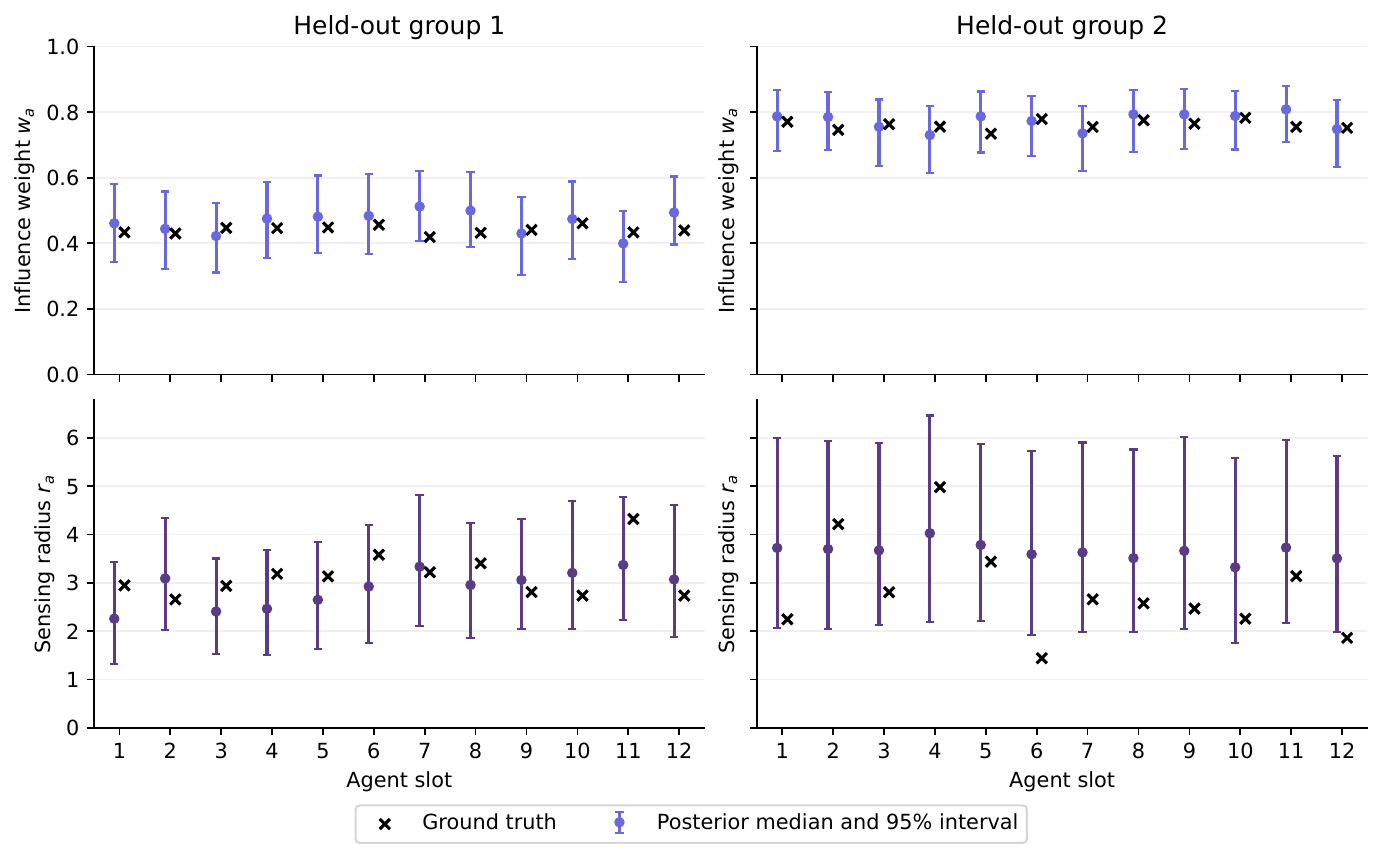}
\caption{Individual posterior estimates for the first two held-out groups in the 12-agent hierarchy, selected by index. The upper row shows influence weights and the lower row sensing radii. Colored points and bars show posterior medians and marginal $95\%$ credible intervals; black crosses show simulated truths.}
    \label{fig:individual-posteriors}
\end{figure}

\begin{table}[htbp]
\centering
\caption{Recovery in the 12-agent partial-pooling model with individual weights and bounded sensing radii. RMSE uses posterior medians; the prior baseline predicts each parameter by its empirical prior median. Individual rows report arithmetic means of the 12 agent-specific metrics; contraction is the mean reduction in posterior variance relative to the prior-reference bank.}
\label{tab:heterogeneity-recovery}
\small
\begin{tabular}{lccc}
\toprule
Parameter & RMSE & Prior-median RMSE & Contraction \\
\midrule
$\mu_w$ & 0.035 & 0.227 & 0.953 \\
$\sigma_w$ & 0.188 & 0.296  & 0.408 \\
$\mu_r$ & 0.488 & 1.021  & 0.736 \\
$\sigma_r$ & 0.161 & 0.175 & 0.155 \\
$v$ & 0.040 & 0.427  & 0.951 \\
$\eta$ & 0.018 & 0.229  & 0.959 \\
\midrule
$w_a$ (mean over agents) & 0.054 & 0.240  & 0.928 \\
$r_a$ (mean over agents) & 0.824 & 1.272  & 0.611 \\
\bottomrule
\end{tabular}
\end{table}

For partial pooling, we simulate $12$ agents over $20$-second trials. \autoref{fig:recovery-heterogeneity} shows overall recovery, while \autoref{fig:individual-posteriors} displays individual estimates and uncertainty within two held-out groups. \autoref{tab:heterogeneity-recovery} reports the aggregate metrics. Individual-weight RMSE averages $0.054$, compared with $0.240$ for the prior-median predictor, with mean marginal $95\%$ coverage of $95.8\%$.

After centering the weights within each group, RMSE is $0.049$ against a zero-difference baseline of $0.088$, correlation is $0.827$, and $95\%$ interval coverage is $96.4\%$. Individual radii also see improved parameter recovery: centered RMSE is $0.685$ against a zero-difference baseline of $0.755$, with correlation $0.422$ and $95\%$ coverage of $91.4\%$. Population radius spread is also weakly constrained, with a weak contraction $0.155$. Thus, these results suggests that the neural estimator is capable of accounting for individual differences in model parameters.

\FloatBarrier
\section{Discussion}
\label{sec:discussion}

In this work, we presented a novel computational model of human motion dynamics during collective navigation of virtual worlds in immersive rooms. At a high level, the model combines a diffusion-based and agent-based modeling approaches to represent individual agents' influences as spatial cognition and social dynamics, where collective motion patterns emerge from the agents' volitional interaction with the virtual world. The immersive room setting, in which the same virtual world experience is shared by multiple agents in the physical world, allows us to situate these influences and contain their observations in a unified manner, thereby highlighting the use of new tools to generate novel concepts and models of human interaction.

Further, we leveraged neural simulation-based inference (SBI). where we use a neural estimator to assess how well agents' internal properties can be identified from their observable motion patterns. As our results suggested, the agents’ exploration dynamics during virtual navigation, captured by the tradeoff between individual agency and the tendency to ``go with the flow'', can be characterized by the neural estimator with reasonable accuracy. In particular, parameters directly linked to an agent’s actions (i.e., the modulation weight $w$ and motion speed $v$) were recovered reliably, and remain sensitive under perturbative conditions, as manifested by the neural estimator's high performance in recovering the agent's rotational noise variance $\eta$. In addition, the neural estimator also detects the limited identifiability of the parameter reflecting agents' local environmental awareness (i.e., sensing radius $r$).

Among the identifiable parameters, the agent's modulation weight $w$, which can be interpreted as a state of the agent's motion governed by its situational awareness, is recovered with greater uncertainty than the drift rate $v$. This uncertainty is similar to that observed in the recovery of $\eta$, which reflects the decision uncertainty for the individual agent under internal influences. In other experiments, we observe that fixing noise variance $\eta$ resulted in poorer estimation of the modulation weight $w$. This suggests some degrees of interaction between the parameters. We interpret this as the result of a mixture effect introduced by the combined influences. In real-world scenarios, agents may not explicitly consider how much they wish to explore independently versus follow others. They are likely agnostic to both approaches and may lean toward one or the other without being self-aware. The same uncertainty can also be explained in data space: when the agent balances the two influences, the resulting data become ambiguous from the perspective of each isolated influence. This compounded ambiguity likely contributes to the variability in the estimated modulation weights.

The auditability of our SBI framework is afforded by the extensive diagnostics inherent in Bayesian workflows \cite{gelman2020bayesian} and made efficient through amortization. As demonstrated in our work, ABI can not only recover the governing parameters explicitly related to the agent's actions, but also detect the non-identifiability of model parameters related to the agents' local awareness of the environment. ABI's effectiveness in the context of this work has a broader implication to the use of immersive room as a tool to study perceptually-driven spatial cognition. On the one hand, we see an increasing adoption of room-scale virtual reality system with spatial audio capabilities in the study of multimodal perception \cite{wierstorf2017, kahl2025assessing}. On the other hand, from the perspective of modeling, the use of diffusion-based approaches in modeling perceptually-driven dynamics of cognition, especially those involving auditory perception, is also a relatively recent \cite{mathias2016unified}. By integrating our model in a tool-driven setup, we reveal the potential to make these two trends converge by developing structured theories of perceptually-driven spatial cognition in multi-user virtual environments. 

\subsection{Limitations}

The focus of our present work is on model development and simulation study, which suggests limitations and avenues for further development. First, the developed model and their corresponding extensions, while being able to simulate agent motion in the immersive rooms, do not suggest that the corresponding behavioral mechanisms can be reliably identified in human motion tracking data. Future work would be dedicated to experimental design and further evaluate the model's reliability and ecological validity using empirical observations on human participants.

Beyond the scope of our model extensions, there remain several valuable directions for further developments. First, the model remains a first-order dynamical system in which trajectories are simulated directly from speed and heading. A second-order formulation that explicitly models acceleration could capture changes in speed and smoother transitions during navigation. Second, the model assumes that all participants in the immersive room experience the virtual navigation in its full duration, and does not take into account the possibility of agent entering and exiting the room at any given time. A tracking mechanism for agent count would allow for partial agent observation. Such extensions would require additional validation of both its behavioral dynamics and parameter recovery.

The model also represents positions and headings in a two-dimensional plane. Thus, our simulation results are only representative within this planar setting. The model does not distinguish gaze from locomotor heading or represent vertical movement and responses to elevated targets. These omissions could affect the interpretation of inferred social influence or directional noise, although such effects have not been demonstrated by the present simulations. A practical extension would first retain floor-constrained locomotion while incorporating three-dimensional gaze and beacon elevation, followed by a full spatial formulation.

Finally, the beacon-salience extension remains a simplified representation of stimulus-driven attention. Richer formulations could represent beacons through auditory and visual features and allow agents to differ in their sensitivity to these features \cite{onat2007, kaya2017}. Integrating eye-tracking with motion-tracking data would also allow the model to distinguish gaze, head orientation, and locomotion, providing a stronger empirical basis for interpreting motion in terms of attention.

\subsection{Conclusion}

We developed a model of human motion dynamics during collective navigation in immersive virtual environments. The model represents agent motion as a mixture of individual and collective influences, and we can use simulation-based inference (SBI) to assess how well the underlying properties of agent locomotion can be recovered from interactions with salient spatial objects in the virtual world. To the best of our knowledge, this work represents one of the first applications of SBI to the study of social dynamics in the built environment. While architecture has only recently emerged as a context for investigating situated human interactions \cite{alavi2019}, the integration of simulation-based methods into the broader field of human–computer interaction also remains relatively nascent \cite{murraysmith2022}. 

At the same time, although prior studies have used deep learning to model task-driven multi-agent motion trajectories in finite spatial domains \cite{omidshafiei2021time}, no previous work in virtual environments has modeled the combined influence of local and global factors. By applying SBI in a simulated immersive room, our study takes a first step toward a computational model of distributed cognition in the built environment, with potential applications to human-centered interfaces in intelligent building systems. In doing so, we aim to contribute to the growing development of immersive rooms and other smart environments that can incorporate models of human behavior and ultimately transform how people experience the virtual world together.

\section*{Statement of consent for publication}

Part of \autoref{fig:dynamical-analogy} was adapted from \cite{beer2000} with permission from Elsevier. The permission was obtained on November 28, 2025.

\section*{Acknowledgement}

MJH and STR are supported by the National Science Foundation under Grant No.~2448380. We would like to acknowledge the Collaborative-Research Augmented Immersive Virtual Environment (CRAIVE-Lab) and the Curtis R.~Priem Experimental Media and Performing Art Center (EMPAC) at Rensselaer Polytechnic Institute for the facility support of this work. In addition, we would like to acknowledge Ted Krueger, Carla Leit\~{a}o, and Jonas Braasch for their thoughtful feedback during the development process.

\section*{Conflict of Interest Statement}

The authors declare no conflict of interest.

\section*{Code Availability}

The code implementation of the present work can be found in the following GitHub repository: \href{https://github.com/jerrymhuang/TogetherFlow}{https://github.com/jerrymhuang/TogetherFlow}.

\bibliographystyle{ieeetr}
\bibliography{references}

\begin{thebibliography}{10}

\bibitem{cruzneira1992}
C.~Cruz-Neira, D.~J. Sandin, T.~A. DeFanti, R.~V. Kenyon, and J.~C. Hart, ``The
  cave: Audio visual experience automatic virtual environment,'' {\em Commun.
  ACM}, vol.~35, p.~64–72, 6 1992.

\bibitem{sharma2017interactions}
G.~Sharma, J.~Braasch, and R.~J. Radke, ``Interactions in a human-scale
  immersive environment: The craive-lab,'' in {\em Cross-Surface 2016, in
  conjunction with the ACM International Conference on Interactive Surfaces and
  Spaces}, 2017.

\bibitem{huang2023rois}
M.~Huang, S.~R.~V. Chabot, C.~Leitão, T.~Krueger, and J.~Braasch,
  ``Spatially-aware group interaction design framework for collaborative
  room-oriented immersive systems,'' {\em Applied Ergonomics}, vol.~113,
  p.~104076, 2023.

\bibitem{kucheramorin2014}
J.~Kuchera-Morin, M.~Wright, G.~Wakefield, C.~Roberts, D.~Adderton, B.~Sajadi,
  T.~H{\"o}llerer, and A.~Majumder, ``Immersive full-surround multi-user system
  design,'' {\em Computers \& Graphics}, vol.~40, pp.~10--21, 2014.

\bibitem{chen2015}
W.~Chen, {\em Collaboration in Multi-user Immersive Virtual Environment}.
\newblock PhD thesis, Universit{\'e} Paris Saclay (COmUE), 2015.

\bibitem{kuhlen2014}
T.~W. Kuhlen and B.~Hentschel, ``Quo vadis cave: does immersive visualization
  still matter?,'' {\em IEEE computer graphics and applications}, vol.~34,
  no.~5, pp.~14--21, 2014.

\bibitem{chabot2020}
S.~Chabot, J.~Drozdal, M.~Peveler, Y.~Zhou, H.~Su, and J.~Braasch, ``A
  collaborative, immersive language learning environment using augmented
  panoramic imagery,'' in {\em 2020 6th international conference of the
  immersive learning research network (iLRN)}, pp.~225--229, IEEE, 2020.

\bibitem{jones2014roomalive}
B.~Jones, R.~Sodhi, M.~Murdock, R.~Mehra, H.~Benko, A.~Wilson, E.~Ofek,
  B.~MacIntyre, N.~Raghuvanshi, and L.~Shapira, ``Roomalive: Magical
  experiences enabled by scalable, adaptive projector-camera units,'' in {\em
  UIST '14 Proceedings of the 27th annual ACM symposium on User interface
  software and technology}, pp.~637--644, ACM, October 2014.

\bibitem{pejsa2016room2room}
T.~Pejsa, J.~Kantor, H.~Benko, E.~Ofek, and A.~Wilson, ``Room2room: Enabling
  life-size telepresence in a projected augmented reality environment,'' in
  {\em CSCW '16 Proceedings of the 19th ACM Conference on Computer-Supported
  Cooperative Work and Social Computing}, pp.~1716--1725, ACM, 2 2016.

\bibitem{sanz2015}
F.~A. Sanz, A.-H. Olivier, G.~Bruder, J.~Pettré, and A.~Lécuyer, ``Virtual
  proxemics: Locomotion in the presence of obstacles in large immersive
  projection environments,'' in {\em 2015 IEEE Virtual Reality (VR)},
  pp.~75--80, 2015.

\bibitem{warren2024crowds}
W.~H. Warren, J.~B. Falandays, K.~Yoshida, T.~D. Wirth, and B.~A. Free, ``Human
  crowds as social networks: Collective dynamics of consensus and
  polarization,'' {\em Perspectives on Psychological Science}, vol.~19, no.~2,
  pp.~522--537, 2024.
\newblock PMID: 37526132.

\bibitem{kirsh2019architects}
D.~Kirsh, ``Do architects and designers think about interactivity
  differently?,'' {\em ACM Transactions on Computer-Human Interaction (TOCHI)},
  vol.~26, no.~2, pp.~1--43, 2019.

\bibitem{kirsh2025reimagining}
D.~Kirsh, ``Reimagining space: how activity space explains human behaviour in
  buildings,'' {\em Architectural Science Review}, pp.~1--11, 2025.

\bibitem{huang2020immersion}
M.~Huang, S.~Chabot, T.~Krueger, C.~Leit{\~a}o, and J.~Braasch, ``From
  immersion to collaborative embodiment: extending collective interactivity in
  built environments through human-scale audiovisual immersive technologies,''
  {\em Digital Creativity}, vol.~31, no.~3, pp.~200--212, 2020.

\bibitem{beer2000}
R.~D. Beer, ``Dynamical approaches to cognitive science,'' {\em Trends in
  Cognitive Sciences}, vol.~4, no.~3, pp.~91--99, 2000.

\bibitem{bonabeau2002agent}
E.~Bonabeau, ``Agent-based modeling: Methods and techniques for simulating
  human systems,'' {\em Proceedings of the national academy of sciences},
  vol.~99, no.~suppl\_3, pp.~7280--7287, 2002.

\bibitem{demarchi2014agent}
S.~De~Marchi and S.~E. Page, ``Agent-based models,'' {\em Annual Review of
  political science}, vol.~17, no.~1, pp.~1--20, 2014.

\bibitem{dignum2025simulation}
E.~Dignum, H.~Choudhary, and M.~Lees, ``Simulation-based inference in
  agent-based models using spatio-temporal summary statistics,'' in {\em
  International Conference on Computational Science}, pp.~239--254, Springer,
  2025.

\bibitem{lavin2021simulation}
A.~Lavin, H.~Zenil, B.~Paige, {\em et~al.}, ``Simulation intelligence: Towards
  a new generation of scientific methods,'' {\em arXiv preprint}, 2021.

\bibitem{cranmer2020frontier}
K.~Cranmer, J.~Brehmer, and G.~Louppe, ``The frontier of simulation-based
  inference,'' {\em Proceedings of the National Academy of Sciences}, 2020.

\bibitem{radev_bayesflow_2023}
S.~T. Radev, M.~Schmitt, L.~Schumacher, L.~Elsemüller, V.~Pratz, Y.~Schälte,
  U.~Köthe, and P.-C. Bürkner, ``{BayesFlow}: {Amortized} {Bayesian}
  {Workflows} {With} {Neural} {Networks},'' {\em Journal of Open Source
  Software}, 2023.

\bibitem{ratcliff2016diffusion}
R.~Ratcliff, P.~L. Smith, S.~D. Brown, and G.~McKoon, ``Diffusion decision
  model: Current issues and history,'' {\em Trends in cognitive sciences},
  vol.~20, no.~4, pp.~260--281, 2016.

\bibitem{warren2018}
W.~H. Warren, ``Collective motion in human crowds,'' {\em Current directions in
  psychological science}, vol.~27, no.~4, pp.~232--240, 2018.

\bibitem{bill2022}
J.~Bill, S.~J. Gershman, and J.~Drugowitsch, ``Visual motion perception as
  online hierarchical inference,'' {\em Nature communications}, vol.~13, no.~1,
  p.~7403, 2022.

\bibitem{vicsek1995}
T.~Vicsek, A.~Czir\'ok, E.~Ben-Jacob, I.~Cohen, and O.~Shochet, ``Novel type of
  phase transition in a system of self-driven particles,'' {\em Phys. Rev.
  Lett.}, vol.~75, pp.~1226--1229, Aug 1995.

\bibitem{bechinger2016}
C.~Bechinger, R.~Di~Leonardo, H.~L\"owen, C.~Reichhardt, G.~Volpe, and
  G.~Volpe, ``Active particles in complex and crowded environments,'' {\em Rev.
  Mod. Phys.}, vol.~88, p.~045006, Nov 2016.

\bibitem{reynolds1987}
C.~W. Reynolds, ``Flocks, herds and schools: A distributed behavioral model,''
  {\em SIGGRAPH Comput. Graph.}, vol.~21, p.~25–34, aug 1987.

\bibitem{yanandkalay2006}
W.~Yan and Y.~E. Kalay, ``{Geometric, Cognitive and Behavioral Modeling of
  Environmental Users},'' in {\em Design Computing and Cognition '06} (J.~S.
  Gero, ed.), (Dordrecht), pp.~61--79, Springer Netherlands, 2006.

\bibitem{helbing1998social}
D.~Helbing and P.~Molnar, ``Social force model for pedestrian dynamics,'' {\em
  arXiv preprint cond-mat/9805244}, 1998.

\bibitem{truong2017toward}
X.-T. Truong and T.~D. Ngo, ``Toward socially aware robot navigation in dynamic
  and crowded environments: A proactive social motion model,'' {\em IEEE
  Transactions on Automation Science and Engineering}, vol.~14, no.~4,
  pp.~1743--1760, 2017.

\bibitem{kretzschmar2016socially}
H.~Kretzschmar, M.~Spies, C.~Sprunk, and W.~Burgard, ``Socially compliant
  mobile robot navigation via inverse reinforcement learning,'' {\em The
  International Journal of Robotics Research}, vol.~35, no.~11, pp.~1289--1307,
  2016.

\bibitem{dachner2022visual}
G.~C. Dachner, T.~D. Wirth, E.~Richmond, and W.~H. Warren, ``The visual
  coupling between neighbours explains local interactions underlying human
  ‘flocking','' {\em Proceedings of the Royal Society B}, vol.~289, no.~1970,
  p.~20212089, 2022.

\bibitem{dachner2014behavioral}
G.~C. Dachner and W.~H. Warren, ``Behavioral dynamics of heading alignment in
  pedestrian following,'' {\em Transportation Research Procedia}, vol.~2,
  pp.~69--76, 2014.

\bibitem{wirth2023}
T.~D. Wirth, G.~C. Dachner, K.~W. Rio, and W.~H. Warren, ``Is the neighborhood
  of interaction in human crowds metric, topological, or visual?,'' {\em PNAS
  Nexus}, vol.~2, p.~118, May 2023.

\bibitem{toner2018walking}
J.~Toner, ``Why walking is easier than pointing: Hydrodynamics of dry active
  matter,'' {\em arXiv preprint arXiv:1812.00310}, 2018.

\bibitem{warren2006dynamics}
W.~H. Warren, ``The dynamics of perception and action.,'' {\em Psychological
  review}, vol.~113, no.~2, p.~358, 2006.

\bibitem{dinocera2014}
D.~di~Nocera, A.~Finzi, S.~Rossi, and M.~Staffa, ``The role of intrinsic
  motivations in attention allocation and shifting,'' {\em Frontiers in
  Psychology}, vol.~5, 2014.

\bibitem{abrams2003}
R.~A. Abrams and S.~E. Christ, ``Motion onset captures attention,'' {\em
  Psychological Science}, vol.~14, no.~5, pp.~427--432, 2003.
\newblock PMID: 12930472.

\bibitem{prime2010}
S.~L. Prime and L.~R. Harris, ``Predicting the position of moving audiovisual
  stimuli,'' {\em Experimental Brain Research}, vol.~203, pp.~249--260, June
  2010.

\bibitem{carlileleung2016}
S.~Carlile and J.~Leung, ``The perception of auditory motion,'' {\em Trends in
  Hearing}, vol.~20, p.~2331216516644254, 2016.

\bibitem{leung2016head}
J.~Leung, V.~Wei, M.~Burgess, and S.~Carlile, ``Head tracking of auditory,
  visual, and audio-visual targets,'' {\em Frontiers in neuroscience}, vol.~9,
  p.~493, 2016.

\bibitem{royden1996}
C.~S. Royden and E.~C. Hildreth, ``Human heading judgments in the presence of
  moving objects,'' {\em Perception \& Psychophysics}, vol.~58, pp.~836--856,
  Jan. 1996.

\bibitem{roggerone2019motion}
V.~Roggerone, J.~Vacher, C.~Tarlao, and C.~Guastavino, ``Auditory motion
  perception emerges from successive sound localizations integrated over
  time,'' {\em Scientific Reports}, vol.~9, no.~1, p.~16437, 2019.

\bibitem{cohenlhyver2020}
B.~Cohen-Lhyver, S.~Argentieri, and B.~Gas, ``Audition as a trigger of head
  movements,'' in {\em The Technology of Binaural Understanding} (J.~Blauert
  and J.~Braasch, eds.), pp.~697--731, Cham: Springer International Publishing,
  2020.

\bibitem{kolarik2016distance}
A.~J. Kolarik, B.~C.~J. Moore, P.~Zahorik, S.~Cirstea, and S.~Pardhan,
  ``Auditory distance perception in humans: a review of cues, development,
  neuronal bases, and effects of sensory loss,'' {\em Attention, Perception, \&
  Psychophysics}, vol.~78, no.~2, pp.~373--395, 2016.

\bibitem{lepelley2016}
M.~E.~L. Pelley, C.~J. Mitchell, T.~Beesley, D.~N. George, and A.~J. Wills,
  ``Attention and associative learning in humans: An integrative review.,''
  {\em Psychological bulletin}, vol.~142 10, pp.~1111--1140, 2016.

\bibitem{fendrich2001}
R.~Fendrich and P.~M. Corballis, ``The temporal cross-capture of audition and
  vision,'' {\em Perception \& Psychophysics}, vol.~63, pp.~719--725, 5 2001.

\bibitem{ratcliff2008diffusion}
R.~Ratcliff and G.~McKoon, ``The diffusion decision model: theory and data for
  two-choice decision tasks,'' {\em Neural computation}, vol.~20, no.~4,
  pp.~873--922, 2008.

\bibitem{marienhagen2025bridging}
J.~Marienhagen, L.~B. Moyse, A.~Schakowski, B.~Kahl, J.~Davidson, A.~El~Hady,
  R.~H. Kurvers, and D.~Deffner, ``Bridging reinforcement-learning and
  drift-diffusion modeling to uncover the cognitive processes underlying
  collective foraging,'' {\em PsyArXiv}, 2025.

\bibitem{grazzini2017}
J.~Grazzini, M.~G. Richiardi, and M.~Tsionas, ``Bayesian estimation of
  agent-based models,'' {\em Journal of Economic Dynamics and Control},
  vol.~77, pp.~26--47, 2017.

\bibitem{shiono2021estimation}
T.~Shiono, ``Estimation of agent-based models using {Bayesian} deep learning
  approach of {BayesFlow},'' {\em Journal of Economic Dynamics and Control},
  vol.~125, p.~104082, 2021.

\bibitem{dyer2024}
J.~Dyer, P.~Cannon, J.~D. Farmer, and S.~M. Schmon, ``Black-box bayesian
  inference for agent-based models,'' {\em Journal of Economic Dynamics and
  Control}, vol.~161, p.~104827, 2024.

\bibitem{youngbloodpassmore2024}
M.~Youngblood and S.~Passmore, ``Simulation-based inference with deep learning
  shows speed climbers combine innovation and copying to improve performance,''
  {\em PsyArXiv}, Dec 2024.

\bibitem{pitocchelli2024}
J.~Pitocchelli, A.~Albina, R.~A. Bentley, D.~Guerra, and M.~Youngblood,
  ``Temporal stability in songs across the breeding range of geothlypis
  philadelphia (mourning warbler) may be due to learning fidelity and
  transmission biases,'' {\em Ornithology}, p.~ukae046, 2024.

\bibitem{angione2022}
C.~Angione, E.~Silverman, and E.~Yaneske, ``Using machine learning as a
  surrogate model for agent-based simulations,'' {\em PLOS ONE}, vol.~17,
  pp.~1--24, 02 2022.

\bibitem{dyer2021approximate}
J.~Dyer, P.~Cannon, and S.~M. Schmon, ``Approximate bayesian computation with
  path signatures,'' {\em arXiv preprint arXiv:2106.12555}, 2021.

\bibitem{daly2022}
A.~J. Daly, L.~{De Visscher}, J.~M. Baetens, and B.~{De Baets}, ``Quo vadis,
  agent-based modelling tools?,'' {\em Environmental Modelling and Software},
  vol.~157, p.~105514, 2022.

\bibitem{said2016applying}
N.~Said, M.~Engelhart, C.~Kirches, S.~K{\"o}rkel, and D.~V. Holt, ``Applying
  mathematical optimization methods to an act-r instance-based learning
  model,'' {\em PloS one}, vol.~11, no.~7, p.~e0158832, 2016.

\bibitem{radev2020bayesflow}
S.~T. Radev, U.~K. Mertens, A.~Voss, L.~Ardizzone, and U.~K{\"o}the,
  ``{BayesFlow}: Learning complex stochastic models with invertible neural
  networks,'' {\em IEEE transactions on neural networks and learning systems},
  2020.

\bibitem{sainsbury2024likelihood}
M.~Sainsbury-Dale, A.~Zammit-Mangion, and R.~Huser, ``Likelihood-free parameter
  estimation with neural bayes estimators,'' {\em The American Statistician},
  vol.~78, no.~1, pp.~1--14, 2024.

\bibitem{elsemuller2024deep}
L.~Elsem{\"u}ller, M.~Schnuerch, P.-C. B{\"u}rkner, and S.~T. Radev, ``A deep
  learning method for comparing bayesian hierarchical models.,'' {\em
  Psychological Methods}, 2024.

\bibitem{clemenson2021}
G.~D. Clemenson, A.~Maselli, A.~Fiannaca, A.~Miller, and M.~Gonzalez~Franco,
  ``Rethinking {GPS} navigation: creating cognitive maps through auditory
  clues,'' {\em Scientific Reports}, vol.~11, pp.~7764--7764, April 2021.

\bibitem{huang2024digitaltwin}
M.~J. Huang and J.~Braasch, ``Integrated digital twin of immersive rooms for
  remote spatial telepresence in shared virtual worlds,'' {\em The Journal of
  the Acoustical Society of America}, vol.~156, 2024.

\bibitem{neuhoff2018}
J.~G. Neuhoff, ``Adaptive biases in visual and auditory looming perception.,''
  in {\em Spatial biases in perception and cognition.}, pp.~180--190, New York,
  NY, US: Cambridge University Press, 2018.

\bibitem{hayhoe2005eye}
M.~Hayhoe and D.~Ballard, ``Eye movements in natural behavior,'' {\em Trends in
  cognitive sciences}, vol.~9, no.~4, pp.~188--194, 2005.

\bibitem{berkhout1993acoustic}
A.~J. Berkhout, D.~de~Vries, and P.~Vogel, ``Acoustic control by wave field
  synthesis,'' {\em The Journal of the Acoustical Society of America}, vol.~93,
  no.~5, pp.~2764--2778, 1993.

\bibitem{lipman2023flow}
Y.~Lipman, R.~T.~Q. Chen, H.~Ben-Hamu, M.~Nickel, and M.~Le, ``Flow matching
  for generative modeling,'' in {\em The 11th International Conference on
  Learning Representations}, 2023.

\bibitem{kuhmichel2026bayesflow}
L.~K{\"u}hmichel, J.~M. Huang, V.~Pratz, J.~Arruda, H.~Olischl{\"a}ger,
  D.~Habermann, S.~Kucharsky, L.~Elsem{\"u}ller, A.~Mishra, N.~Bracher, {\em
  et~al.}, ``Bayesflow 2.0: Multi-backend amortized bayesian inference in
  python,'' {\em arXiv preprint arXiv:2602.07098}, 2026.

\bibitem{wen2022transformers}
Q.~Wen, T.~Zhou, C.~Zhang, W.~Chen, Z.~Ma, J.~Yan, and L.~Sun, ``Transformers
  in time series: A survey,'' {\em arXiv preprint arXiv:2202.07125}, 2022.

\bibitem{talts2018validating}
S.~Talts, M.~Betancourt, D.~Simpson, A.~Vehtari, and A.~Gelman, ``Validating
  {Bayesian} inference algorithms with simulation-based calibration,'' {\em
  arXiv preprint}, 2018.

\bibitem{sailynoja2022}
T.~Säilynoja, P.-C. Bürkner, and A.~Vehtari, ``Graphical test for discrete
  uniformity and its applications in goodness-of-fit evaluation and multiple
  sample comparison,'' {\em Statistics and Computing}, vol.~32, 3 2022.

\bibitem{yao2023discriminative}
Y.~Yao and J.~Domke, ``Discriminative calibration: Check bayesian computation
  from simulations and flexible classifier,'' {\em Advances in Neural
  Information Processing Systems}, vol.~36, pp.~36106--36131, 2023.

\bibitem{schad2021toward}
D.~J. Schad, M.~Betancourt, and S.~Vasishth, ``Toward a principled bayesian
  workflow in cognitive science.,'' {\em Psychological methods}, vol.~26,
  no.~1, p.~103, 2021.

\bibitem{lueckmann2021benchmarking}
J.-M. Lueckmann, J.~Boelts, D.~Greenberg, P.~Goncalves, and J.~Macke,
  ``Benchmarking simulation-based inference,'' in {\em Proceedings of The 24th
  International Conference on Artificial Intelligence and Statistics}
  (A.~Banerjee and K.~Fukumizu, eds.), vol.~130 of {\em Proceedings of Machine
  Learning Research}, pp.~343--351, PMLR, 13--15 Apr 2021.

\bibitem{gelman2020bayesian}
A.~Gelman, A.~Vehtari, D.~Simpson, {\em et~al.}, ``Bayesian workflow,'' {\em
  arXiv preprint}, 2020.

\bibitem{wierstorf2017}
H.~Wierstorf, A.~Raake, and S.~Spors, ``Assessing localization accuracy in
  sound field synthesis,'' {\em The Journal of the Acoustical Society of
  America}, vol.~141, no.~2, pp.~1111--1119, 2017.

\bibitem{kahl2025assessing}
B.~Kahl, ``Assessing the viability of wave field synthesis in vr-based
  cognitive research,'' {\em arXiv preprint arXiv:2507.03797}, 2025.

\bibitem{mathias2016unified}
S.~R. Mathias, ``Unified analysis of accuracy and reaction times via models of
  decision making,'' in {\em Proceedings of Meetings on Acoustics}, vol.~26,
  AIP Publishing, 2016.

\bibitem{onat2007}
S.~Onat, K.~Libertus, and P.~König, ``Integrating audiovisual information for
  the control of overt attention,'' {\em Journal of Vision}, vol.~7,
  pp.~11--11, July 2007.

\bibitem{kaya2017}
E.~M. Kaya and M.~Elhilali, ``Modelling auditory attention,'' {\em
  Philosophical Transactions of the Royal Society B: Biological Sciences},
  vol.~372, no.~1714, p.~20160101, 2017.

\bibitem{alavi2019}
H.~S. Alavi, E.~F. Churchill, M.~Wiberg, D.~Lalanne, P.~Dalsgaard, A.~Fatah~gen
  Schieck, and Y.~Rogers, ``Introduction to human-building interaction (hbi):
  Interfacing hci with architecture and urban design,'' {\em ACM Trans.
  Comput.-Hum. Interact.}, vol.~26, Mar. 2019.

\bibitem{murraysmith2022}
R.~Murray-Smith, A.~Oulasvirta, A.~Howes, J.~M\"{u}ller, A.~Ikkala,
  M.~Bachinski, A.~Fleig, F.~Fischer, and M.~Klar, ``What simulation can do for
  hci research,'' {\em Interactions}, vol.~29, p.~48–53, Nov. 2022.

\bibitem{omidshafiei2021time}
S.~Omidshafiei, D.~Hennes, M.~Garnelo, E.~Tarassov, Z.~Wang, R.~Elie, J.~T.
  Connor, P.~Muller, I.~Graham, W.~Spearman, {\em et~al.}, ``Time-series
  imputation of temporally-occluded multiagent trajectories,'' {\em arXiv
  preprint arXiv:2106.04219}, 2021.

\bibitem{schumacher2023neural}
L.~Schumacher, P.-C. B{\"u}rkner, A.~Voss, U.~K{\"o}the, and S.~T. Radev,
  ``Neural superstatistics for bayesian estimation of dynamic cognitive
  models,'' {\em Scientific Reports}, vol.~13, no.~1, p.~13778, 2023.

\bibitem{wright2005open}
M.~Wright, ``Open sound control: an enabling technology for musical
  networking,'' {\em Organised Sound}, vol.~10, no.~3, pp.~193--200, 2005.

\bibitem{carpentier2021spat}
T.~Carpentier, ``Spat\~{}: a comprehensive toolbox for sound spatialization in
  max,'' {\em Ideas S{\'o}nicas}, vol.~13, no.~24, pp.~12--23, 2021.

\bibitem{pulkki2001spatial}
V.~Pulkki, {\em Spatial sound generation and perception by amplitude panning
  techniques}.
\newblock Helsinki University of Technology, 2001.

\end{thebibliography}
\clearpage
\appendix

\section{Model Extensions}
\label{sec:model-extensions}

\noindent The movement model formulated in \autoref{sec:model} made many static assumptions about the agents and the beacons. While we use this model as a baseline, its expressivity is inherently limited. In this appendix, we consider multiple scenarios where the model can be extended relaxing some of its assumptions. The scenarios are:

\begin{enumerate}
\item \textbf{Collision avoidance}: promotes separation between nearby agents by an added repulsive term.
\item \textbf{Salience-weighted beacons}: agents approach a beacon based on both its proximity and perceived significance.
\item \textbf{Time-varying influence}: agents' internal and external influences modulate dynamically during navigation.
\item \textbf{Beacon switching}: time-varying beacon salience motivates the agent to approach different beacons.
\end{enumerate}

For each extension, we followed the same SBI workflow as outlined in \autoref{sec:sim-study}. For Scenario 1 and 2, we train the corresponding neural estimators using BayesFlow \citep{kuhmichel2026bayesflow} over $300$ epochs, $50$ steps per epoch, and $32$ generated simulation batches per step, totaling $480{,}000$ simulations. We evaluate each neural estimator using $300$ simulations with $1{,}000$ posterior draws per simulation. For Scenario 3, we leverage neural superstatistics \citep{schumacher2023neural} to estimate the dynamics of the agents' modulation weight, and perform offline training on $30$-second trials, $4{,}000$ simulations, and $250$ posterior draws per test simulation. We report normalized root-mean-square error (NRMSE), posterior contraction, and expected calibration error (ECE). NRMSE is normalized by prior-only error. NRMSE and posterior contraction use median aggregation across test simulations. 

\subsection{Collision Avoidance}
\label{app:ext-collision}

\noindent As described in \autoref{ssec:internal-influence}, we used the Vicsek model as our base formulation of internal influence in \autoref{eq:vicsek}. This interaction aligns agents with their neighbors but does not enforce their separation. When collision avoidance is of interest, we add a repulsive term to the orientation update.

Let $\mathcal{N}^{\rho}_{a,t}=\{n\ne a: 0<\lVert\mathbf{x}_{a,t}-\mathbf{x}_{n,t}\rVert\leq\rho\}$ collect neighbors within a personal-space radius $\rho$. This radius is specified separately from the sensing radius $r$. The separation vector sums unit vectors directed away from nearby neighbors, weighted by their proximity,
\begin{equation}\label{eq:separation}
    \mathbf{s}_{a,t}
    = \sum_{n\in\mathcal{N}^{\rho}_{a,t}}
      q_{a,n,t}
      \frac{\mathbf{x}_{a,t}-\mathbf{x}_{n,t}}
           {\lVert\mathbf{x}_{a,t}-\mathbf{x}_{n,t}\rVert},
    \qquad
    q_{a,n,t}=1-\frac{\lVert\mathbf{x}_{a,t}-\mathbf{x}_{n,t}\rVert}{\rho}.
\end{equation}
The weight vanishes at the personal-space boundary. For $\lVert\mathbf{s}_{a,t}\rVert>0$, let $\psi_{a,t}$ be its bearing. The orientation update acquires the additional separation drift
\begin{equation}\label{eq:heading-separation}
    \Delta\theta^{\mathrm{sep}}_{a,t}
    =\kappa\lVert\mathbf{s}_{a,t}\rVert
      \mathcal{W}(\psi_{a,t}-\theta_{a,t})\,\Delta t,
\end{equation}
where the operator $\mathcal{W}$ indicates wrapping within a full circle so that the corresponding angular difference is confined in $[-\pi,\pi)$. This contribution is zero when the separation vector is zero and is added before the simulator's turn-rate cap. It sits outside the modulation by $w$, allowing separation to counteract both beacon attraction and neighbor alignment. We fix $\rho=0.4$ and $\kappa=0.6$ in the experiment below.

\begin{figure}[htbp]
    \begin{center}
    \includegraphics[width=\textwidth]{figures/ext-recovery-m0.png}\\[4pt]
    \includegraphics[width=\textwidth]{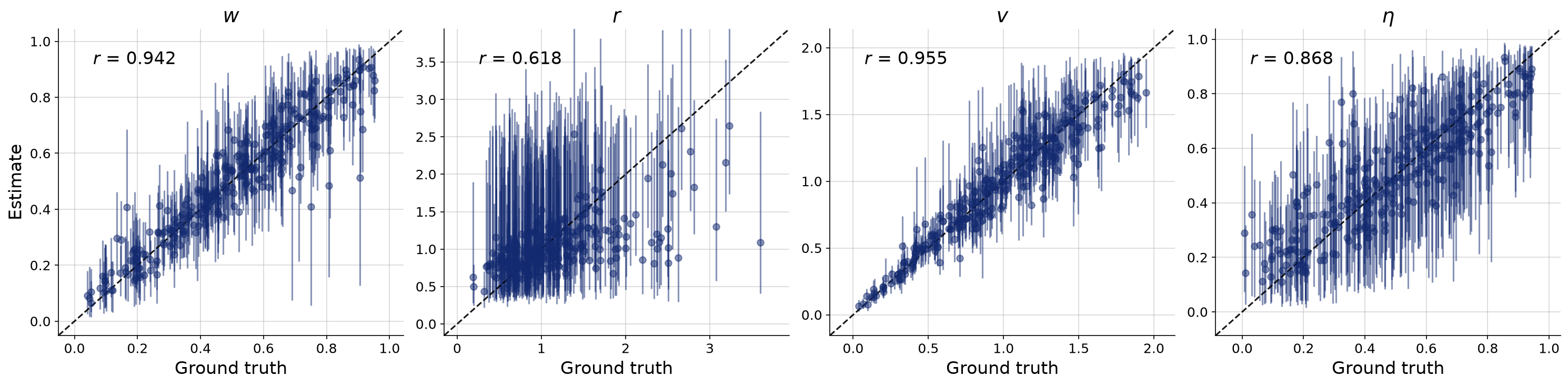}
    \end{center}
\caption{Parameter recovery for the baseline model (\textit{top}) and the collision-avoidance extension (\textit{bottom}), with $\rho=0.4$ and $\kappa=0.6$.}
    \label{fig:recovery-collision}
\end{figure}

\begin{figure}[htbp]
    \begin{center}
    \includegraphics[width=\textwidth]{figures/ext-coverage-m0.png}\\[4pt]
    \includegraphics[width=\textwidth]{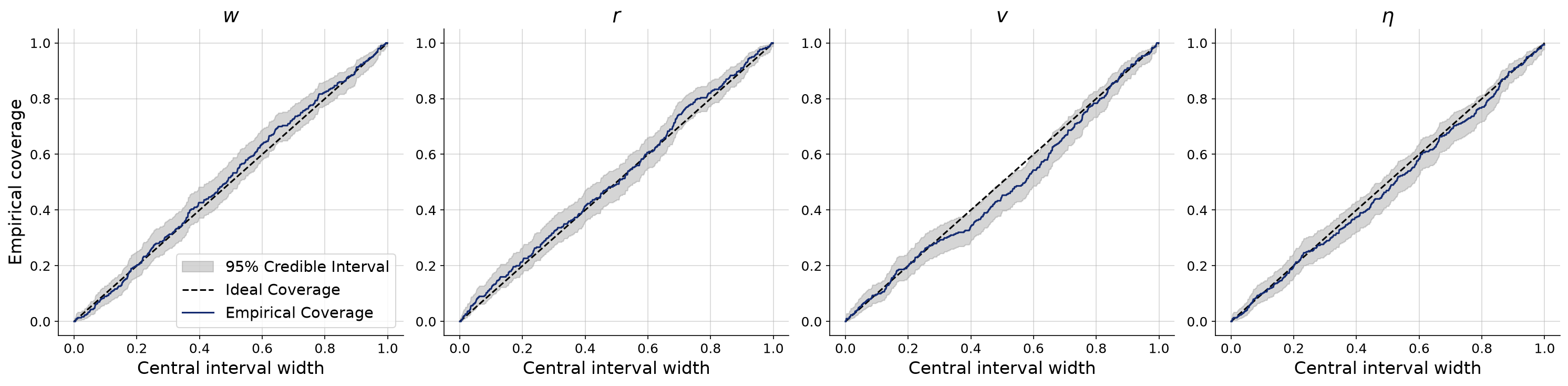}
    \end{center}
\caption{Empirical coverage for the baseline model (\textit{top}) and the collision-avoidance extension (\textit{bottom}), with $\rho=0.4$ and $\kappa=0.6$.}
    \label{fig:coverage-collision}
\end{figure}

\begin{table}[htbp]
\centering
\caption{Parameter recovery and calibration for the baseline model and the collision-avoidance extension with $\rho=0.4$ and $\kappa=0.6$.}
\label{tab:collision-recovery}
\small
\begin{tabular}{llcccc}
\toprule
& & $w$ & $r$ & $v$ & $\eta$ \\
\midrule
\multirow{3}{*}{Reference} & NRMSE & 0.307 & 0.827 & 0.258 & 0.423 \\
 & Posterior contraction & 0.910 & 0.302 & 0.936 & 0.826 \\
 & ECE & 0.032 & 0.009 & 0.016 & 0.016 \\
\midrule
\multirow{3}{*}{Collision avoidance} & NRMSE & 0.282 & 0.839 & 0.265 & 0.459 \\
 & Posterior contraction & 0.923 & 0.337 & 0.942 & 0.777 \\
 & ECE & 0.020 & 0.010 & 0.017 & 0.009 \\
\bottomrule
\end{tabular}
\end{table}

\paragraph{Simulation Study}
We compare the extension with the baseline scenario with no repulsion term added. Parameter recovery and empirical coverage are shown in \autoref{fig:recovery-collision} and \autoref{fig:coverage-collision}, with metrics in \autoref{tab:collision-recovery}. In general, we observe that adding the repulsion term does not affect the performance of parameter estimates. In this scenario, the recovery of the agents' sensing radius $r$ remains limited. 

\subsection{Beacon Salience}
\label{ssec:ext-salience}

\noindent The external influence of \autoref{ssec:external-influence} selects the nearest beacon, so the approach target is determined by proximity. In more realistic scenarios, beacons can differ in their strength in guiding agent movement. For example, a large, vivid landmark or a louder bird chirp may compete for attention with closer, less salient objects.

To account for beacon salience, we assign each beacon $b\in B$ a salience score $s_b>0$ and replace the selection rule with
\begin{equation}\label{eq:selection}
    b_{a,t}=\arg\max_{b\in B}
    \frac{s_b^{\,\alpha}}
         {\lVert\mathbf{x}_{a,t}-\mathbf{x}_{b,t}\rVert},
\end{equation}
where $\alpha\sim\mathrm{Gamma}(2,1)$ governs how strongly salience overrides proximity. At $\alpha=0$, the rule reduces to nearest-beacon selection; at $\alpha=1$, the score is salience divided by distance. Larger values give greater weight to salience. 

\begin{figure}[htbp]
    \begin{center}
    \includegraphics[width=\textwidth]{figures/ext-recovery-m0.png}\\[4pt]
    \includegraphics[width=\textwidth]{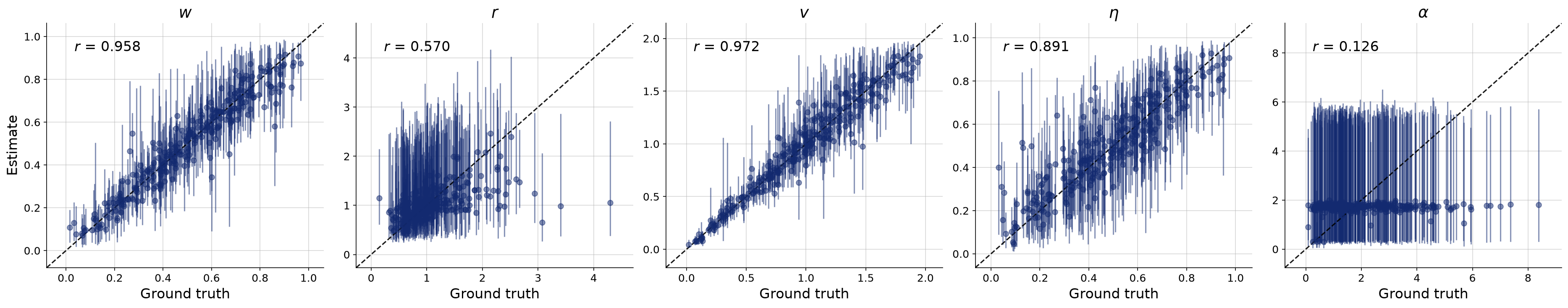}
    \end{center}
\caption{Parameter recovery for the baseline model (\textit{top}) and the beacon-salience extension (\textit{bottom}).}
    \label{fig:recovery-salience}
\end{figure}

\begin{figure}[htbp]
    \begin{center}
    \includegraphics[width=\textwidth]{figures/ext-coverage-m0.png}\\[4pt]
    \includegraphics[width=\textwidth]{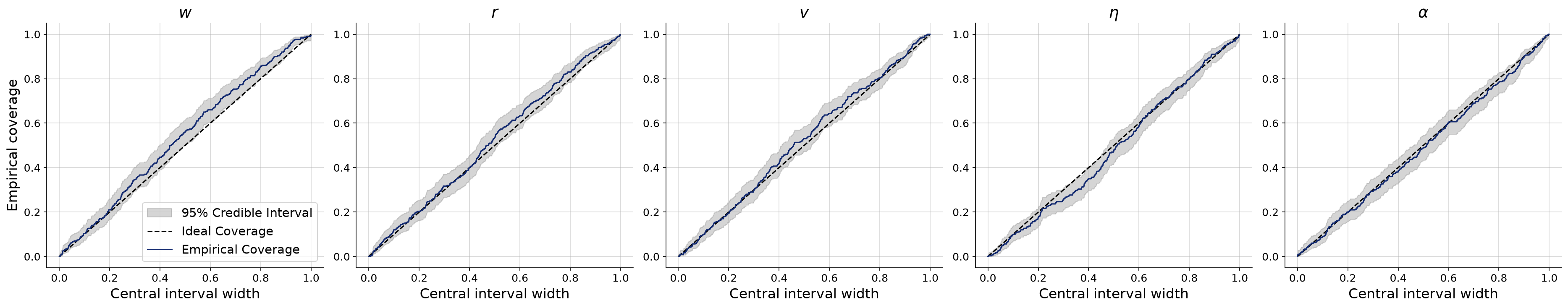}
    \end{center}
\caption{Empirical coverage for the baseline model (\textit{top}) and the beacon-salience extension (\textit{bottom}).}
    \label{fig:coverage-salience}
\end{figure}

\begin{table}[htbp]
\centering
\caption{Parameter recovery and calibration for the baseline model and the beacon-salience extension.}
\label{tab:salience-recovery}
\small
\begin{tabular}{llccccc}
\toprule
& & $w$ & $r$ & $v$ & $\eta$ & $\alpha$ \\
\midrule
\multirow{3}{*}{Reference} & NRMSE & 0.307 & 0.827 & 0.258 & 0.423 & --- \\
 & Posterior contraction & 0.910 & 0.302 & 0.936 & 0.826 & --- \\
 & ECE & 0.032 & 0.009 & 0.016 & 0.016 & --- \\
\midrule
\multirow{3}{*}{Salience} & NRMSE & 0.264 & 0.790 & 0.205 & 0.404 & 0.957 \\
 & Posterior contraction & 0.930 & 0.338 & 0.958 & 0.848 & 0.105 \\
 & ECE & 0.041 & 0.021 & 0.011 & 0.011 & 0.010 \\
\bottomrule
\end{tabular}
\end{table}

\paragraph{Simulation Study}
We compare the extension with the baseline with only proximity-based approach rule. For the extension, we add the salience strength $\alpha$ as an estimation target. Parameter recovery, empirical coverage, and metrics can be found in \autoref{fig:recovery-salience}, \autoref{fig:coverage-salience}, and \autoref{tab:salience-recovery}, respectively. In general, we see a noticeable improvement in the estimation of $w$ and $v$, while the sensing radius $r$ becomes less identifiable. However, The salience strength $\alpha$ is non-identifiable, with NRMSE $0.957$ and contraction $0.105$, despite an ECE of $0.010$. Thus, allowing salience to affect beacon selection does not result in the tradeoff between its strength and the other estimation targets.

\subsection{Time-Varying Influence}
\label{ssec:ext-timevarying}

\noindent In the baseline model, we assume that the influence weight remains constant throughout a trial. However, the balance between beacon-driven and neighbor-driven movement may change as agents navigate through the environment. To accommodate this possibility, we replace the constant weight $w$ with a time-varying weight $w_t$.

Let $z_t=\operatorname{logit}(w_t)$ be the agents' influence weight at logit scale. We model changes in influence using a random walk on the logit scale,
\begin{equation}
\label{eq:timevarying-influence}
    z_{t+\Delta t}=z_t+\tau\sqrt{\Delta t}\,\epsilon_t,
    \qquad \epsilon_t\sim\mathcal{N}(0,1),
    \qquad w_t=\frac{1}{1+\exp(-z_t)},
\end{equation}
with $z_0=\operatorname{logit}(w_0)$. Here, $w_0\sim\mathrm{Beta}(2,2)$ is the initial influence weight and and $\tau\sim\mathrm{Beta}(1,3)$ controls the magnitude of its fluctuations. The inverse-logit transformation keeps the weight within range $[0, 1]$. Setting $\tau=0$ recovers a constant influence weight.

This formulation allows influence to fluctuate without prescribing a direction of change. In particular, it does not impose an sudden transition from exploration to following. As a result, the varying shared influence weight over time can align closely with the smooth navigation experience of the agents. 

\begin{figure}[htbp]
    \begin{center}
    \includegraphics[width=\textwidth]{figures/ext-recovery-m0.png}\\[4pt]
    \includegraphics[width=\textwidth]{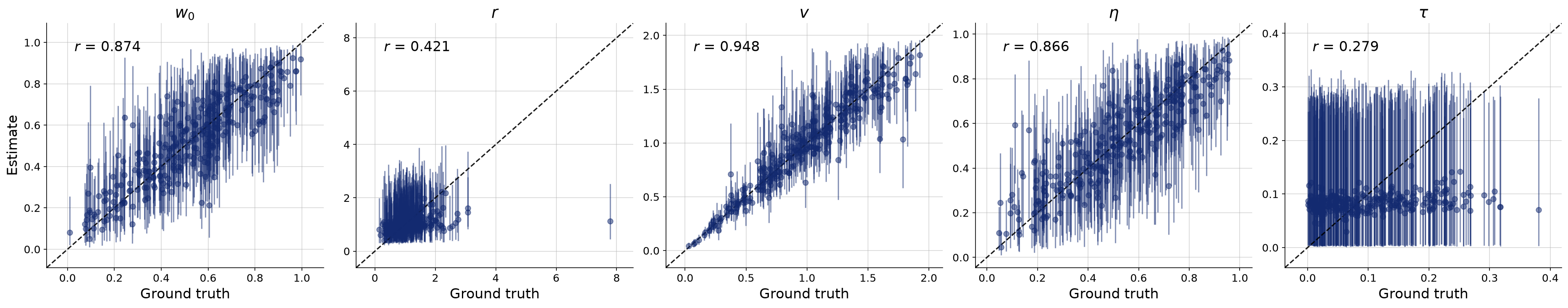}
    \end{center}
\caption{Parameter recovery for the baseline model (\textit{top}) and the time-varying influence extension (\textit{bottom}). The reference estimates a constant $w$, whereas the extension estimates the initial weight $w_0$ and volatility $\tau$ alongside the shared radius, speed, and heading noise.}
    \label{fig:recovery-timevarying}
\end{figure}

\begin{figure}[p]
    \centering
    \includegraphics[width=\textwidth,height=0.68\textheight,keepaspectratio]{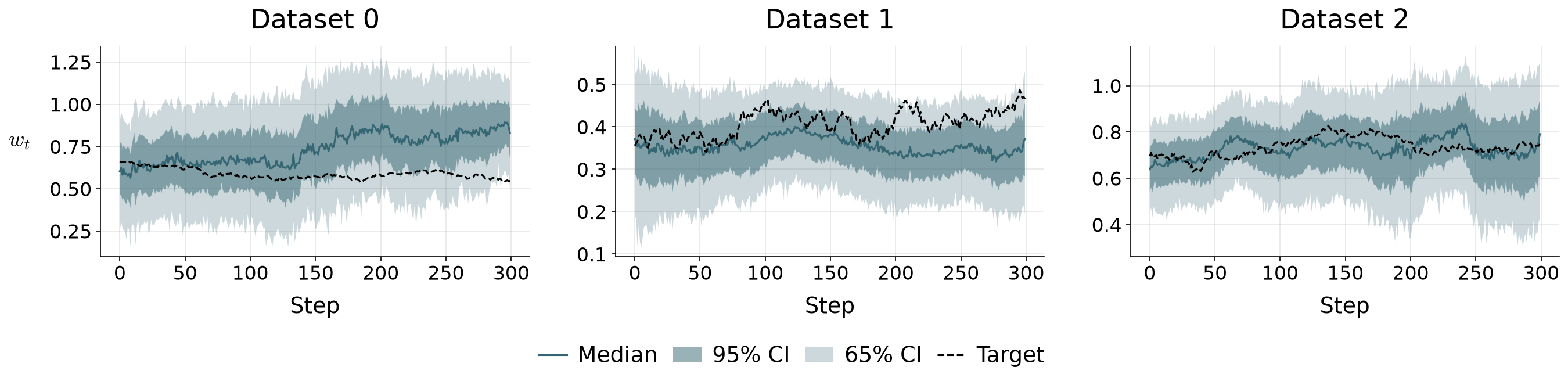}
\caption{Posterior estimates for the first three held-out datasets. Solid lines show posterior medians, shaded bands pointwise $95\%$ credible intervals, and dashed lines simulated targets. Only $w_t$ varies in the generator; $r$, $v$, $\eta$, and $\tau$ are constant targets predicted at each step. The horizontal axis is simulation step, with $\Delta t=0.1$ seconds.}
    \label{fig:timevarying-paths}
\end{figure}

\begin{table}[htbp]
\centering
\caption{Parameter recovery and calibration for the time-varying influence extension and the baseline model. Both infer $r$; the reference estimates a constant $w$, whereas the extension estimates $w_0$ and $\tau$.}
\label{tab:timevarying-recovery}
\small
\begin{tabular}{llccccc}
\toprule
& & $w$ or $w_0$ & $r$ & $v$ & $\eta$ & $\tau$ \\
\midrule
\multirow{3}{*}{Reference} & NRMSE & 0.307 & 0.827 & 0.258 & 0.423 & --- \\
 & Posterior contraction & 0.910 & 0.302 & 0.936 & 0.826 & --- \\
 & ECE & 0.032 & 0.009 & 0.016 & 0.016 & --- \\
\midrule
\multirow{3}{*}{Time-varying} & NRMSE & 0.466 & 0.752 & 0.280 & 0.478 & 0.941 \\
 & Posterior contraction & 0.797 & 0.445 & 0.914 & 0.767 & 0.090 \\
 & ECE & 0.014 & 0.017 & 0.029 & 0.020 & 0.027 \\
\bottomrule
\end{tabular}
\end{table}

\paragraph{Simulation Study}
For this study, we the time-varying influence $w_t$ along with the time-invariant parameters. The initial weight has contraction $0.797$, compared with $0.910$ for the reference's constant weight. The recovery of volatility $\tau$ is weaker, with NRMSE $0.941$ and contraction $0.090$. The experiment therefore provides less information about the magnitude of fluctuations than about the initial weight.

\autoref{fig:timevarying-paths} shows posterior trajectories of the modulation weights $w_t$ based on 3 simulated datasets. Mean per-step contraction for $w_t$ is $0.794$, with mean per-step ECE $0.0315$, but the examples also show tracking errors and broad intervals. The path can be informative while its volatility remains weakly recovered; these results concern simulated trajectories under a separate inference setup.

\subsection{Beacon Switching}
\label{ssec:ext-switching}

\noindent As an immersive room travels in the virtual world, the agents' attention can shift between different beacons due to the changes of their salience. Here, we consider an additional model extension where beacon salience changes over time. This extension represents settings in which there is a transition of perceptual significance that motivates the agent to switch targets to approach.

Each beacon receives a positive salience trajectory shared by all agents. Let $z_{b,t}=\log s_{b,t}$ denote the log-salience of beacon $b$. We model $z_{b,t}$ as a zero-mean Ornstein--Uhlenbeck (OU) process, using its exact transition over a simulation step $\Delta t$:
\begin{equation}
\label{eq:switching-ou}
\begin{aligned}
    z_{b,t+\Delta t} &= \rho z_{b,t}
        + \sigma_s\sqrt{1-\rho^2}\,\epsilon_{b,t},
    &\rho &= \exp\!\left(-\frac{\Delta t}{\tau_s}\right),\\
    s_{b,t} &= \exp(z_{b,t}),
    &\epsilon_{b,t} &\overset{\mathrm{iid}}{\sim}\mathcal{N}(0,1).
\end{aligned}
\end{equation}
We initialize $z_{b,0}\sim\mathcal{N}(0,\sigma_s^2)$ independently across beacons and independently of subsequent innovations, so each trajectory begins in its stationary distribution. Here, $\sigma_s$ is the stationary standard deviation of log-salience and $\tau_s$ is its correlation timescale: log-salience values separated by a lag $h\geq0$ have correlation $\exp(-h/\tau_s)$. The simulations use $\sigma_s=0.7$, $\tau_s=40\,\mathrm{s}$, and $\Delta t=0.1\,\mathrm{s}$. Beacon trajectories are independent of one another, while all agents in a trial experience the same realization for each beacon.

Beacons are scored by salience divided by distance, with the salience exponent fixed at $\alpha=1$. To prevent rapid switching between similarly attractive beacons, an agent retains its current target unless another beacon's score exceeds the current target's score by a factor of $1.5$. The salience-process parameters and switching threshold are fixed design settings, rather than additional inference targets.

\begin{figure}[htbp]
    \centering
    \includegraphics[width=\textwidth]{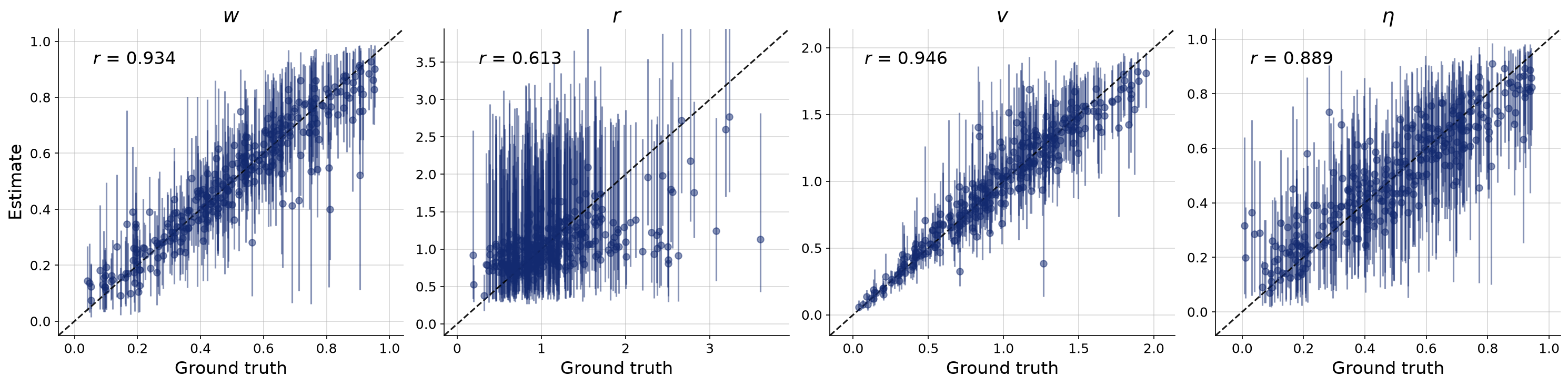}\\[4pt]
    \includegraphics[width=\textwidth]{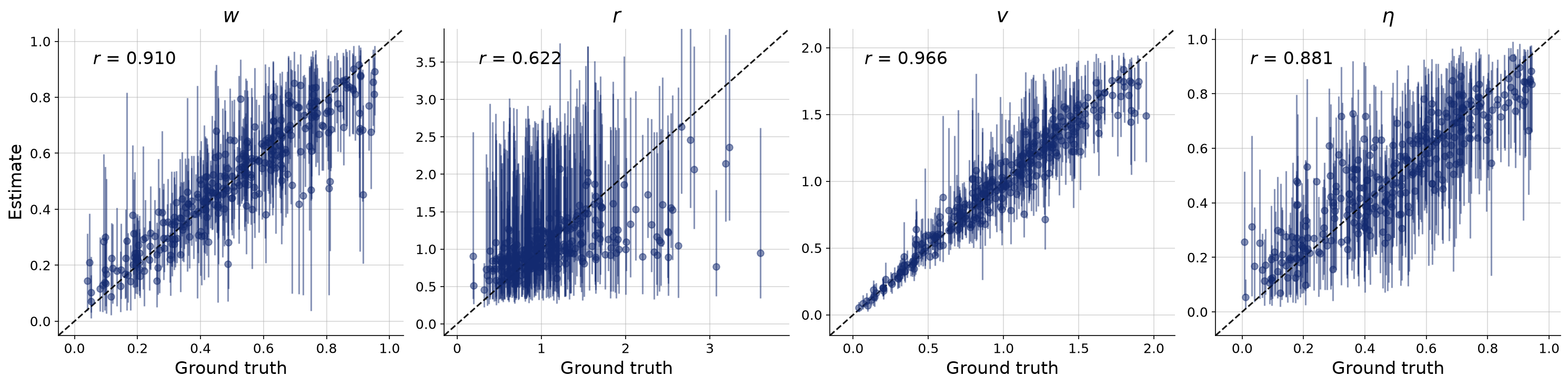}
    \caption{Parameter recovery for the baseline model (\textit{top}) and the beacon-switching extension (\textit{bottom}). Each estimator is trained and evaluated under its own generative model, using a separate set of $300$ held-out datasets.}
    \label{fig:recovery-switching}
\end{figure}

\begin{figure}[htbp]
    \centering
    \includegraphics[width=\textwidth]{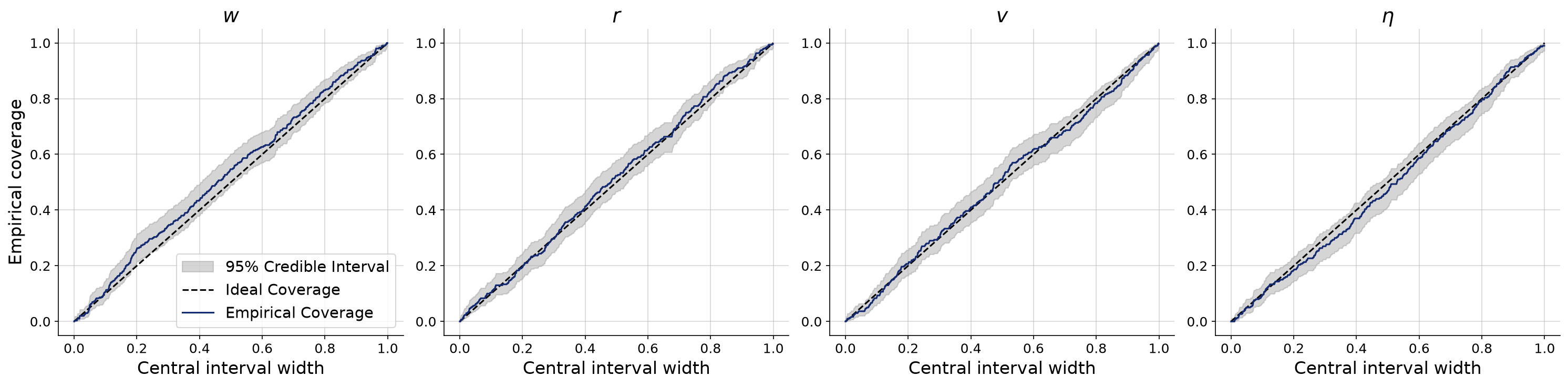}\\[4pt]
    \includegraphics[width=\textwidth]{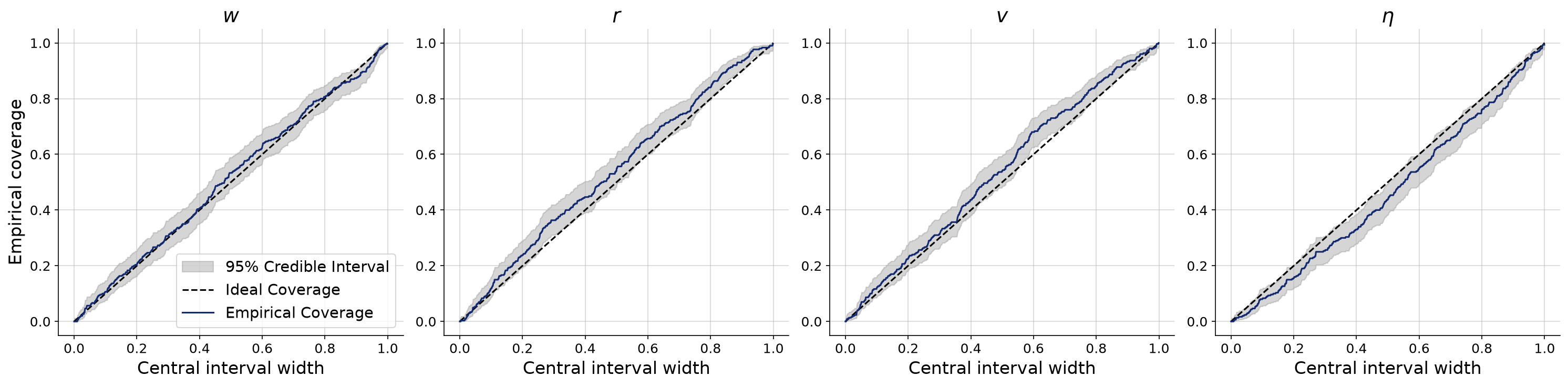}
    \caption{Empirical coverage for the baseline model (\textit{top}) and the beacon-switching extension (\textit{bottom}). Curves compare empirical coverage with the nominal probability of central credible intervals; the dashed diagonal denotes ideal coverage. Evaluation datasets are the same as in \autoref{fig:recovery-switching}.}
    \label{fig:coverage-switching}
\end{figure}

\begin{figure}[htbp]
    \centering
    \includegraphics[width=\textwidth]{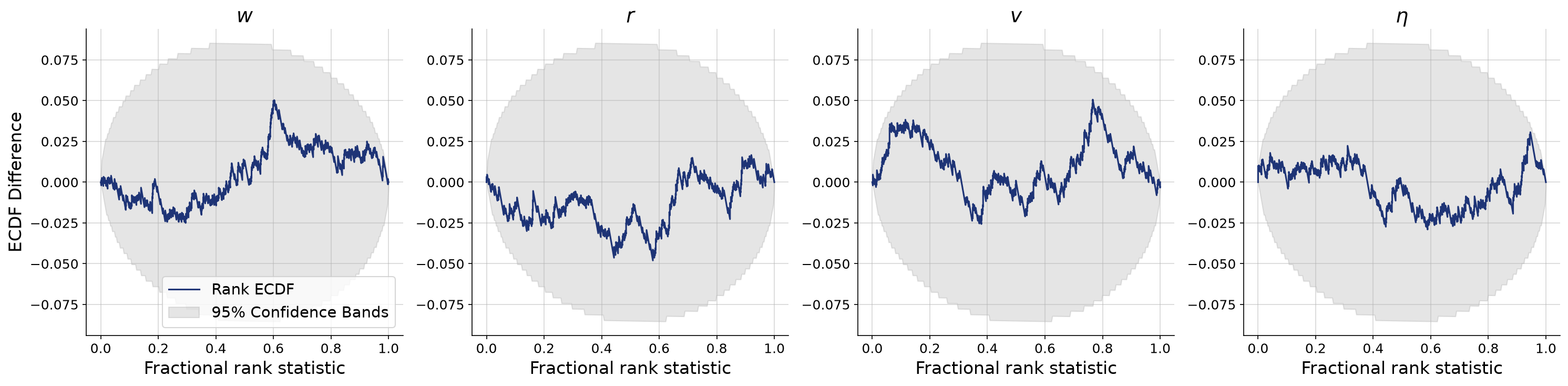}\\[4pt]
    \includegraphics[width=\textwidth]{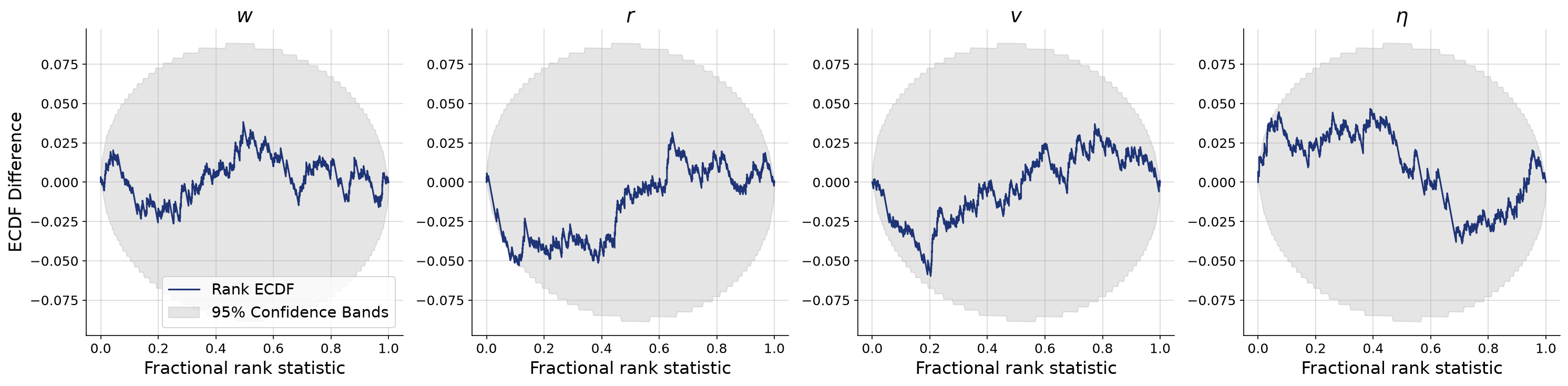}
    \caption{Marginal rank calibration for the baseline model (\textit{top}) and the beacon-switching extension (\textit{bottom}). Curves show the difference between the empirical cumulative distribution of posterior ranks and the uniform cumulative distribution. Shading indicates $95\%$ simultaneous confidence bands for each parameter.}
    \label{fig:calibration-switching}
\end{figure}

\begin{table}[htbp]
\centering
\caption{Parameter recovery and calibration for the baseline model and the beacon-switching extension. Each row block evaluates an estimator under its own generative model. NRMSE uses BayesFlow's prior normalization; ECE denotes coverage-calibration error. Lower NRMSE and ECE are preferable. Posterior contraction measures concentration relative to prior uncertainty and should be interpreted alongside recovery and calibration. Values are rounded to three decimals.}
\label{tab:switching-recovery}
\small
\begin{tabular}{llcccc}
\toprule
Model & Metric & $w$ & $r$ & $v$ & $\eta$ \\
\midrule
\multirow{3}{*}{Reference} & NRMSE & 0.307 & 0.827 & 0.258 & 0.423 \\
 & Posterior contraction & 0.910 & 0.302 & 0.936 & 0.826 \\
 & ECE & 0.032 & 0.009 & 0.016 & 0.016 \\
\midrule
\multirow{3}{*}{Switching} & NRMSE & 0.349 & 0.820 & 0.238 & 0.410 \\
 & Posterior contraction & 0.888 & 0.334 & 0.947 & 0.827 \\
 & ECE & 0.009 & 0.041 & 0.029 & 0.037 \\
\bottomrule
\end{tabular}
\end{table}

\paragraph{Simulation Study}
We assess whether the model parameters remain recoverable when beacon salience varies over time and agents can switch their approach targets. We train separate estimators for the proximity-based reference and the switching extension, using a transformer summary and flow matching. Each estimator is trained for $300$ online epochs of $50$ batches of $32$ simulations and evaluated on $300$ held-out datasets from its own generative model. Trials contain $49$ agents and four beacons over $60\,\mathrm{s}$, with a simulation step of $0.1\,\mathrm{s}$. Both estimators infer $(w,r,v,\eta)$ from positions, headings, angular velocities, and neighbor counts at fixed reference distances; the realized salience trajectories are not supplied to the estimator. Parameter recovery, empirical coverage, rank calibration, and numerical metrics are shown in \autoref{fig:recovery-switching}, \autoref{fig:coverage-switching}, \autoref{fig:calibration-switching}, and \autoref{tab:switching-recovery}, respectively.

The switching extension retains informative recovery of $w$, $v$, and $\eta$, with posterior contraction of $0.888$, $0.947$, and $0.827$, respectively. Relative to the reference, NRMSE increases for $w$ from $0.307$ to $0.349$, while it decreases modestly for $v$ from $0.258$ to $0.238$ and for $\eta$ from $0.423$ to $0.410$. Radius recovery remains weak in both models: under switching, $r$ has NRMSE $0.820$ and contraction $0.334$, compared with $0.827$ and $0.302$ in the reference. These results indicate that the additional beacon dynamics preserve recovery of the influence weight, speed, and heading-noise parameters, while providing little improvement in radius recovery.

Coverage-calibration errors for the switching extension range from $0.009$ to $0.041$. Calibration of $w$ improves relative to the reference, with ECE decreasing from $0.032$ to $0.009$, whereas ECE increases for $r$, $v$, and $\eta$. The coverage curves show some overcoverage for $r$ and $v$ and undercoverage for $\eta$ at intermediate nominal levels. The rank-calibration plots provide a complementary check of the posterior distributions. Thus, the switching extension supports informative inference for several parameters without uniform gains in recovery or calibration; low calibration error for $r$ should not be interpreted as precise radius estimation.

\FloatBarrier
\section{Sensitivity Analyses}
\label{app:sensitivity}

In addition to model extension, we also performed sensitivity analyses along two aspects:

\begin{enumerate}
    \item \textbf{Sensing radius identification}: compare parameter recovery when the shared sensing radius is inferred or fixed.
    \item \textbf{Prior sensitivity}: assess parameter recovery and calibration under alternative prior distributions.
\end{enumerate}

\subsection{Sensing radius identification}
\label{ssec:ext-radius-identification}

\begin{table}[htbp]
\centering
\caption{Parameter recovery and calibration with inferred or fixed sensing radius. The inferred-radius model uses $r\sim\mathrm{LogNormal}(0,0.5)$ and neighborhood summaries at fixed reference distances; fixed-radius models use summaries at their specified radius. Metrics evaluate each model under its own specification, so differences include the observation definitions and radius distribution as well as inference dimension. Dashes indicate that radius is not inferred.}
\label{tab:radius-identification}
\small
\begin{tabular}{llcccc}
\toprule
& & $w$ & $r$ & $v$ & $\eta$ \\
\midrule
\multirow{3}{*}{Inferred radius} & NRMSE & 0.307 & 0.827 & 0.258 & 0.423 \\
 & Posterior contraction & 0.910 & 0.302 & 0.936 & 0.826 \\
 & ECE & 0.032 & 0.009 & 0.016 & 0.016 \\
\midrule
\multirow{3}{*}{Fixed $r=2.5$} & NRMSE & 0.185 & --- & 0.203 & 0.385 \\
 & Posterior contraction & 0.968 & --- & 0.957 & 0.847 \\
 & ECE & 0.012 & --- & 0.016 & 0.009 \\
\midrule
\multirow{3}{*}{Fixed $r=1.25$} & NRMSE & 0.208 & --- & 0.168 & 0.335 \\
 & Posterior contraction & 0.954 & --- & 0.968 & 0.884 \\
 & ECE & 0.009 & --- & 0.042 & 0.011 \\
\midrule
\multirow{3}{*}{Fixed $r=4$} & NRMSE & 0.173 & --- & 0.217 & 0.409 \\
 & Posterior contraction & 0.970 & --- & 0.954 & 0.820 \\
 & ECE & 0.015 & --- & 0.019 & 0.013 \\
\bottomrule
\end{tabular}
\end{table}

\autoref{tab:radius-identification} shows limited recovery of the unknown shared radius, with NRMSE $0.827$ and contraction $0.302$, despite ECE $0.009$. With $r=2.5$ fixed, NRMSE decreases from $0.307$ to $0.185$ for $w$, from $0.258$ to $0.203$ for $v$, and from $0.423$ to $0.385$ for $\eta$, with greater contraction for all three parameters.

All three fixed-radius specifications yield lower NRMSE for the remaining parameters than the inferred-radius model, but no constant performs best across all diagnostics: $r=1.25$ gives the lowest speed NRMSE and the highest speed ECE ($0.042$). Fixing radius is therefore a useful working specification whose performance remains conditional on the chosen value. Because observation definitions and test populations also differ, the gains cannot be attributed solely to removing radius from inference; these experiments do not establish structural non-identifiability of $r$.

\subsection{Prior Sensitivity}
\label{ssec:ext-priors}

\noindent We assess sensitivity to the priors for $w$, $v$, and $\eta$ using the fixed-radius specification with $r=2.5$. \autoref{tab:prior-specs} lists the reference and five alternatives. The wide and concentrated configurations change all three priors together; the remaining configurations change only the speed or noise prior. The radius is a specified constant throughout this analysis; sensitivity to its value is reported in \autoref{ssec:ext-radius-identification}.

\begin{table}[htbp]
\centering
\caption{Prior specifications for the fixed-radius sensitivity analysis. All configurations fix $r=2.5$ room units.}
\label{tab:prior-specs}
\small
\begin{tabular}{lccc}
\toprule
Configuration & $w$ & $v$ & $\eta$ \\
\midrule
Reference & $\mathrm{Beta}(2,2)$ & $2\,\mathrm{Beta}(2,2)$ & $\mathrm{Beta}(2,2)$ \\
Wide & $\mathrm{Beta}(1,1)$ & $2\,\mathrm{Beta}(1,1)$ & $\mathrm{Beta}(1,1)$ \\
Concentrated & $\mathrm{Beta}(5,5)$ & $2\,\mathrm{Beta}(5,5)$ & $\mathrm{Beta}(5,5)$ \\
Slower speed & $\mathrm{Beta}(2,2)$ & $0.05+0.95\,\mathrm{Beta}(2,2)$ & $\mathrm{Beta}(2,2)$ \\
Lower noise & $\mathrm{Beta}(2,2)$ & $2\,\mathrm{Beta}(2,2)$ & $\mathrm{Beta}(2,5)$ \\
Higher noise & $\mathrm{Beta}(2,2)$ & $2\,\mathrm{Beta}(2,2)$ & $1.5\,\mathrm{Beta}(2,2)$ \\
\bottomrule
\end{tabular}
\end{table}

\begin{figure}[p]
    \begin{center}
    \includegraphics[width=\textwidth]{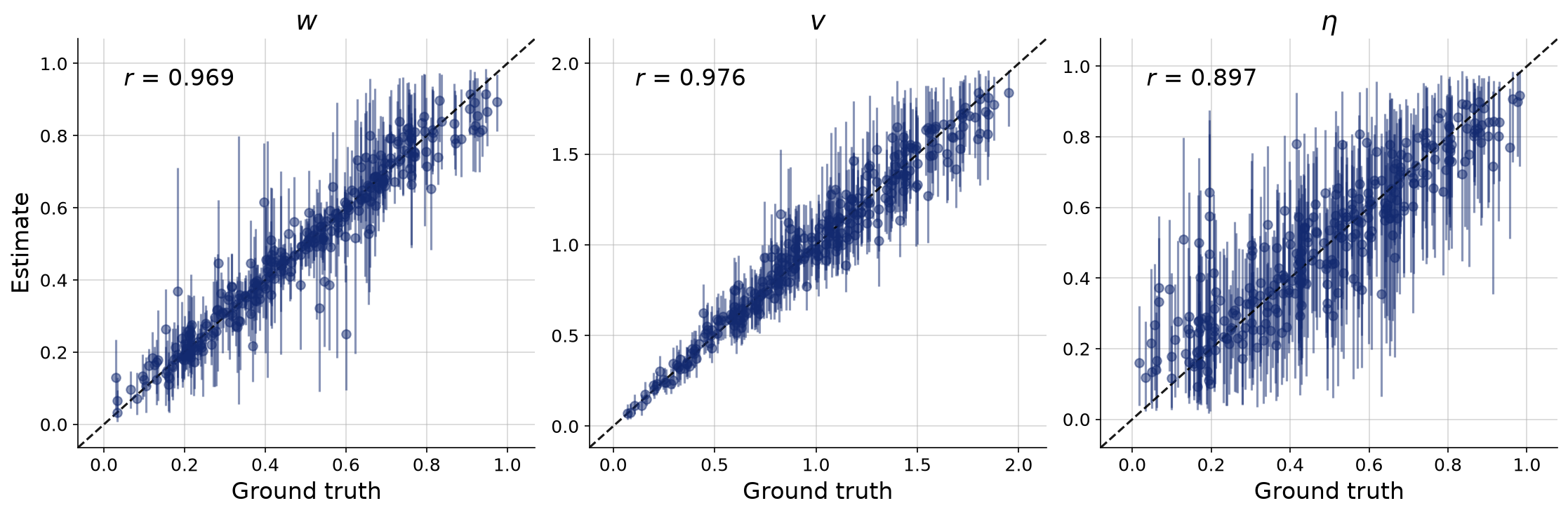}\\[4pt]
    \includegraphics[width=\textwidth]{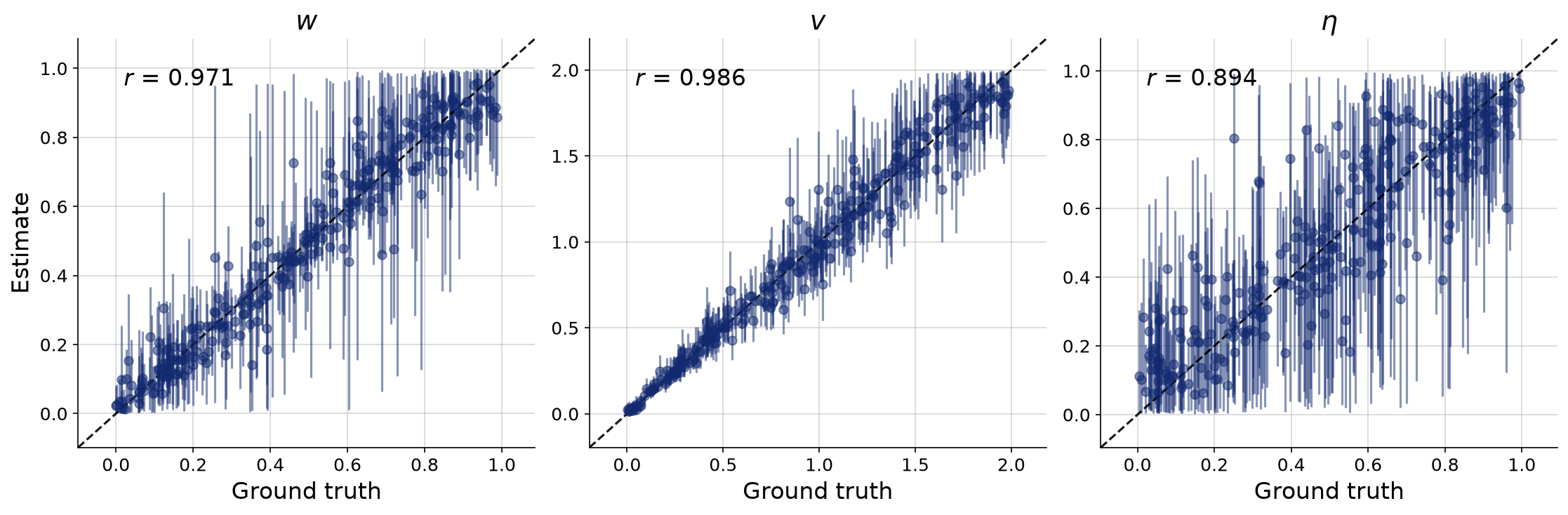}\\[4pt]
    \includegraphics[width=\textwidth]{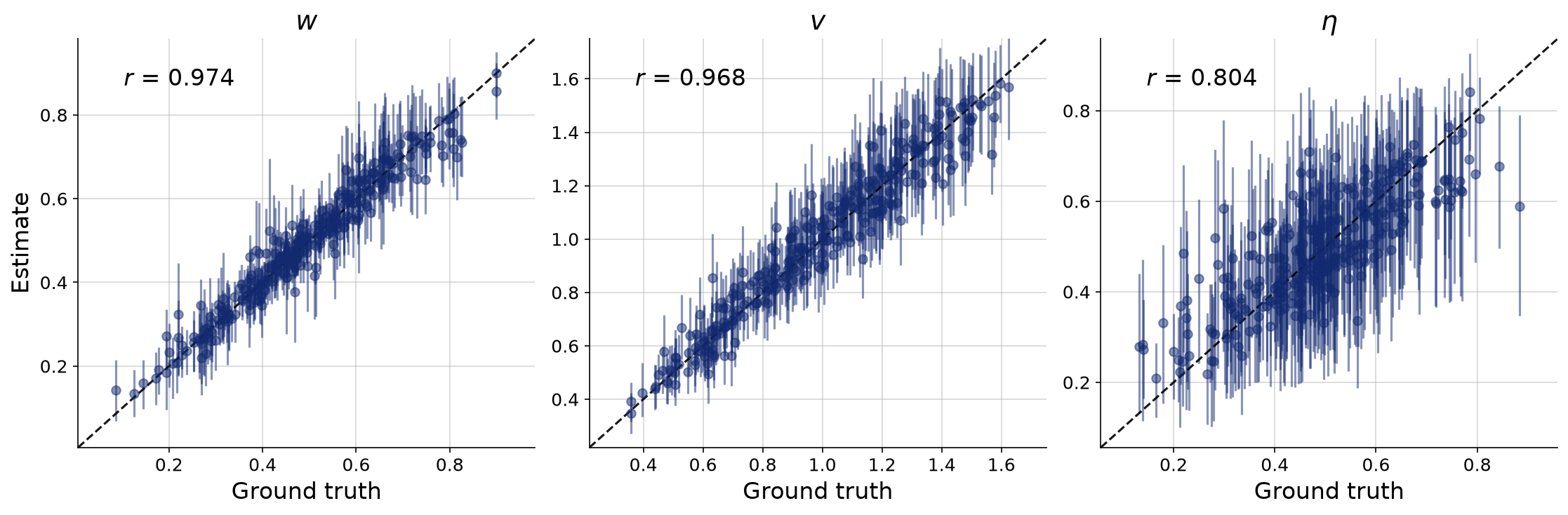}
    \end{center}
\caption{Parameter recovery under the fixed-radius reference priors (\textit{top}), wide priors (\textit{middle}), and concentrated priors (\textit{bottom}). Each row shows $w$, $v$, and $\eta$. Axis ranges reflect the different prior populations; each estimator is evaluated under its own prior specification.}
    \label{fig:recovery-priors}
\end{figure}

\begin{table}[htbp]
\centering
\caption{Parameter recovery and calibration under alternative priors, conditional on $r=2.5$. Each model is evaluated on simulations from its own prior specification. NRMSE is normalized by the corresponding prior-only prediction error; contraction is measured relative to that prior.}
\label{tab:prior-sensitivity}
\small
\begin{tabular}{llccc}
\toprule
& & $w$ & $v$ & $\eta$ \\
\midrule
\multirow{3}{*}{Reference} & NRMSE & 0.185 & 0.203 & 0.385 \\
 & Posterior contraction & 0.968 & 0.957 & 0.847 \\
 & ECE & 0.012 & 0.016 & 0.009 \\
\midrule
\multirow{3}{*}{Wide} & NRMSE & 0.190 & 0.146 & 0.358 \\
 & Posterior contraction & 0.961 & 0.974 & 0.866 \\
 & ECE & 0.014 & 0.026 & 0.006 \\
\midrule
\multirow{3}{*}{Concentrated} & NRMSE & 0.201 & 0.253 & 0.577 \\
 & Posterior contraction & 0.960 & 0.938 & 0.707 \\
 & ECE & 0.013 & 0.014 & 0.009 \\
\midrule
\multirow{3}{*}{Slower speed} & NRMSE & 0.224 & 0.207 & 0.422 \\
 & Posterior contraction & 0.948 & 0.960 & 0.823 \\
 & ECE & 0.016 & 0.010 & 0.015 \\
\midrule
\multirow{3}{*}{Lower noise} & NRMSE & 0.215 & 0.217 & 0.583 \\
 & Posterior contraction & 0.959 & 0.950 & 0.677 \\
 & ECE & 0.008 & 0.023 & 0.015 \\
\midrule
\multirow{3}{*}{Higher noise} & NRMSE & 0.198 & 0.183 & 0.274 \\
 & Posterior contraction & 0.959 & 0.963 & 0.923 \\
 & ECE & 0.028 & 0.021 & 0.023 \\
\bottomrule
\end{tabular}
\end{table}

\autoref{fig:recovery-priors} compares recovery under the reference, wide, and concentrated priors; \autoref{tab:prior-sensitivity} summarizes all configurations. Across the six specifications, ECE remains below $0.028$, while NRMSE ranges from $0.185$ to $0.224$ for $w$, $0.146$ to $0.253$ for $v$, and $0.274$ to $0.583$ for $\eta$. Noise is the least accurately recovered parameter by this metric in every configuration.

Noise contraction varies from $0.677$ under the lower-noise prior to $0.923$ under the higher-noise prior, compared with $0.847$ in the reference. These scores depend on both the prior variance and the simulated parameter population; larger contraction alone does not establish greater absolute precision. Restricting speed gives NRMSE $0.207$ versus $0.203$ in the reference, despite lower ECE ($0.010$ versus $0.016$).

These experiments support low marginal calibration error across the tested priors, while recovery varies with the specification. They assess inference over different prior-predictive populations and do not establish that posteriors for identical observations are insensitive to prior choice. The conclusions are conditional on the fixed radius and do not establish robustness to alternative priors on an unknown $r$.

\subsection{Effect of Simulation Horizon}

\begin{table}[htbp]
\centering
\caption{Parameter recovery and calibration across simulation horizons.
Lower normalized root mean squared error (NRMSE) and expected calibration
error (ECE) indicate better recovery and calibration, respectively.
Higher posterior contraction indicates greater reduction in uncertainty
relative to the prior.}
\label{tab:horizon-metrics}
\small
\begin{tabular}{rlrrrr}
\toprule
$T$ (s) & Metric & $w$ & $r$ & $v$ & $\eta$ \\
\midrule
20 & NRMSE       & 0.3712 & 0.0634 & 0.1979 & 0.4574 \\
   & Contraction & 0.8811 & 0.9966 & 0.9605 & 0.7840 \\
   & ECE         & 0.0289 & 0.0126 & 0.0236 & 0.0047 \\
\midrule
40 & NRMSE       & 0.3543 & 0.0602 & 0.2077 & 0.5118 \\
   & Contraction & 0.8844 & 0.9967 & 0.9531 & 0.7280 \\
   & ECE         & 0.0084 & 0.0375 & 0.0208 & 0.0098 \\
\midrule
60 & NRMSE       & 0.3424 & 0.0591 & 0.1963 & 0.4415 \\
   & Contraction & 0.8896 & 0.9967 & 0.9616 & 0.7954 \\
   & ECE         & 0.0153 & 0.0288 & 0.0128 & 0.0082 \\
\midrule
80 & NRMSE       & 0.2660 & 0.0535 & 0.1777 & 0.3934 \\
   & Contraction & 0.9352 & 0.9974 & 0.9659 & 0.8411 \\
   & ECE         & 0.0122 & 0.0326 & 0.0216 & 0.0212 \\
\midrule
100 & NRMSE       & 0.3280 & 0.0577 & 0.2120 & 0.5598 \\
    & Contraction & 0.8930 & 0.9969 & 0.9536 & 0.6868 \\
    & ECE         & 0.0064 & 0.0540 & 0.0204 & 0.0123 \\
\bottomrule
\end{tabular}
\end{table}

For the baseline model, we also compared varying time horizons of $T \in \{20,40,60,80,100\}$ seconds using the transformer summary network and flow-matching estimator. Table~\ref{tab:horizon-metrics} shows no monotonic improvement with increasing duration of observation. The 80-second run achieved the lowest NRMSE and greatest posterior contraction for all four parameters. Extending the window to 100 seconds worsened recovery relative to 80 seconds, particularly for $\eta$, whose NRMSE increased from $0.3934$ to $0.5598$ and contraction decreased from $0.8411$ to $0.6868$. Similar recovery performance is also observed in shorter time horizons. Meanwhile, network calibration shows varying performance between different parameters across the observation windows. These results provide no consistent evidence that longer time horizons alone improve inference. 

\section{Neural Estimator Architecture}
\label{app:architecture}

The neural estimator combines a summary network, which compresses the observed trajectories, with an inference network, which approximates the posterior conditional on that summary. We compare a convolutional bidirectional LSTM summary and a time-series transformer summary \citep{wen2022transformers}, each paired with either flow matching \citep{lipman2023flow} or a diffusion model. This $2\times2$ comparison allows us to justify the choice of the neural estimator architecture.

The recurrent summary comprises two strided convolutional layers with group normalization, a bidirectional LSTM with $256$ units per direction, and a projection to a $12$-dimensional summary. The transformer uses 3 attention blocks with embedding width $64$ and four heads per block, an 8-dimensional Time2Vec embedding, and the same summary dimension. Counts from the saved models are $637{,}900$ parameters for the recurrent summary and $230{,}463$ for the transformer. Both inference networks use a time-conditioned subnet with four hidden layers of width $256$ and time-embedding dimension $32$; each contains $996{,}116$ parameters. Matching subnet size controls capacity, but the flow-matching and diffusion objectives and sampling dynamics differ.

\begin{table}[htbp]
\centering
\caption{Recovery and calibration in the matched architecture comparison. NRMSE uses BayesFlow's prior normalization; ECE denotes coverage-calibration error. Lower NRMSE and ECE are preferable. Posterior contraction measures concentration relative to prior uncertainty and must be interpreted alongside recovery and calibration: a concentrated but miscalibrated posterior is not evidence of successful inference. Values are rounded to three decimals.}
\label{tab:architecture}
\small
\begin{tabular}{lllcccc}
\toprule
Summary & Inference & Metric & $w$ & $r$ & $v$ & $\eta$\\
\midrule
\multirow{3}{*}{Conv+BiLSTM} & \multirow{3}{*}{Flow matching} & NRMSE & 0.707 & 0.644 & 1.092 & 0.838\\
 &  & Contraction & 0.899 & 0.813 & 0.862 & 0.758\\
 &  & ECE & 0.367 & 0.229 & 0.400 & 0.271\\
\midrule
\multirow{3}{*}{Transformer} & \multirow{3}{*}{Flow matching} & NRMSE & 0.332 & 0.736 & 0.242 & 0.431\\
 &  & Contraction & 0.900 & 0.460 & 0.939 & 0.802\\
 &  & ECE & 0.011 & 0.039 & 0.032 & 0.009\\
\midrule
\multirow{3}{*}{Conv+BiLSTM} & \multirow{3}{*}{Diffusion} & NRMSE & 0.698 & 0.895 & 1.337 & 0.873\\
 &  & Contraction & 0.871 & 0.298 & 0.851 & 0.504\\
 &  & ECE & 0.315 & 0.035 & 0.412 & 0.165\\
\midrule
\multirow{3}{*}{Transformer} & \multirow{3}{*}{Diffusion} & NRMSE & 0.371 & 0.768 & 0.291 & 0.484\\
 &  & Contraction & 0.871 & 0.435 & 0.913 & 0.746\\
 &  & ECE & 0.018 & 0.026 & 0.019 & 0.007\\
\bottomrule
\end{tabular}
\end{table}

\begin{figure}[htbp]
\centering
\includegraphics[width=\textwidth]{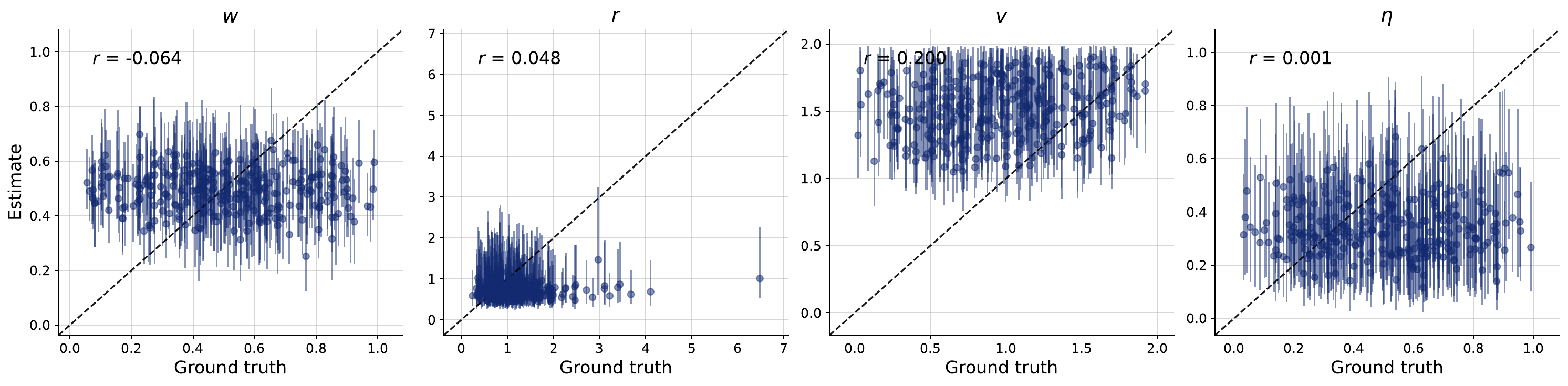}\\[4pt]
\includegraphics[width=\textwidth]{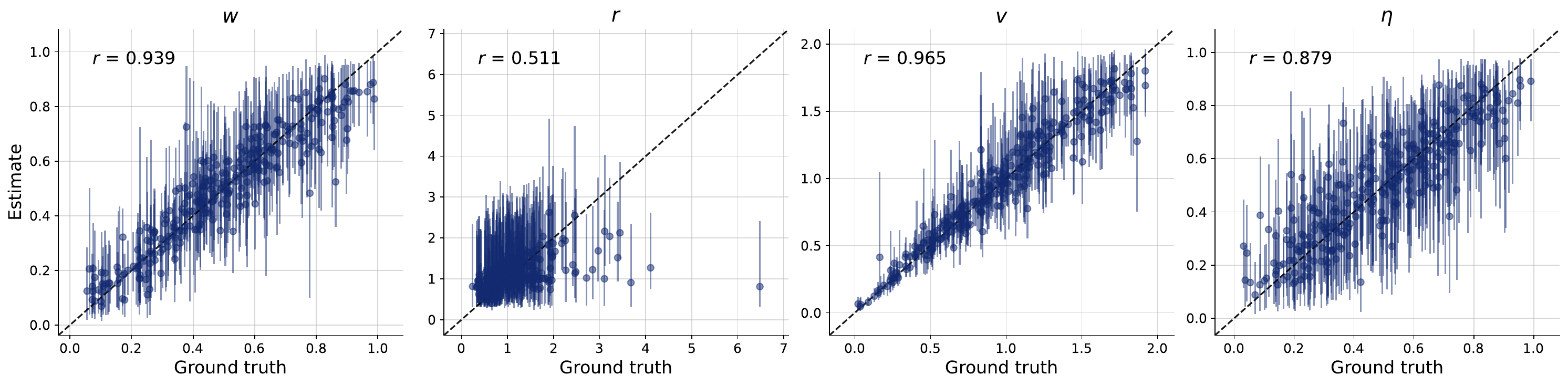}
\caption{\textit{Parameter recovery with flow matching.} Top: convolutional--recurrent summary. Bottom: transformer summary. Dots show posterior medians and vertical lines show $95\%$ equal-tailed credible intervals for the same $300$ held-out datasets. The recurrent configuration's narrow intervals do not compensate for its recovery errors.}
\label{fig:arch-recovery-fm}
\end{figure}

\begin{figure}[htbp]
\centering
\includegraphics[width=\textwidth]{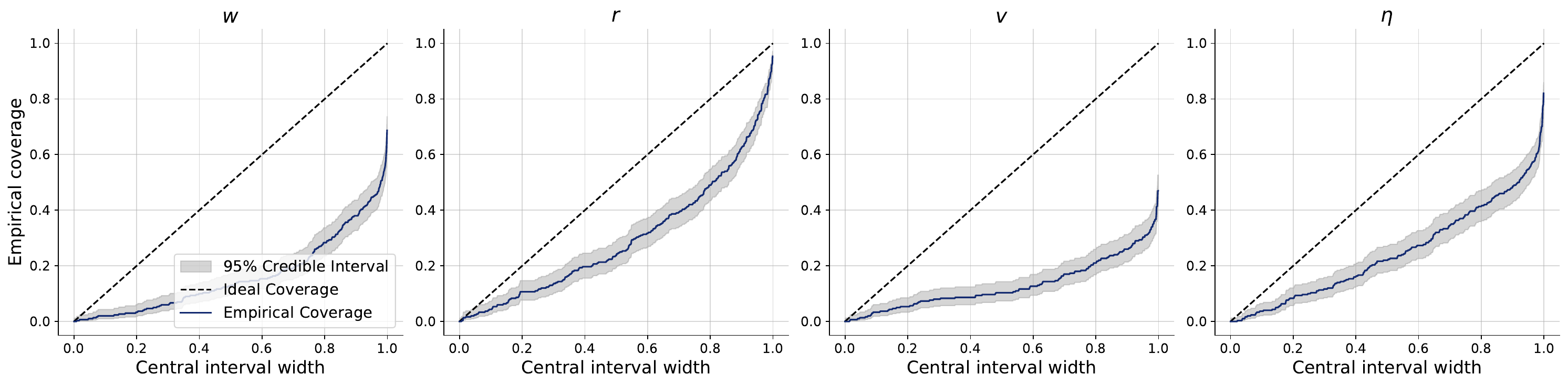}\\[4pt]
\includegraphics[width=\textwidth]{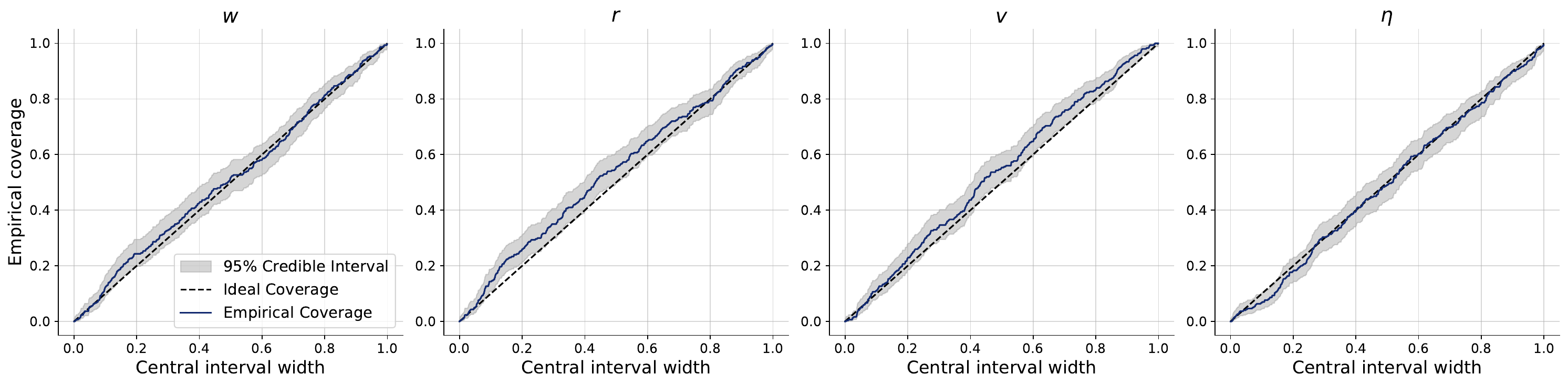}
\caption{\textit{Empirical coverage with flow matching.} Rows correspond to \autoref{fig:arch-recovery-fm}. The recurrent configuration substantially undercovers, while the transformer configuration is much closer to nominal coverage.}
\label{fig:arch-coverage-fm}
\end{figure}

\begin{figure}[htbp]
\centering
\includegraphics[width=\textwidth]{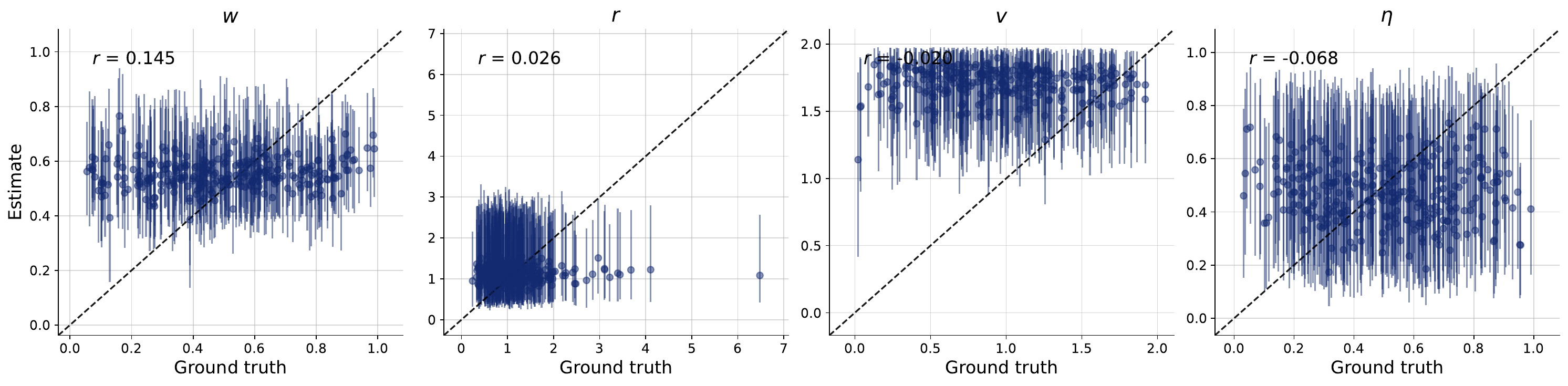}\\[4pt]
\includegraphics[width=\textwidth]{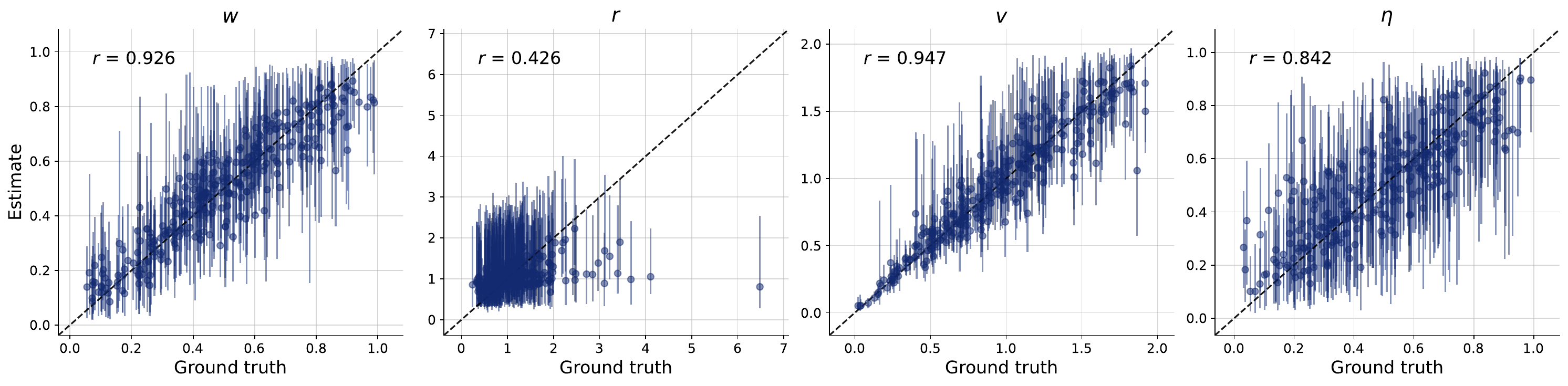}
\caption{\textit{Parameter recovery with diffusion.} Top: convolutional--recurrent summary. Bottom: transformer summary. The transformer improves recovery for all four targets in this configuration; the recurrent configuration remains particularly unreliable for $w$ and $v$.}
\label{fig:arch-recovery-diff}
\end{figure}

\begin{figure}[htbp]
\centering
\includegraphics[width=\textwidth]{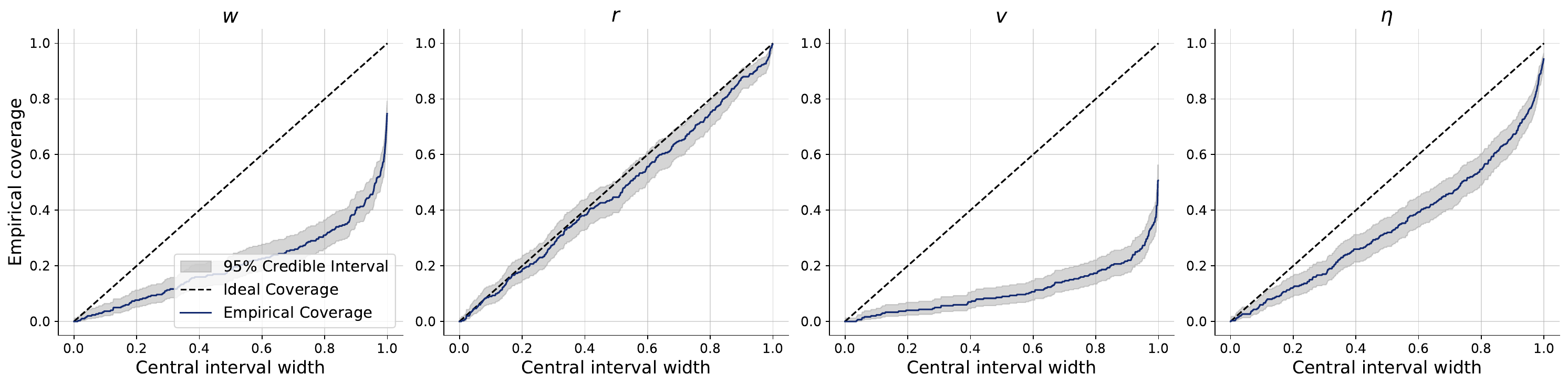}\\[4pt]
\includegraphics[width=\textwidth]{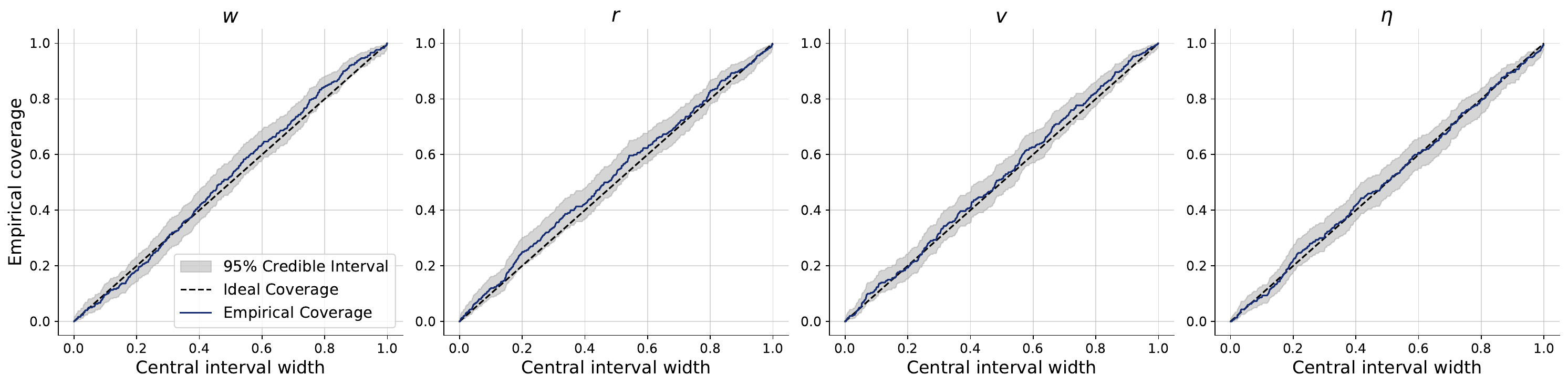}
\caption{\textit{Empirical coverage with diffusion.} Rows correspond to \autoref{fig:arch-recovery-diff}. The recurrent configuration substantially undercovers $w$ and $v$, whereas the transformer configuration is much closer to nominal coverage.}
\label{fig:arch-coverage-diff}
\end{figure}

The transformer gives substantially better recovery of $w$, $v$, and $\eta$ under both inference networks, with substantially improved accuracy compared with the recurrent summary (\autoref{fig:arch-recovery-fm}). In addition, the recurrent networks produces markedly overconfident posteriors when compared to the transformer (\autoref{fig:arch-coverage-fm}). Therefore, the transformer is the preferred summary network for both inference networks.

With the transformer as summary network, flow matching gives better accuracy across all estimation targets than the diffusion model (\autoref{tab:architecture}). Both inference networks are well-calibrated for the estimates. Consequently, flow matching is the preferred inference network when using the transformer summary.

\FloatBarrier
\section{Further detail about the immersive room digital twin}
\label{app:digital-twin}

The digital twin for the immersive room provides the basic rendering framework for the collective virtual navigation in the present work. In Unity, each immersive room is represented by an interactive, full-scale geometry made in 3D modeling tools, and is attached to a built-in first-person (FPS) controller with only horizontal rotation and 2D movement enabled, with a panoramic camera calibrated to the room’s display and the controller's forward direction. Each digital twin $M$ maintains the room’s world position $\mathbf{x}_M$ and rotation $\theta_M$. Nearby virtual sound sources are selected using a room-centered distance detection threshold $R_{M}$, and their relative spatial information is streamed in real time through Open Sound Control (OSC) \citep{wright2005open} to a local soundscape renderer implemented in the Ircam Spat spatial audio rendering tool \citep{carpentier2021spat} using vector-based amplitude panning \citep{pulkki2001spatial}. Beacon distance to the room $||\mathbf{x}_{b, t} - \mathbf{x}_{M, t}||$ and salience $s_{b, t}$ determine loudness and spatial rendering parameters, connecting the virtual environment to the sensory cues available to participants. When available, motion tracking data is streamed from the immersive room into the digital twin, tracking the agents' positions $\mathbf{X}$ and rotation $\boldsymbol{\Theta}$. Agent identity is not represented and recorded. As with any FPS controllers, the digital twin can be placed within any designed or prepared virtual environments as a scene object, and shipped as an executable accordingly.

\end{document}